\documentclass[twocolumn,twocolappendix]{aastex701}
\usepackage{fontspec}

\graphicspath{{./}}

\newcommand\sersic{S\'ersic}

\newcommand\gampen{GaMPEN}
\newcommand\lenstronomy{LENSTRONOMY}
\newcommand\gb{\textit{g}}
\newcommand\rb{\textit{r}}
\newcommand\ib{\textit{i}}

\shorttitle{HSC Multi-Band Morphologies}
\shortauthors{Sarkar et al.}

\usepackage{soul}
\usepackage{enumitem}
\usepackage{bm}
\usepackage{amsmath}
\usepackage{gensymb}
\usepackage{multirow}
\usepackage{array}
\usepackage{hhline}
\usepackage{fvextra} 
\usepackage{float}
\newfloat{listing}{tbp}{lol}
\floatname{listing}{Listing}

\PassOptionsToPackage{hyphens}{url}\usepackage{hyperref}
\hypersetup{hypertexnames=false} 

\usepackage{xcolor} 
\definecolor{LightGray}{gray}{0.95}

\font\bngxi=bang10 scaled 1100

\def\*#1*#2{o\null{#2}{#1}}

\def\d#1{\oalign{\smash{#1}\crcr\hidewidth{$\!$\rm.}\hidewidth}}

\def\sh#1{\setbox0=\hbox{#1}%
     \kern-.02em\copy0\kern-\wd0
     \kern.04em\copy0\kern-\wd0
     \kern-.02em\raise.0433em\box0 }
\newfontfamily\zhfont{FandolSong-Regular}[Extension=.otf, SmallCapsFont=*] 

\begin{document}
\title{Complete Multi-Band Structural Characterization for 10 Million Galaxies in the Hyper Suprime-Cam Wide Survey using Deep Learning}

\author[0009-0007-0050-9762]{Milind Sarkar}
\affiliation{Department of Physical Sciences, Indian Institute of Science Education and Research (IISER), Mohali, Punjab, India}
\email[show]{milindsarkaryt@gmail.com}
\correspondingauthor{Milind Sarkar}

\author[0000-0002-2525-9647]{Aritra Ghosh ({\bngxi Airt/r \*gh*eaSh})}
\altaffiliation{LSST-DA Catalyst Fellow}
\affil{DiRAC Institute and the Department of Astronomy, University of Washington, Seattle, WA, USA}
\affil{eScience Institute, University of Washington, Seattle, WA, USA}
\email{aritraghsh09@gmail.com}

\author[0000-0001-5576-8189]{Andrew Connolly}
\affil{DiRAC Institute and the Department of Astronomy, University of Washington, Seattle, WA, USA}
\affil{eScience Institute, University of Washington, Seattle, WA, USA}
\email{}

\author[0000-0002-0745-9792]{C. Megan Urry}
\affil{Yale Center for Astronomy and Astrophysics, New Haven, CT, USA}
\affil{Department of Physics, Yale University, New Haven, CT, USA}
\email{}

\author{Andrew Coli}
\affil{Department of Statistics and Data Science, Yale University, New Haven, CT, USA}
\email{}

\author[0000-0003-2573-9832]{Joshua S. Speagle ({\zhfont 沈佳士})}
\affiliation{David A. Dunlap Department of Astronomy \& Astrophysics, University of Toronto, 50 St. George St, Toronto, ON M5S 3H4, Canada}
\affiliation{Department of Statistical Sciences, University of Toronto, 9th Floor, 700 University Ave, Toronto, ON M5S 3G3, Canada}
\affiliation{Dunlap Institute for Astronomy \& Astrophysics, University of Toronto, 50 St. George St, Toronto, ON M5S 3H4, Canada}
\affiliation{Data Sciences Institute, University of Toronto, 10th Floor, 700 University Ave, Toronto, ON M7A 2S4, Canada}
\email{}

\author[0000-0003-4442-2750]{Rhythm Shimakawa}
\affil{Waseda Institute for Advanced Study (WIAS), Waseda University, Shinjuku, Tokyo, Japan}
\email{}

\author[0000-0001-5822-5622]{Charles Willard}
\affil{DiRAC Institute and the Department of Astronomy, University of Washington, Seattle, WA, USA}
\email{}

\author[0009-0009-9628-1655]{Diego Lockyer}
\affil{Department of Physics \& Astronomy, San Francisco State University, San Francisco, CA, USA}
\affil{Department of Physics, California Polytechnic State University San Luis Obispo, San Luis Obispo, CA, USA}
\email{}

\begin{abstract}

Galaxy structure is wavelength-dependent, as different photometric bands trace distinct stellar populations and structural components. We present per-galaxy Bayesian posterior distributions of the bulge-to-total light ratio ($L_B/L_T$), effective radius ($R_e$), and flux ($F$) in the $g$, $r$, and $i$ bands for nearly $10$ million galaxies across $\sim$1400\,deg$^2$ of the Hyper Suprime-Cam Subaru Strategic Program (HSC-SSP) Wide survey at $z \leq 0.75$ and $m < 23$. To our knowledge, this is the largest multi-band catalog of bulge--disk structural parameters with per-galaxy posteriors. $\sim$3.9 million of the above galaxies have full multi-band coverage. Measurements are generated using the Galaxy Morphology Posterior Estimation Network (\gampen{}), a Bayesian deep-learning framework. We demonstrate that we can adapt \gampen{} from previously-trained single-band models to the new bands by using transfer learning on only $\sim$0.5\% of our dataset. Despite this extremely small training set, \gampen{} successfully reproduces light-profile fitting measurements and its output posteriors are extremely well-calibrated with $\lesssim6\%$ mean deviation. This approach offers a scalable pathway to structural parameter measurements with robust uncertainties for Rubin, Euclid, and Roman data, which we will pursue in forthcoming work. While structural measurements remain robust across most bands and redshift ranges, performance degrades significantly for the $g$-band at $z \gtrsim 0.5$, where it increasingly probes rest-frame ultraviolet emission; these measurements are therefore excluded from the public release. This public catalog will enable large-scale statistical studies of morphological color gradients, wavelength-dependent galaxy structure, and galaxy evolution.

\end{abstract}


\section{Introduction} \label{sec:intro}

The structure of galaxies is closely connected to many of their fundamental physical properties, including stellar mass, star-formation activity, stellar kinematics, merger history, environment, and the growth of supermassive black holes \citep[e.g.,][]{Bender1992DynamicallyProperties,Tremaine2002TheCorrelation,pozzetti_10,wuyts_11,Huertas-Company2016MassCANDELS,powell_17,shimakawa_2021,Dimauro2022CoincidenceGrowth, Ghosh_2024_env, igel2026environmentalsculptinggalaxystructure}. Consequently, quantitative measurements of galaxy structure across large cosmological volumes have become a cornerstone of modern studies of galaxy formation and evolution. Structural parameters such as galaxy size, concentration, bulge prominence, and luminosity provide powerful constraints on competing evolutionary scenarios and, when combined with other observables, can reveal the physical processes responsible for shaping galaxies over cosmic time \citep[e.g.,][]{Kauffmann2004TheGalaxies,Weinmann2006PropertiesMass,Schawinski2007TheGalaxies,vanderWel2008TheMass,schawinski_14}. Robust measurements of these quantities, together with realistic estimates of their uncertainties, are particularly important for deriving scaling relations and for confronting theoretical models with observations \citep[e.g.,][]{Bernardi2013TheProfile,vanderWel20143D-HST+CANDELS:3, Ghosh_2024_env}.

While galaxy structure is often treated as a single observable, it is fundamentally wavelength-dependent. Different photometric bands trace different stellar populations and are affected differently by dust attenuation, causing the apparent structure of a galaxy to vary with wavelength \citep[e.g.,][]{giavalisco1996,papovich2003}. Numerous studies have shown that structural parameters such as effective radius and \sersic{} index exhibit systematic trends across the optical and near-infrared spectrum, reflecting underlying gradients in stellar age, metallicity, and star-formation activity \citep[e.g.,][]{kelvin_12_gama,vulcani2014_gama,kennedy2016_gama}. Bulges generally become more prominent at longer wavelengths, while shorter wavelengths increasingly emphasize young star-forming regions and clumpy structures. As a result, measurements obtained in a single photometric band provide only a partial description of galaxy structure. Multi-wavelength structural measurements are therefore essential for understanding the connection between stellar populations and galaxy morphology, quantifying morphological color gradients, and interpreting the effects of morphological $k$-corrections at higher redshift.

Traditional approaches to quantifying galaxy structure rely on fitting analytic light profiles to imaging data \citep[e.g.,][]{vdw_12,tarsitano_18}. Although such methods have been enormously successful, their computational cost becomes prohibitive for the hundreds of millions of galaxies expected from next-generation surveys. Furthermore, extending profile fitting consistently across multiple photometric bands often requires either independent fits in each filter or sophisticated multi-band fitting frameworks such as MegaMorph \citep[e.g.,][]{haubler_2013_megamorph,vika2013_megamorph}. Such approaches have produced multi-wavelength structural catalogs for a range of surveys. Catalogs covering large areas have typically been restricted to brighter or more nearby galaxies than our sample. These include bulge+disk decompositions of 1.12 million Sloan Digital Sky Survey (SDSS) galaxies at $r\leq18$ \citep{simard_11}, and single-\sersic{} fits for $\sim$25 million galaxies from the Dark Energy Survey (DES) that are reliable to $i\lesssim21.5$ \citep{tarsitano_18}. Other catalogs that reach greater depth or distance have typically been restricted to at most tens of square degrees \citep[e.g.,][]{nedkova_2024,quilley_2025_euclid_q1, merchant_2026_canucs, yang_2026_cosmosweb}. To our knowledge, the catalog presented here is the first to provide multi-band bulge+disk decomposition with per-galaxy posteriors over a wide area ($\sim$1400\,deg$^{2}$), down to $m < 23$ and out to $z \sim 0.75$.


Over the last decade, machine-learning (ML) techniques have emerged as a powerful alternative for extracting information from large astronomical datasets \citep[e.g.,][]{ml_pz,ml_sz,Shallue2018IdentifyingKepler-90,Sharma2020ApplicationClassification,Natarajan2021QuasarNet:Holes}. See \citet{Huertas-Company2023TheSurveys, ting_2026_review} for recent reviews. Unsurprisingly, they have also become increasingly popular for determining galaxy morphology \citep[e.g.,][]{Dieleman2015Rotation-invariantPrediction,huertas_company2015,Tuccillo2018DeepFitting,gamornet_paper,Hausen2020MorpheusData,Walmsley2020GalaxyLearning,Cheng2021GalaxyNetworks,Vega-Ferrero2021PushingSurvey,tanoglidis_2022_lsbg_bnn, Tarsitano2022ImageLearning, walmsley_2023_gz_desi}. However, most of these previous ML-based studies have focused on broad morphological classifications or point estimates of structural parameters, often without providing calibrated uncertainties. 
In \citet{gampen_software_paper}, we introduced the Galaxy Morphology Posterior Estimation Network (\gampen{}), a Bayesian deep-learning framework designed to predict full posterior distributions for a galaxy's bulge-to-total light ratio ($L_B/L_T$), effective radius ($R_e$), and flux ($F$). 
What distinguishes \gampen{} from other approaches is that it predicts a joint posterior over all three parameters, with their full covariance matrix included in the loss function, and that the calibration of these posteriors has been validated with coverage tests on both simulated and real data \citep[][]{gampen_software_paper, Morpho_Single_Band}. This enables rigorous statistical analyses of galaxy structure at scale.

Building upon this framework, \citet{Morpho_Single_Band} used \gampen{} to provide Bayesian structural measurements for approximately eight million galaxies in the Hyper Suprime-Cam (HSC) Subaru Strategic Program (HSC-SSP) Wide survey. To ensure that structure was measured consistently across redshift, that study selected a single observed band within each redshift interval such that all measurements approximately corresponded to rest-frame $g$-band. 

In this work, we extend the framework of \citet{Morpho_Single_Band} to generate homogeneous estimates of $L_B/L_T$, $R_e$, and $F$ in the HSC $g$, $r$, and $i$ bands over the redshift range $z\leq0.75$. Combined with the catalog of \citet{Morpho_Single_Band}, this provides what is, to our knowledge, the largest publicly available multi-band catalog of bulge--disk structural parameters with calibrated per-galaxy posterior distributions, complementing the catalogs described previously. 

A key challenge in constructing such a catalog is the lack of large training sets with pre-existing morphological measurements in every band. In this work, we demonstrate that transfer learning can efficiently overcome this limitation. Starting from the trained \gampen{} models of \citet{Morpho_Single_Band}, we adapt them to new photometric bands using only a small subset of galaxies with structural parameters measured through traditional light-profile fitting. This approach allows us to generate Bayesian posterior distributions for nearly ten million galaxies while requiring detailed morphological measurements for only a tiny fraction of the total sample. More broadly, it provides a practical pathway for extending morphology frameworks to future imaging surveys such as the NSF-DOE Vera C. Rubin Observatory’s Legacy Survey of Space \& Time \citep[Rubin-LSST;][]{lsst}, the Euclid mission \citep{laureijs2011eucliddefinitionstudyreport,euclid}, the Nancy Grace
Roman Space Telescope \citep[NGRST;][]{ngrst,akeson2019widefieldinfraredsurvey}, and the
Chinese Space Station Survey Telescope \citep[CSST;][]{csst}. 

In this paper, we present structural measurements for $\sim$9.6 million unique HSC galaxies, combining new measurements in the bands not analyzed by \citet{Morpho_Single_Band} with those of that work. Of these, $\sim$3.9 million are measured in every band released for their redshift (\gb{}, \rb{}, and \ib{} at $z \leq 0.5$; \rb{} and \ib{} at $0.5 < z \leq 0.75$), enabling direct studies of wavelength-dependent bulge+disk structure. We describe the HSC data and sample selection in \S\ref{sec:data}, the \gampen{} framework and our transfer-learning strategy in \S\ref{sec:method}, and the light-profile fitting used to generate training and validation data in \S\ref{sec:galfitting}. In \S\ref{sec:results}, we evaluate the performance of the resulting models and the calibration of their posteriors, and in \S\ref{sec:compare} we compare our predictions with two independent external catalogs. We summarize our results in \S\ref{sec:discussions}.

\section{Data} \label{sec:data}


The primary goal of this study is to apply \gampen{} to \gb, \rb, and \ib-band images from the HSC-SSP Public Data Release 2 \citep[PDR2;][]{hsc_pdr2} at $z \leq 0.75$, focusing on bands not analyzed in \citet{Morpho_Single_Band}. The Subaru Strategic Program, begun in 2014, employs the Hyper Suprime-Cam prime-focus camera on the 8.2 m Subaru Telescope, whose combination of wide field of view, high spatial resolution, and sensitivity makes it ideally suited for obtaining the high-quality imaging required for this study.

\begin{figure}[htb]
    \centering
    \includegraphics[width=\columnwidth]{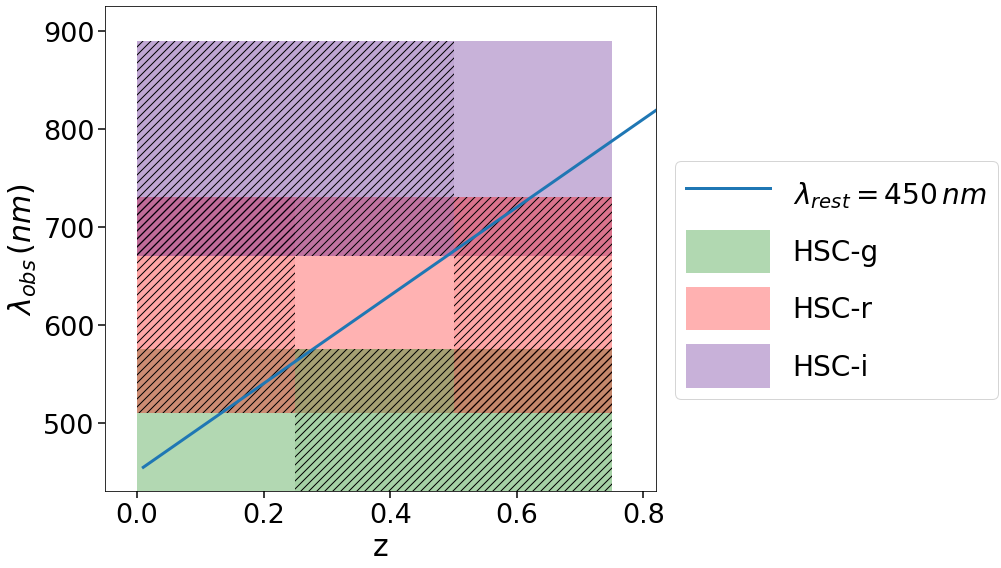}
    \caption{For each redshift bin, we indicate the corresponding effective wavelength coverage of the HSC filters (\gb, \rb, and \ib). The diagonal blue line traces the observed position of rest-frame 450 nm emission as a function of redshift. The striped regions mark the additional filter coverage adopted in this study, while the solid shaded regions show the filters already used in \citet{Morpho_Single_Band}. Whereas \citet{Morpho_Single_Band} restrict their analysis to specific filters to ensure morphology is consistently measured in the rest-frame \gb-band, our study employs all available filters over the entire redshift range to extract the full morphological color information.}
    \label{fig:hsc_response}
\end{figure}

To construct a large yet homogeneous sample of HSC PDR2 galaxies, we base our analysis on the Wide layer, which spans $\sim1400$ deg$^2$ to full survey depth across all filters and includes over 450 million primary sources. To maintain consistency in rest-frame \gb-band morphology measurements across redshift, \citet{Morpho_Single_Band} adopted a filter combination that varies with redshift (see Figure~\ref{fig:hsc_response}): \gb-band for $z\leq0.25$, \rb-band for $0.25<z\leq0.50$, and \ib-band for $0.50<z\leq0.75$.
Our study fills in the missing color coverage in each redshift bin: \rb\ and \ib\ for $z \leq 0.25$, \gb\ and \ib\ for $0.25 < z \leq 0.50$, and \gb\ and \rb\ for $0.50 < z \leq 0.75$. When combined with \citet{Morpho_Single_Band}, this allows us to extract the complete morphological color information for galaxies across the entire redshift range considered.
Given the typical HSC seeing, galaxies with physical sizes $\lesssim 5$ kpc are unresolved beyond $z \sim 0.75$. As light profiles at higher redshift are dominated by the point spread function (PSF), we therefore restrict our analysis to $z \leq 0.75$.

Our galaxy sample is drawn from the PDR2 catalog, which is based on forced photometry performed on the coadded images. The HSC data were processed with a reduction pipeline derived from the prototype pipeline of the Rubin Observatory LSST Data Management system; we refer the reader to \citet{hsc_pipeline} for further details. To separate galaxies from stars, we use the \texttt{extendedness\_value} flag, which compares the Composite Model (CModel) and PSF magnitudes. This classifier becomes increasingly contaminated by stars at $m > 23$ under typical Wide-layer seeing conditions. We therefore restrict our sample in each redshift bin to galaxies with $m < 23$ in the relevant filter.

\begin{deluxetable}{cccccc}[htbp]
\tabletypesize{\footnotesize}
\setlength{\tabcolsep}{3pt}
\tablecaption{Data Characteristics  \label{tab:data}}
\tablecolumns{6}
\tablehead{
\colhead{Sample} & \colhead{Imaging} & \colhead{Redshift} & \colhead{Number} & \colhead{Common} & \colhead{Released} \\
\colhead{Name}  & \colhead{Band} & \colhead{} & \colhead{of Galaxies} & \colhead{Galaxies} & \colhead{Common} }
\startdata
    \hline
    \hline
    Low-z  & \gb{} & $z\leq0.25$ & 820,566 \\
    Low-z  & \rb{} & $z\leq0.25$ & 895,838  \\
    Low-z  & \ib{} & $z\leq0.25$ & 898,354 & 482,096 & 482,096 \\
    Mid-z  & \gb{} & $0.25 < z\leq0.50$ & 1,696,071  \\
    Mid-z  & \rb{} & $0.25 < z\leq0.50$ & 2,937,871 \\
    Mid-z  & \ib{} & $0.25 < z\leq0.50$ & 2,894,764 & 1,140,835 & 1,140,835 \\
    High-z & \gb{} & $0.50 < z\leq0.75$ & 883,532$^{a}$ \\
    High-z & \rb{} & $0.50 < z\leq0.75$ & 2,867,732 \\
    High-z & \ib{} & $0.50 < z\leq0.75$ & 4,247,208 & 614,369 & 2,322,117$^{b}$ \\
    \hline
    \hline
    Total & & $z\leq0.75$ &  9,587,923 & 2,237,300 & 3,945,048$^{c}$ \\
\enddata
\tablecomments{The “Number of Galaxies” column gives the number of galaxies in each band and redshift bin (equivalently, the number of galaxy images in that band). Summed across all bands, this amounts to 18,141,936 galaxy images, while the total number of unique galaxies is 9,587,923 since many appear in multiple bands. The “Common Galaxies” column lists the number of galaxies with imaging available in all three bands within a given redshift range. The ``Released Common'' column lists the corresponding numbers in the public catalog. $^{a}$Excluded from the public release (\S\ref{sec:gampen_v_galfit}). $^{b}$High-$z$ galaxies are released in the \rb{} and \ib{} bands only; this is the number with both. $^{c}$Released galaxies measured in every band released for their redshift bin: 1,622,931 with all three bands (low- and mid-$z$) and 2,322,117 with the \rb{} and \ib{} bands (high-$z$). Galaxies measured in fewer bands are also included in the public catalog. Of the 18,141,936 measurements, 8,005,645 (low-$z$ \gb, mid-$z$ \rb, and high-$z$ \ib) are from \citet{Morpho_Single_Band}; the remaining 10,136,291 are new to this work, of which 9,252,759 are released.}
\end{deluxetable}

\begin{figure}[htb]
    \centering
    \includegraphics[width=\columnwidth]{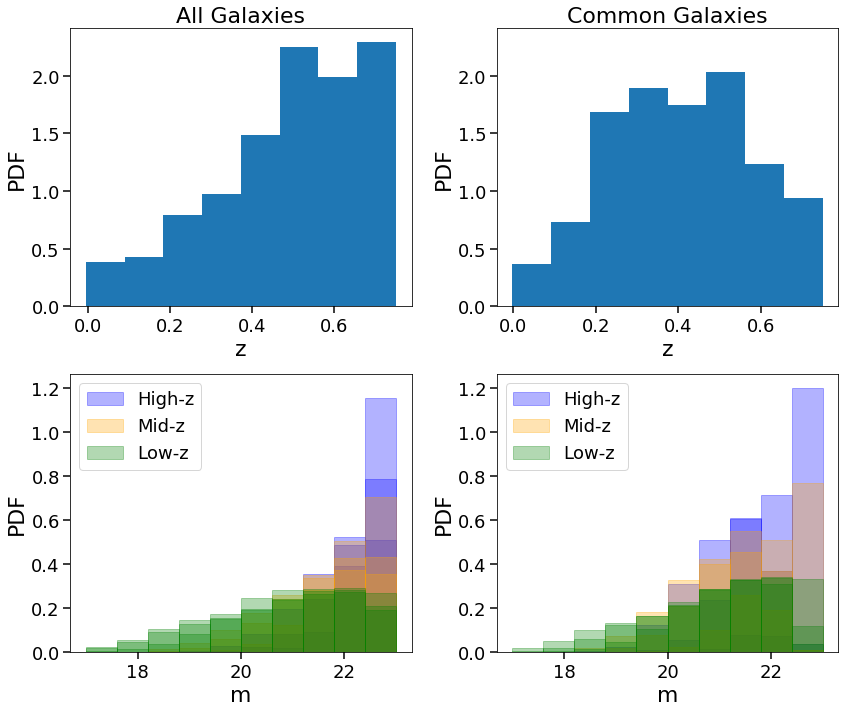}
    \caption{Redshift (top) and magnitude (bottom) distributions of the galaxy samples. The left panels show the full set of $\sim 9.6$ million unique galaxies from this work combined with \citet{Morpho_Single_Band}, while the right panels show the $\sim$2.2 million galaxies
    with coverage in all three HSC bands ($g$, $r$, and $i$) at $z \leq 0.75$, referred to as the ``common galaxies'' sample; of these, $\sim$1.6 million (the low- and mid-$z$ galaxies) are in the public release, since the high-$z$ \gb-band measurements are excluded. Spectroscopic redshifts are used where available, and high-quality photometric redshifts otherwise. A full summary of the datasets used in this study is provided in Table~\ref{tab:data}.}
    \label{fig:z_mag_distr}
\end{figure}

Approximately $3.1\%$ of our galaxies have spectroscopic redshifts, with the remainder assigned high-quality photometric redshifts. The spectroscopic compilation assembled by the HSC-SSP team draws on numerous surveys. Further details are provided in \citet{photoz_hsc_pdr2}. Building on this reference set, the HSC PDR2 photometric redshifts were produced with two independent methods, and we use the results from Mizuki \citep{mizuki}, which employs Bayesian template fitting.
To retain only secure estimates, we required \texttt{photoz\_risk\_best} $< 0.1$, a parameter that measures the likelihood of a redshift falling outside $z_{\mathrm{true}} \pm 0.15(1+z_{\mathrm{true}})$. In addition, galaxies flagged with \texttt{cleanflags\_any} were excluded to avoid imaging-related systematics. We verified that these selection criteria do not significantly alter the overall redshift or magnitude distributions of the sample.

\begin{figure*}[htb]
    \centering
    \includegraphics[width = 0.75\textwidth]{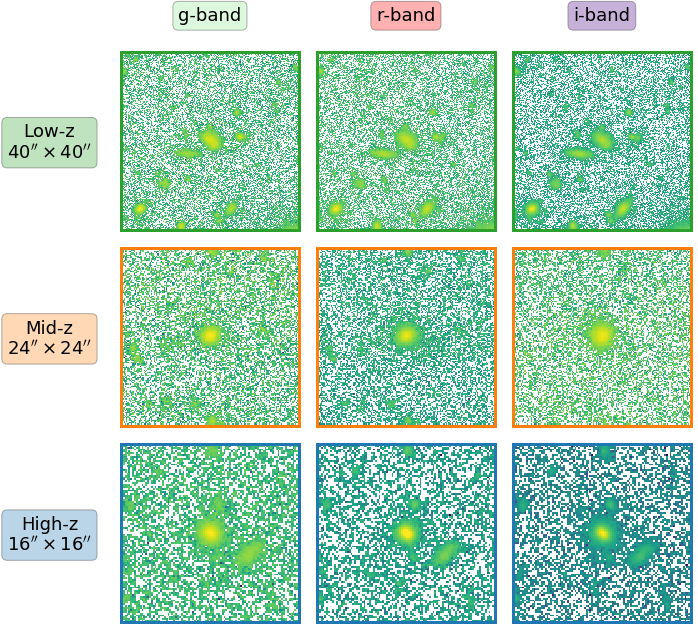}
    \caption{For each redshift bin, we show cutouts of a single randomly selected galaxy that is imaged in all three bands (\gb, \rb, and \ib), with the target object centered in each frame. These examples illustrate that many cutouts include secondary sources, which can lead machine-learning models to produce spurious classifications. In \gampen{}, this issue is mitigated by the Spatial Transformer Network, which automatically crops out most secondary objects
    (see \S\ref{sec:gampen}).}
    \label{fig:real_data_eg}
\end{figure*}

Our selection yields $\sim$820k galaxies in the \gb-band, $\sim$895k in the \rb-band, and $\sim$898k in the \ib-band within the low-$z$ bin. In the mid-$z$ bin, the numbers rise to $\sim$1.69 million (\gb-band), $\sim$2.93 million (\rb-band), and $\sim$2.89 million (\ib-band), while the high-$z$ bin contains $\sim$883k (\gb-band), $\sim$2.86 million (\rb-band), and $\sim$4.24 million (\ib-band). The overlap across all three filters corresponds to $\sim$482k galaxies in the low-$z$, $\sim$1.14 million in the mid-$z$, and $\sim$614k in the high-$z$ bin (see Table~\ref{tab:data}). 
The corresponding redshift and magnitude distributions are shown in Figure~\ref{fig:z_mag_distr}. Galaxy image cutouts were obtained using the \href{https://hsc-release.mtk.nao.ac.jp/das_cutout/pdr2/}{HSC Image Cutout Service}\footnote{\href{https://hsc-release.mtk.nao.ac.jp/das_cutout/pdr2/}{https://hsc-release.mtk.nao.ac.jp/das\_cutout/pdr2/}}, with sizes of $\sim40\arcsec$, $\sim24\arcsec$, and $\sim16\arcsec$ for the low-, mid-, and high-$z$ bins, respectively. These sizes are sufficiently large to encompass the targets in each bin. Based on light-profile fitting tests performed on a representative sub-sample (§\ref{sec:galfitting}), we estimate that $\sim99.5\%$ of galaxies have cutouts at least ten times larger than their major axis. Examples of galaxy cutouts in each band from each redshift bin are displayed in Figure~\ref{fig:real_data_eg}.

\section{Cross-Band Training with \gampen{}} \label{sec:method}

Since most of the galaxies in our sample (\S\ref{sec:data}) lack prior morphological measurements, we adopt a transfer-learning strategy that minimizes the need for large training sets. Instead of training on simulated galaxies as in \citet{Morpho_Single_Band}, we make use of the trained models provided in that work. For each redshift bin, we begin with the model trained on the band used by \citet{Morpho_Single_Band}---for instance, the \gb-band model at $z \leq 0.25$---and fine-tune it to the remaining bands in the same redshift interval. A similar procedure is carried out in the $0.25 < z \leq 0.50$ and $0.50 < z \leq 0.75$ ranges, where the available trained models are adapted to the bands not analyzed previously. Fine-tuning in every case is performed using a small subset of real galaxies with morphologies measured using GALFIT \citep{galfit}. Another subset of GALFIT-ed galaxies is used to validate and characterize the performance of our band-specific fine-tuned models.

\begin{figure*}[htb]
    \centering
    \includegraphics[width=0.5\linewidth, trim=2 0 3 0, clip]{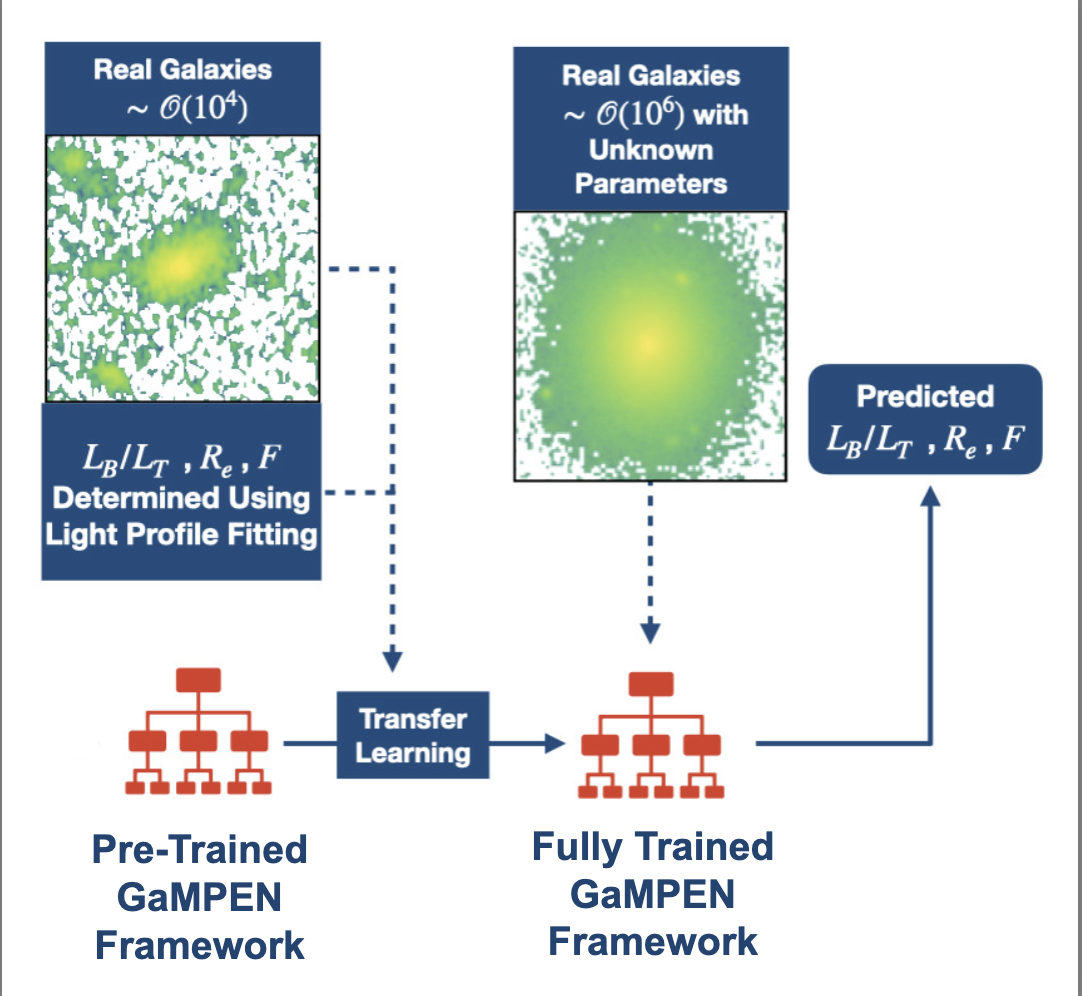}
    \caption{Diagram outlining the different stages of training \gampen{}. We begin by using pretrained models from \citet{Morpho_Single_Band}. Thereafter, we fine-tune each pretrained model for the new bands within the same redshift slice (see \S \ref{sec:data} for band information), using $0.5\%$ of our real data sample. For these galaxies used in fine-tuning, we pre-determine their morphological parameters using light-profile fitting, as described in \S \ref{sec:galfitting}. Finally, we process all the remaining galaxies in our dataset through the trained \gampen{} framework to obtain estimates of their morphological parameters and associated uncertainties.}

    \label{fig:full_gampen_workflow}
\end{figure*}

The process of training \gampen{} in this work proceeds as follows, and is summarized in Figure \ref{fig:full_gampen_workflow}:

\begin{itemize}
    \item Start from the trained models of \citet{Morpho_Single_Band} for the available band in each redshift bin.
    \item Apply transfer learning to adapt these models to the remaining bands within the same redshift bin, using a small fraction of real galaxies ($\sim$75,000 in total, all bands combined, of which $\sim$53,000 are used for training and $\sim$11,000 for validation; the remaining $\sim$11,000 form the test set below) with GALFIT-derived morphological parameters.
    \item Validate the performance of the fine-tuned models on an independent test set of real galaxies ($\sim$11,000 for all bands combined) with GALFIT-derived parameters.
    \item Apply the trained models to the remainder of the dataset ($\sim$99\%) to generate posterior predictions.
\end{itemize}

We describe each of these steps in detail below.

\subsection{Brief Introduction to \gampen{}} \label{sec:gampen}

\gampen{} \citep{gampen_software_paper} is a Bayesian deep-learning framework designed to infer posterior distributions of galaxy structural parameters. In this work, we use \gampen{} to predict the bulge-to-total light ratio, effective radius, and flux. Since the architecture, training procedure, and uncertainty formalism of \gampen{} have been described in detail by \citet{gampen_software_paper} and its application to HSC galaxy morphology was presented in \citet{Morpho_Single_Band}, we provide only a brief overview here and refer the reader to those works and Appendix~\ref{ap:gampen} for a complete technical description.

\begin{figure*}[htb]
    \centering
    \includegraphics[width=\textwidth]{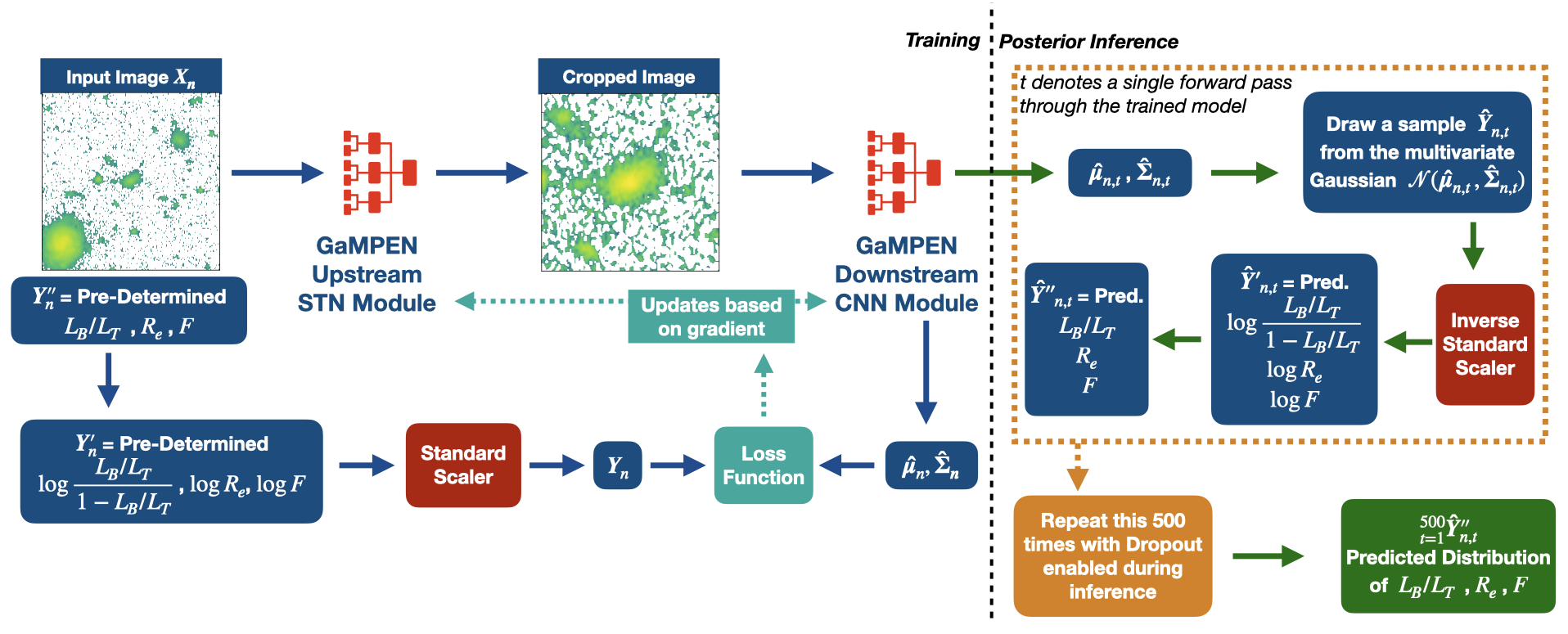}
    \caption{Overview of the \gampen{} workflow. \textit{Left:} training phase, where galaxies with known parameter values are processed by the STN and CNN, and the loss is minimized with stochastic gradient descent. \textit{Right:} inference phase, where trained networks generate posterior predictions. During inference, variables are rescaled back into their physical units. Posterior estimates are obtained by running 500 forward passes with dropout enabled, each yielding a draw from the predicted multivariate Gaussian. The ensemble of draws forms the posterior distribution for each galaxy. Figure adapted from \citet{Morpho_Single_Band}.}
    \label{fig:posterior_pred_workflow}
\end{figure*}

\gampen{} consists of an upstream Spatial Transformer Network \citep[STN;][]{jaderberg_2015_stn} followed by a Convolutional Neural Network \citep[CNN;][]{lecun_1998_cnn}. The STN automatically crops and centers the target galaxy through a learned affine transformation, thereby reducing preprocessing requirements and mitigating contamination from nearby sources. 
The transformed image is subsequently passed to the CNN, which predicts a multivariate Gaussian distribution describing the galaxy's structural parameters. The entire procedure used to estimate posteriors is summarized in Figure~\ref{fig:posterior_pred_workflow}.

The framework accounts for both aleatoric and epistemic uncertainties. Aleatoric uncertainties, arising from limitations in the imaging data such as noise and PSF effects, are encoded in the covariance matrix of the predicted multivariate Gaussian and learned during training through minimization of the negative log-likelihood \citep{gampen_software_paper}. Epistemic uncertainties, associated with limitations of the model itself, are estimated using Monte Carlo dropout \citep{gal_2016}, which reinterprets dropout regularization \citep{Srivastava2014Dropout:Overfitting} as approximate Bayesian inference. During inference, dropout remains active and each galaxy is processed 500 times, producing an ensemble of predictions whose variation captures epistemic uncertainty. Together, these components yield the full posterior distributions used throughout this work. 

\subsection{Data Transformations} \label{sec:transformations}

Following \citet{gampen_software_paper, Morpho_Single_Band}, we apply a series of transformations to both the input images and target variables to improve numerical stability during training. Input images are passed through an $\operatorname{arsinh}$ transformation, which reduces the dynamic range of pixel values and has been shown to improve neural network convergence \citep[e.g.,][]{zanisi_21, walmsley_decals, tanaka_22}.

The target variables are transformed using a logit function for $L_B/L_T$ and logarithmic functions for $R_e$ and $F$:

\begin{equation}
\boldsymbol{Y_n'} = f''(\boldsymbol{Y}_n'') = \left( \log \frac{L_B/L_T}{1 - L_B/L_T}, \log R_e, \log F \right),
\label{eq:transformation_f''}
\end{equation}

\noindent
where $\boldsymbol{Y}_n''=[L_B/L_T,R_e,F]$ denotes the untransformed parameter vector. Each transformed parameter is subsequently standardized using the mean and variance of the training set. During inference, the standardization and parameter transformations are reversed (Figure~\ref{fig:posterior_pred_workflow}), restoring the parameters to their physical units and enforcing the constraints $0\leq L_B/L_T\leq1$, $R_e>0$, and $F>0$. 

\subsection{Cross-Band Transfer Learning using Real Data} \label{sec:tl_sample}

Transfer learning has been a standard concept in data science since the late 1990s \citep[e.g.,][]{blum_1998}. The idea is to begin with a network trained for a specific dataset and task, and then fine-tune its weights and biases for a related but distinct problem. In \citet{gamornet_paper, Morpho_Single_Band}, we showed how this approach can be applied to galaxy morphology/structure determination by first training on simulated light profiles and then adapting the network with a small amount of real data. In this study, rather than repeating that procedure, we take advantage of the pretrained models released in \citet{Morpho_Single_Band}, since we are using the same survey as that work. For each redshift bin, we begin with the model trained on the band used in that work---for example, the \gb-band model at $z \leq 0.25$---and adapt it to the other filters in the same redshift slice (e.g., the \rb{} and \ib{} bands for $z \leq 0.25$). The same strategy is applied at higher redshifts, where the available pretrained models are fine-tuned to the bands not previously analyzed. For this fine-tuning process, we use $\sim$14,000 real HSC-Wide galaxies for each band in each redshift bin ($\sim$6,000 for the high-$z$ \gb-band; see below). 

\begin{figure*}[htb]
    \centering
    \includegraphics[width = 0.9\textwidth]{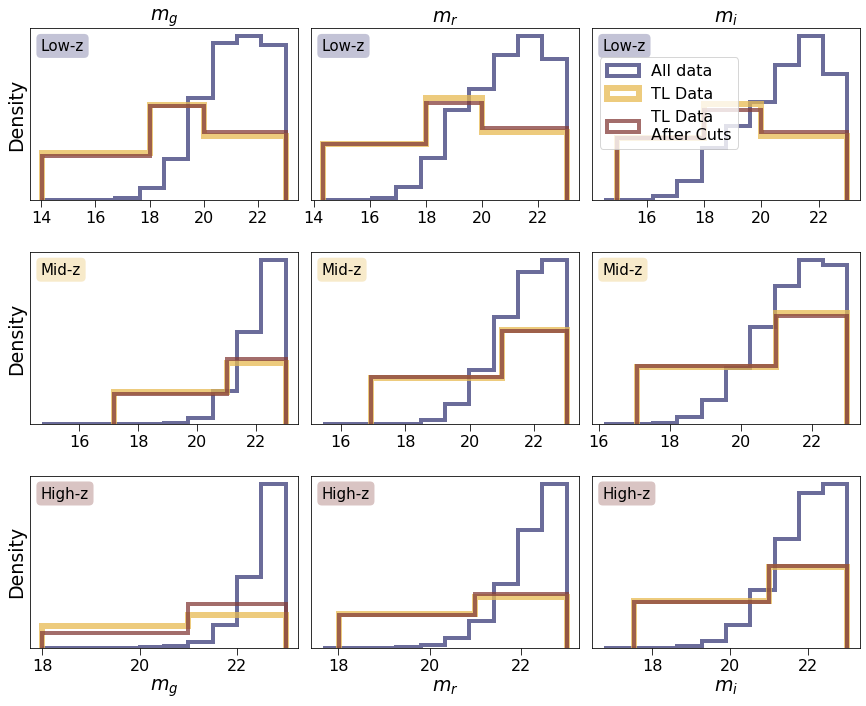}
    \caption{Magnitude distributions for galaxies in each band and redshift bin, shown together with the subsets used for transfer learning (both before and after applying the quality cuts described in \S\ref{sec:galfitting}). The first column corresponds to the \gb-band across the different redshift ranges, the second to the \rb-band, and the third to the \ib-band. The y-axis shows density rather than raw counts. Overall, the transfer-learning sample represents $\sim$0.5\% of the full dataset. Some magnitude bins appear denser than others (e.g., $18 < m \leq 20$ in the low-$z$ bin) because they span narrower ranges while containing a similar number of galaxies.}
    \label{fig:tl_para_dist}
\end{figure*}

To construct a sample for transfer learning, we begin with the galaxies summarized in Table~\ref{tab:data}. When training a CNN, the critical requirement is not reproducing the observed parameter distributions, but rather ensuring that the full parameter space is covered with sufficient fidelity. Over-representing certain galaxy types---even if that reflects the real data---can reduce performance on rarer populations \citep[e.g.,][]{gamornet_paper}. As expected and shown in Figure~\ref{fig:z_mag_distr}, the distributions in all three redshift bins are strongly skewed toward fainter galaxies. To guarantee adequate coverage of the bright population, we divide the mid- and high-$z$ samples into two bins ($m \leq 21$ and $m > 21$), and the low-$z$ sample into three bins ($m \leq 18$, $18 < m \leq 20$, and $m > 20$). From these, we select $\sim$20,000 galaxies for each band in each redshift bin ($\sim$13,000 for the high-$z$ \gb-band; $\sim$113,000 in total), sampling equally across the magnitude intervals. Morphological parameters for these galaxies are then derived from light-profile fitting, as described in \S\ref{sec:galfitting}. After removing unreliable fits with the quality cuts listed in Table~\ref{tab:galfit_quality_cuts}, $\sim$14,000 galaxies remain for each band in each redshift bin ($\sim$6,000 for the high-$z$ \gb-band), or $\sim$75,000 in total. We refer to this as the transfer-learning sample. The resulting magnitude distributions are shown in Figure~\ref{fig:tl_para_dist}, demonstrating that the transfer-learning sample covers the relevant parameter space well and thus provides \gampen{} with sufficient examples across the full range of galaxy morphologies. For completeness, Figure~\ref{fig:tl_para_dist} also shows the distributions of the transfer-learning sample before the quality cuts were applied. 

Of the $\sim$14,000 galaxies per band per redshift bin ($\sim$6,000 for the high-$z$ \gb-band), 70\% are used for fine-tuning the pretrained \gampen{} models, 15\% for hyperparameter selection, and the remaining 15\% for testing the performance of the trained networks (\S\ref{sec:results}). To expand the effective transfer-learning sample, we augment the data by applying random rotations and horizontal/vertical flips. This increases the transfer-learning sample by a factor of six, from $\sim$14,000 to an effective $\sim$84,000 galaxies per band per redshift bin (from $\sim$6,000 to $\sim$36,000 for the high-$z$ \gb-band). Using the light-profile fitting results as labels, we fine-tune the three pretrained \gampen{} models. Hyperparameters are selected based on validation loss, and the final values adopted for transfer learning are reported in Appendix~\ref{ap:sec:trained_models}. 

\subsubsection{Fine-Tuning the Dropout Rate} \label{sec:fine_tuning_dropout}

\begin{figure*}[htb]
    \centering
    \includegraphics[width = 0.9\textwidth]{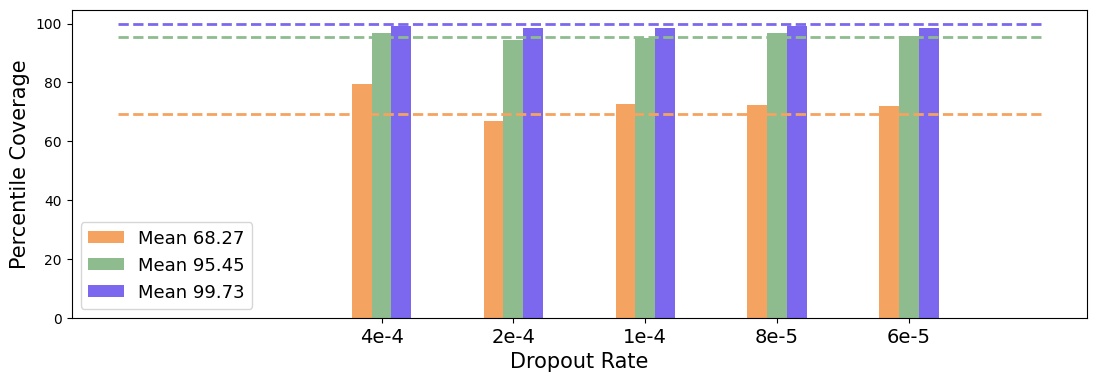}
    \caption{Percentile coverage probabilities corresponding to different dropout rates in the \rb-band sample of the low-$z$ bin. The values shown are the average over the three output parameters. Coverage probability is defined as the fraction of validation galaxies whose light-profile-fitted value falls within a given confidence interval of the predicted posterior. A dropout rate of $6\times10^{-5}$ produces coverage probabilities that closely match the nominal confidence levels. The full search spanned $10^{-5}$ to $10^{-3}$ (\S\ref{sec:fine_tuning_dropout}); only the range around the optimum is shown.}
    \label{fig:dropout_calibration}
\end{figure*}

\begin{deluxetable}{c|c|ccc}[htbp]
\tablecaption{Coverage Probabilities on the Test Set Across Bands in this study}
\tablecolumns{5}
\tablehead{
\colhead{Sample} & \colhead{Parameter} & \colhead{$68.27\%$} & \colhead{$95.45\%$} & \colhead{$99.73\%$}
}
\startdata
\hline
\hline
\multirow[c]{2}{*}{Low-$z$ } & $L_B/L_T$ & 72.11 & 96.43 & 99.03 \\
\multirow[c]{2}{*}{\rb-band} & $R_e$     & 70.26 & 94.79 & 98.39 \\
                             & $F$       & 73.81 & 95.85 & 98.63 \\
                             \hline
                             & Mean      & 72.06 & 95.69 & 98.68 \\
\hline
\multirow[c]{2}{*}{Low-$z$ } & $L_B/L_T$ & 72.16 & 96.19 & 99.06 \\
\multirow[c]{2}{*}{\ib-band} & $R_e$     & 72.56 & 94.52 & 98.90 \\
                             & $F$       & 74.67 & 94.89 & 98.08 \\
                             \hline
                             & Mean      & 73.13 & 95.25 & 98.68 \\
\hline
\multirow[c]{2}{*}{Mid-$z$ } & $L_B/L_T$ & 74.42 & 96.71 & 99.30 \\
\multirow[c]{2}{*}{\gb-band} & $R_e$     & 73.35 & 94.61 & 98.32 \\
                             & $F$       & 69.49 & 97.77 & 99.66 \\
                             \hline
                             & Mean      & 72.42 & 96.37 & 99.09 \\
\hline
\multirow[c]{2}{*}{Mid-$z$ } & $L_B/L_T$ & 75.85 & 97.05 & 99.06 \\
\multirow[c]{2}{*}{\ib-band} & $R_e$     & 69.72 & 93.62 & 98.38 \\
                             & $F$       & 77.29 & 97.59 & 99.37 \\
                             \hline
                             & Mean      & 74.29 & 96.08 & 98.94 \\
\hline
\multirow[c]{2}{*}{High-$z$ } & $L_B/L_T$ & 66.23 & 92.88 & 97.88 \\
\multirow[c]{2}{*}{\gb-band}  & $R_e$     & 70.65 & 92.95 & 97.24 \\
                              & $F$       & 86.86 & 97.82 & 99.61 \\
                              \hline
                              & Mean      & 74.58 & 94.55 & 98.24 \\
\hline
\multirow[c]{2}{*}{High-$z$ } & $L_B/L_T$ & 68.35 & 96.01 & 98.95 \\
\multirow[c]{2}{*}{\rb-band}  & $R_e$     & 78.47 & 98.74 & 99.87 \\
                              & $F$       & 70.52 & 96.09 & 98.24 \\
                              \hline
                              & Mean      & 72.44 & 96.94 & 99.02 \\
\enddata
\tablecomments{
Coverage probabilities (in percent) for each structural parameter and photometric band
in the galaxy sample. They represent the fraction of test galaxies whose GALFIT-derived parameter
values lie within a given confidence interval of the predicted distribution.
All bands with the exception of the high-$z$ \gb-band exhibit comparable coverage.
}
\label{tab:cov_prob_all_z}
\end{deluxetable}

We train \gampen{} by minimizing its loss function with stochastic gradient descent. The hyperparameters that require tuning are the learning rate (the step size during gradient descent), momentum (an acceleration term to speed convergence), the strength of L2 regularization (which penalizes large weight values), batch size (the number of images processed before each weight update), and the dropout rate, which sets the level of epistemic uncertainty estimated by the network (\S\ref{sec:gampen}). To determine suitable values, we trained models across a grid of hyperparameter combinations and selected those that minimized the loss on the validation set. The final set of adopted hyperparameters is listed in Appendix~\ref{ap:sec:trained_models}.

Because the dropout rate directly influences the predicted uncertainties, we explored its effect separately. Networks were trained with dropout rates spanning $10^{-3}$ to $10^{-5}$, while keeping momentum, learning rate, and batch size fixed at their optimized values. To compare models, we computed coverage probabilities, defined as the fraction of validation galaxies whose GALFIT-derived parameters fall within a given confidence interval of the posterior predicted by \gampen{}. Coverage was evaluated at the $68.27\%$, $95.45\%$, and $99.73\%$ levels, corresponding to the $1\sigma$, $2\sigma$, and $3\sigma$ intervals of a Gaussian distribution. For each posterior, the $68.27\%$ interval was defined as the region containing the highest-probability $68.27\%$ of the integrated distribution. Since the predicted posteriors are discrete, we reconstructed them with kernel density estimation, and then defined the $95.45\%$ and $99.73\%$ intervals analogously. We then measured the fraction of validation examples with GALFIT parameters lying inside each interval. An unbiased estimator should return coverage values consistent with the nominal confidence levels (e.g., 68.27\% of cases inside the $68.27\%$ interval). For each redshift bin, we adopted the dropout rate that produced the closest agreement with these targets. Because the dropout rate is tuned to match the nominal coverage on the validation set, the coverage probabilities reported in \S\ref{sec:gampen_v_galfit} are measured on a separate test set that played no part in this tuning; they therefore provide an independent check of calibration rather than restating the tuning target.  

Figure~\ref{fig:dropout_calibration} illustrates this procedure for the low-$z$ \rb-band sample, showing the coverage probabilities averaged across the three output parameters. At large dropout rates, \gampen{} systematically overestimates epistemic uncertainty, driving coverage probabilities above the expected values. Conversely, very small dropout rates lead to underestimation. At $6\times10^{-5}$, the coverage probabilities align closely with the nominal confidence levels, yielding well-calibrated posteriors. The same calibration procedure is repeated for our entire sample, and the final chosen dropout rates are listed in Table~\ref{tab:hyper_para} (Appendix~\ref{ap:sec:trained_models}), while the coverage probabilities attained for our entire sample are shown in Table~\ref{tab:cov_prob_all_z}.

\section{GALFIT-ing Galaxies for Transfer Learning \& Validation} \label{sec:galfitting}

\begin{figure*}[htb]
    \centering
    \includegraphics[width = 0.83\textwidth]{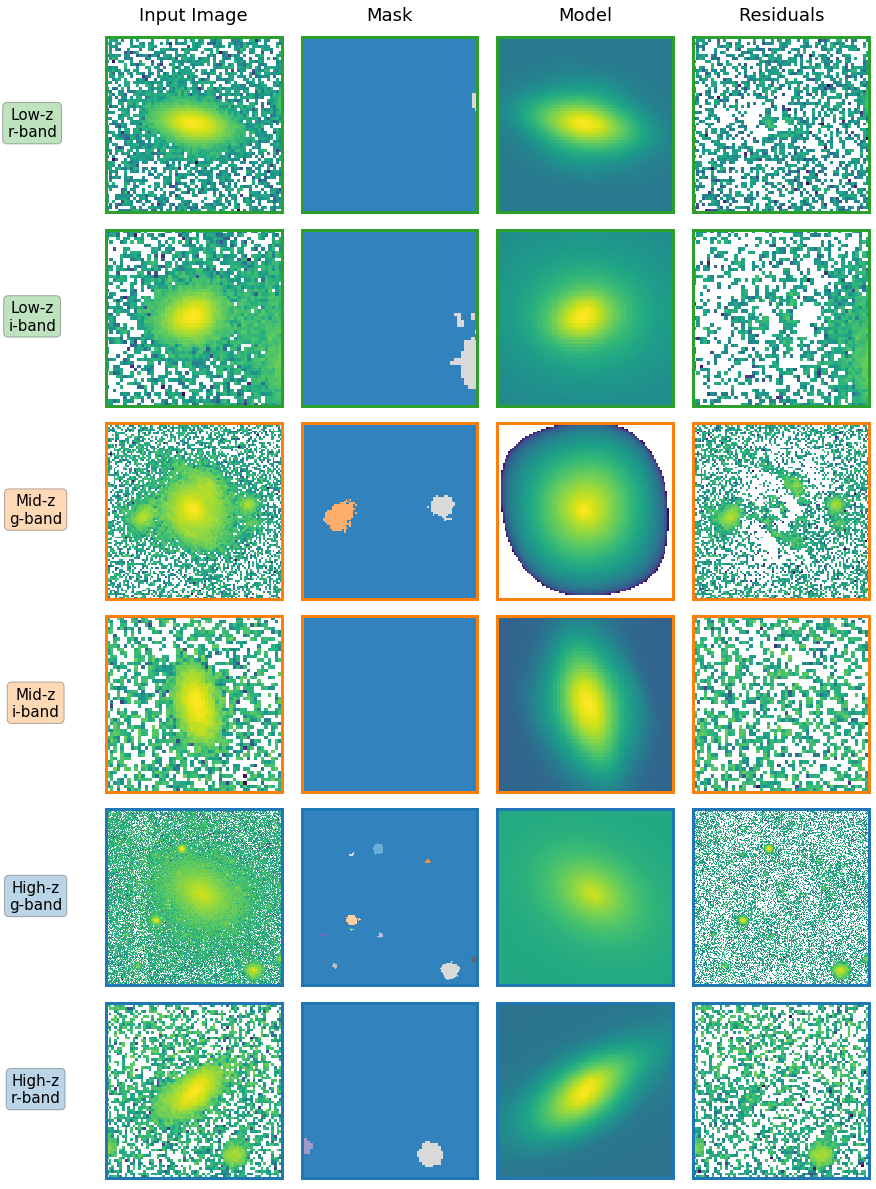}
    \caption{
     Steps in our light-profile fitting pipeline used to measure galaxy structural parameters. For each redshift bin and band shown here, we present (from left to right) the input cutout, the Source Extractor segmentation mask, the best-fit GALFIT model, and the residual image. Because our models do not include Fourier bending modes or coordinate twists, non-axisymmetric features (e.g., bars or spiral structure) are not modeled and remain in the residuals. For bands not included in this figure, the reader can refer to Figure 11 of \citet{Morpho_Single_Band} for an analogous illustration.
    }
    \label{fig:galfit_example_fits}
\end{figure*}

In this section, we describe the semi-automated pipeline adapted from \citet{Morpho_Single_Band} to measure the morphological parameters of $\sim$113,000 galaxies ($\sim$20,000 per band per redshift bin, and $\sim$13,000 for the high-$z$ \gb-band), of which $\sim$75,000 pass the quality cuts described below (see \S\ref{sec:tl_sample} for the full breakdown). These measurements provide the labels for transfer learning and allow us to test the performance of the trained \gampen{} models. 

Parameter estimation was carried out using GALFIT \citep{galfit}, a two-dimensional light-profile fitting algorithm. Prior to running GALFIT, we used Source Extractor \citep{s_extract} on all the image cutouts to generate segmentation maps. These maps were used to mask secondary sources within each cutout during the fitting stage. Source Extractor outputs also served as initial guesses for morphological parameters and were used to define the cutout size for each galaxy, which we set to ten times the Source Extractor estimate of the effective radius.

Each galaxy was modeled with two \sersic{} components: a disk component fixed to $n = 1$ and a bulge component with $3.5 < n < 5.0$. Fitting proceeded in two rounds. In the first, we applied broad constraints to facilitate convergence: radii were limited to $0.5$ -- $90$ pixels ($0.85$ -- $15.12$ arcsec), magnitudes restricted to within $\pm 7.5$ of the Source Extractor guess, the difference between bulge and disk magnitudes bounded between $-7.5$ and $+7.5$ ($0.001 < L_B/L_T < 0.999$), and the centroids of the two components required to lie within 30 pixels ($5.04$ arcsec) of each other. These constraints are intentionally generous for galaxies in this redshift range and are consistent with previous studies \citep[e.g.,][]{vdw_12, Tuccillo2018DeepFitting}. A second round of fitting was then performed with only the \sersic{} index and centroid separation constraints retained, allowing the full parameter space to be explored. The effective radius $R_e$ of each galaxy was defined as the radius enclosing 50\% of the total light in the combined bulge $+$ disk model. Examples of some fits---including cutouts, masks, models, and residuals---for representative galaxies from each redshift bin are shown in Figure~\ref{fig:galfit_example_fits}.


Following the two-stage fitting, we excluded galaxies for which GALFIT failed to converge in each band. The fraction of excluded galaxies was less than 0.2\% in all bins. We also performed a visual inspection of $\sim$300 randomly selected galaxies from each bin (a total of $\sim$1800 systems). This quality check revealed three characteristic failure modes: (i) one component assigned an unphysically small axis ratio, producing a lopsided bulge or disk; (ii) the centroids of the two components were displaced by an implausibly large amount; or (iii) the fits produced excessively large residuals. Examples of these cases are presented in Appendix~\ref{ap:sec:galfit_failures}. 

\begin{deluxetable}{c|c|c}[htbp]
\tablecaption{GALFIT Quality Cuts \label{tab:galfit_quality_cuts}}
\tablecolumns{3}
\tablehead{
\colhead{Sample} & \colhead{Criteria} & \colhead{Excluded \%} }
\startdata
    \hline
    \hline
    \multirow{3}{*}{Low-z} & (\texttt{problematic\_value\_flags == True})  & \multirow{4}{*}{31.74\%} \\
    \multirow{3}{*}{\rb-band}& OR (Sep. Dist. $> 2''$) & \\
    & OR (\texttt{max\_iters\_flag == True} & \\ 
    & AND Red. $\chi^2 > 2.25$) & \\
    \hline
    \multirow{3}{*}{Low-z} & (\texttt{problematic\_value\_flags == True})  & \multirow{4}{*}{31.13\%} \\
    \multirow{3}{*}{\ib-band}& OR (Sep. Dist. $> 2''$) & \\
    & OR (\texttt{max\_iters\_flag == True} & \\ 
    & AND Red. $\chi^2 > 2.25$) & \\
    \hline
    \multirow{3}{*}{Mid-z} & (\texttt{problematic\_value\_flags == True})  & \multirow{4}{*}{36.21\%} \\
    \multirow{3}{*}{\gb-band}& OR (Sep. Dist. $> 1''$) & \\
    & OR (\texttt{max\_iters\_flag == True} & \\ 
    & AND Red. $\chi^2 > 1.25$) & \\
    \hline
    \multirow{3}{*}{Mid-z} & (\texttt{problematic\_value\_flags == True})  & \multirow{4}{*}{31.73\%} \\
    \multirow{3}{*}{\ib-band}& OR (Sep. Dist. $> 1''$) & \\
    & OR (\texttt{max\_iters\_flag == True} & \\ 
    & AND Red. $\chi^2 > 1.25$) & \\
    \hline
    \multirow{3}{*}{High-z} & (\texttt{problematic\_value\_flags == True})  & \multirow{4}{*}{51.36\%} \\
    \multirow{3}{*}{\gb-band}& OR (Sep. Dist. $> 0''.7$) & \\
    & OR (\texttt{max\_iters\_flag == True} & \\ 
    & AND Red. $\chi^2 > 1.25$) & \\
    \hline
    \multirow{3}{*}{High-z} & (\texttt{problematic\_value\_flags == True})  & \multirow{4}{*}{34.13\%} \\
    \multirow{3}{*}{\rb-band}& OR (Sep. Dist. $> 0''.7$) & \\
    & OR (\texttt{max\_iters\_flag == True} & \\ 
    & AND Red. $\chi^2 > 1.25$) & \\
\enddata
\end{deluxetable}

To remove problematic fits from the transfer-learning and validation samples, we combine GALFIT diagnostic flags with a set of empirically motivated quality criteria. Specifically, we exclude galaxies that satisfy any of the conditions summarized in Table~\ref{tab:galfit_quality_cuts}. The $\texttt{problematic\_value\_flags}$ flag identifies fits with unphysical parameter values (e.g., extremely small axis ratios or radii) or numerical convergence issues.\footnote{Refer to Bullet 8 at \url{https://users.obs.carnegiescience.edu/peng/work/galfit/TOP10.html}.} The $\texttt{max\_iters\_flag}$ identifies fits that terminated after reaching the maximum allowed number of iterations. In addition, we reject fits with excessively large separations between the bulge and disk centroids or poor reduced-$\chi^2$ values.

Table \ref{tab:galfit_quality_cuts} shows that the adopted thresholds vary across redshift bins and bands. This variation accounts for the fact that galaxies at higher redshift are preferentially smaller, fainter, and have lower signal-to-noise ratios. For the high-$z$ \gb-band sample, where the observed band increasingly probes rest-frame ultraviolet (UV) emission, we find that a larger fraction of fits fail the quality criteria, consistent with the increased morphological complexity and lower reliability of bulge--disk decompositions in this regime.

The specific thresholds were chosen by balancing the need to exclude clearly problematic fits against maintaining a sufficiently large and representative training sample. Although the exact values are necessarily somewhat subjective, visual inspection of $\sim300$ randomly selected galaxies from each sample confirmed that the adopted criteria effectively remove the dominant failure modes identified above while preserving the overall distribution of galaxy properties. Nevertheless, the cuts remove a non-trivial fraction of each sample ($\sim30$\% for most cases; Table~\ref{tab:galfit_quality_cuts}). Because they act on the quality of the fit, they might preferentially exclude galaxies that a bulge+disk model describes poorly, such as mergers and irregular or clumpy systems, for which bulge+disk parameters are themselves least well defined. The validation presented in \S\ref{sec:gampen_v_galfit} therefore applies most directly to galaxies reasonably described by a bulge+disk model, a limitation shared by any training set built from parametric light-profile fits. To facilitate future studies using alternative quality criteria, we publicly release the complete catalog of GALFIT measurements and associated quality flags.

\begin{figure*}[htb]
    \centering
    \includegraphics[width = 0.95\textwidth]{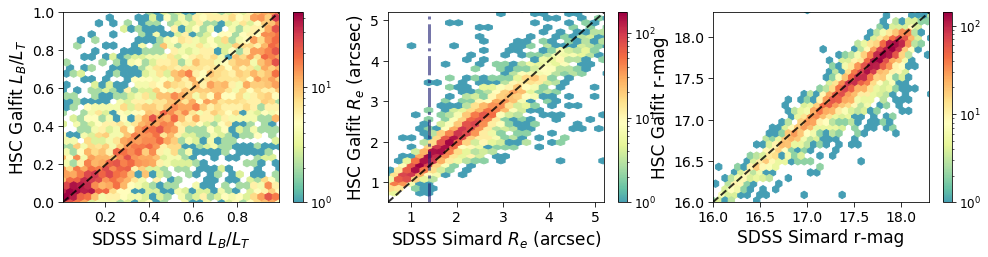}
    \caption{
    Comparison of morphological parameters derived from our light-profile fitting of HSC imaging against those measured for the same galaxies by \citet{simard_11} using SDSS data. The black dashed diagonal indicates the one-to-one relation. In the middle panel, the vertical line marks the median SDSS \rb-band seeing. The two sets of measurements agree well overall; the largest disagreements, particularly in $R_e$, occur for compact galaxies with sizes comparable to or smaller than the SDSS seeing, where disagreement is expected.
    }
    \label{fig:gfit_vs_simard}
\end{figure*}

To assess the reliability of the GALFIT-derived structural parameters used for transfer learning and validation, we compare our measurements with the bulge--disk decomposition catalog of \citet{simard_11}, derived from SDSS imaging. A detailed comparison between HSC-based GALFIT measurements and the \citet{simard_11} catalog was presented in \citet{Morpho_Single_Band} using the low-$z$ \gb-band sample. Here, we repeat this exercise for the low-$z$ \rb-band sample to verify the consistency of the GALFIT measurements used in the present multi-wavelength analysis.

Using a positional cross-match radius of $0.15^{\prime\prime}$, we identify 6047 galaxies common to both catalogs. Figure~\ref{fig:gfit_vs_simard} compares the resulting structural parameters. We note that HSC and SDSS differ substantially in image quality, with HSC providing significantly better seeing and finer pixel sampling than SDSS. Consequently, exact agreement between the two analyses is not expected. Furthermore, differences in filter transmission curves, image-processing pipelines, and survey depth can introduce additional systematic offsets between the measured parameters.

Despite these differences, the overall agreement between the two catalogs is encouraging. The flux and effective radius measurements exhibit strong correlations and closely follow the one-to-one relation over most of the parameter range. The scatter is noticeably larger for $L_B/L_T$, which is expected given that bulge--disk decomposition requires simultaneously constraining multiple structural components and is therefore more sensitive to differences in image quality, spatial resolution, and fitting methodology. The improved image quality of HSC is particularly relevant for compact systems, where structural measurements approach the resolution limit of SDSS. Consistent with the findings of \citet{Morpho_Single_Band}, most discrepancies in $R_e$ occur for galaxies with sizes comparable to or smaller than the median SDSS seeing. 
The overall agreement demonstrates that the GALFIT measurements used for transfer learning are broadly consistent with previously published bulge--disk decompositions and provides an independent validation of our fitting pipeline.

\section{Evaluating \gampen{}'s performance} \label{sec:results}
After completing both rounds of training and evaluation, we apply the final \gampen{} models to all the remaining galaxies in our sample. In this section, we assess the performance of \gampen{} and examine the reliability of the predicted parameters.

\subsection{Inspecting the Predicted Posteriors}

\begin{figure*}[htb]
    \centering
    \includegraphics[width = 0.95\textwidth]{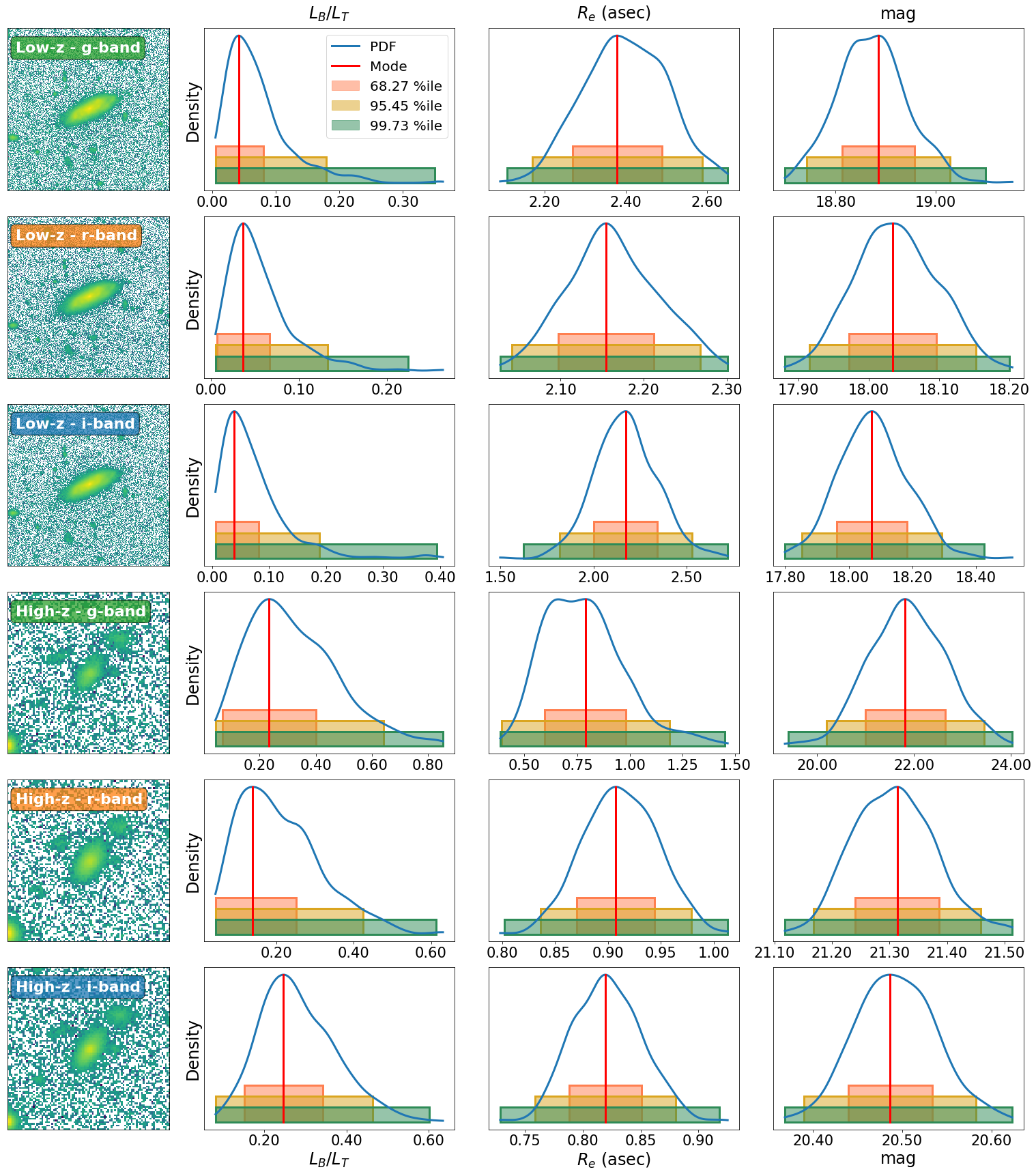}
    \caption{
    Examples of predicted posterior distributions for two randomly selected galaxies from the low- and high-redshift bins. The first three rows show the predicted posteriors for the same low-$z$ galaxy in three different bands, and the last three rows show a second galaxy from the high-$z$ sample. The solid blue curves denote the probability density functions obtained via kernel density estimation, from which the confidence intervals (pink, yellow, and green boxes) are derived. The mode of each distribution (solid red line) marks the most probable value of the corresponding parameter.}
    \label{fig:example_pred_dists}
\end{figure*}

Using the procedure outlined in \S\ref{sec:method} and shown in Figure~\ref{fig:posterior_pred_workflow}, we use the trained \gampen{} models to obtain the joint posterior distributions of all output parameters for each of the $\sim 10$ million galaxies. Figure~\ref{fig:example_pred_dists} shows the marginalized posteriors for two randomly selected galaxies from our low-$z$ and high-$z$ samples across all three bands, along with the corresponding input cutouts that were fed into \gampen{}. The predicted posteriors are generally unimodal, smooth, and resemble Gaussian or skewed Gaussian distributions. For each case, we indicate the regions that enclose $68.27\%$, $95.45\%$, and $99.73\%$ of the total probability. We estimate the probability density function (PDF) using kernel density estimation (blue line in the figure), and we adopt the mode of this PDF as the ``predicted value''  throughout the paper. The figure also highlights that \gampen{} naturally produces posteriors of different widths depending on the information content of the input image (for instance, the broader $L_B/L_T$ distribution in the sixth row compared to the first). We explore this behavior further in \S\ref{sec:uncertainties}.

By design, \gampen{} predicts only physically allowed values. This is most obvious in the $L_B/L_T$ distributions in rows 4 and 5 of Figure~\ref{fig:example_pred_dists}, which remain confined between 0 and 1 without any artificial truncation. This arises from the data transformations described in \S\ref{sec:transformations}. The same approach ensures that the predicted $R_e$ and flux values are always positive.

\begin{figure*}[htb]
    \centering
    \includegraphics[width=0.6\textwidth]{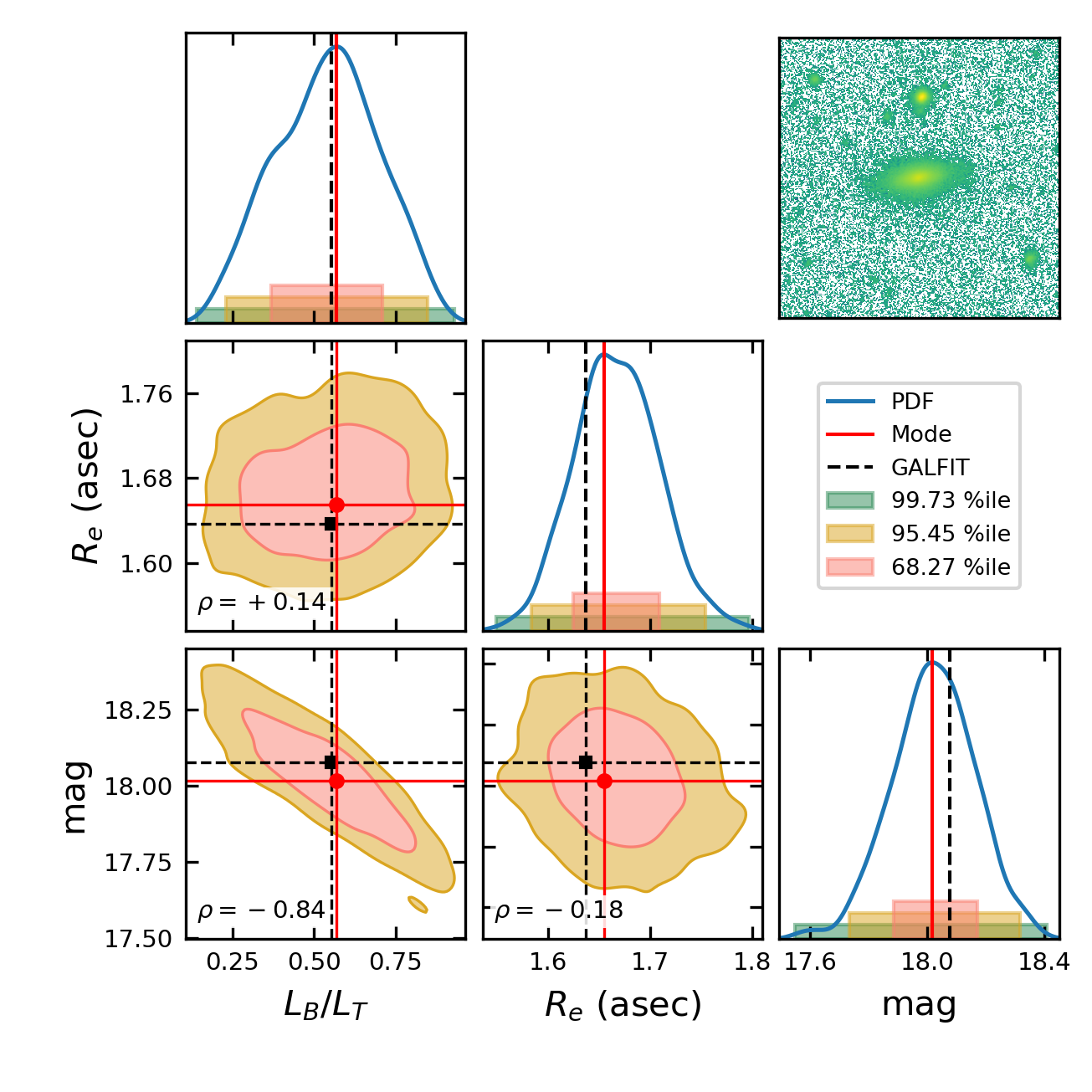}
    \caption{Joint and marginal posterior distributions predicted by \gampen{} for a galaxy from the low-$z$ \rb-band test set, whose input cutout is shown in the top-right panel. Diagonal panels show the marginal posteriors (blue curves, from kernel density estimation), with shaded bars marking the $68.27\%$, $95.45\%$, and $99.73\%$ confidence intervals. Off-diagonal panels show the joint posterior for each pair of parameters, with contours enclosing $68.27\%$ and $95.45\%$ of the probability and the Pearson correlation coefficient $\rho$ given in the lower-left corner. In all panels, the mode of the posterior is marked in red and the GALFIT value in dashed black.}
    \label{fig:corner_plot}
\end{figure*}

Because \gampen{} predicts the full covariance matrix of its outputs, the predicted joint posteriors also encode correlations between parameters, which the marginal distributions in Figure~\ref{fig:example_pred_dists} do not show. Figure~\ref{fig:corner_plot} shows the joint and marginal posteriors for a galaxy from the low-$z$ \rb-band test set. For this galaxy, $L_B/L_T$ and magnitude are strongly anti-correlated (Pearson $\rho = -0.84$), whereas $R_e$ is only weakly correlated with either parameter ($\rho = +0.14$ with $L_B/L_T$ and $\rho = -0.18$ with magnitude). Such correlations matter when propagating uncertainties into derived quantities.

While checking the predicted posteriors for all $\sim 10$ million galaxies, we occasionally found cases where \gampen{} predicted $R_e$ or flux values outside the parameter range covered by the training data. Since machine-learning models are generally reliable for interpolation but can behave unpredictably when pushed into regimes not represented in the training set \citep[e.g.,][]{candela_09, recht_19, taori_20}, we apply a conservative cut. For each redshift bin, we remove galaxies whose predicted $R_e$ or $F$ values lie more than 0.5 arcsec or 0.5 mag beyond the training limits. This step excludes $\sim5.9\%$, $\sim5.32\%$, $\sim0.2\%$, $\sim1.82\%$, $\sim0.74\%$, and $\sim1.21\%$ of the objects in the low-$z$ \rb-band, low-$z$ \ib-band, mid-$z$ \gb-band, mid-$z$ \ib-band, high-$z$ \gb-band, and high-$z$ \rb-band, respectively.

\subsection{Comparing \gampen{} predictions to GALFIT predictions} \label{sec:gampen_v_galfit}

Of the $\sim75,000$ galaxies that pass the \textsc{GALFIT} quality cuts in \S\ref{sec:galfitting}, we use $85\%$ for training and validation, and reserve the remaining $15\%$---which the trained \gampen{} models never see---for evaluating the performance of the predictions. We refer to this subset as the ``test set'' throughout. 
In Table~\ref{tab:cov_prob_all_z}, we show the coverage probabilities achieved by \gampen{} on the test set. Recall that in \S\ref{sec:fine_tuning_dropout}, the dropout rate was tuned using the validation set, whereas the values reported in Table~\ref{tab:cov_prob_all_z} are based solely on the test set, so they are not simply the quantity optimized during tuning. Ideally, the empirical coverage would match the nominal confidence levels (e.g., the true value would lie within the $68.27\%$ highest-probability region of the predicted posterior $68.27\%$ of the time). In practice, the coverage probabilities obtained with \gampen{} remain close to the stated confidence levels, both when averaged over the three output parameters and when considered separately. The mean coverage never differs from the nominal value by more than $6.3$ percentage points, and for individual parameters the deviation is at most $10.2$ percentage points excluding the anomalous high-$z$ \gb-band. 

These small deviations follow a consistent pattern across all six samples. The mean coverage exceeds the nominal value at the $68.27\%$ level (by 3.8--6.3 percentage points), remains close to nominal at the $95.45\%$ level, and falls marginally below it at the $99.73\%$ level (by $\lesssim$1 percentage point, excluding the high-$z$ \gb-band). The uncertainties predicted by \gampen{} are therefore conservative at the $1\sigma$ level, even in the most discrepant case (the $68.27\%$ flux confidence interval for the high-$z$ \gb-band model), while the true errors have slightly heavier tails than the predicted posteriors. If a specific application required near-perfect calibration of coverage, users could post-process the predicted distributions using techniques such as importance sampling \citep{importance_sampling}. We emphasize that incorporating the full parameter covariance in the loss function was essential for achieving simultaneous calibration across all three outputs.



\begin{figure*}[htb]
    \centering
    \includegraphics[width = 0.80\textwidth]{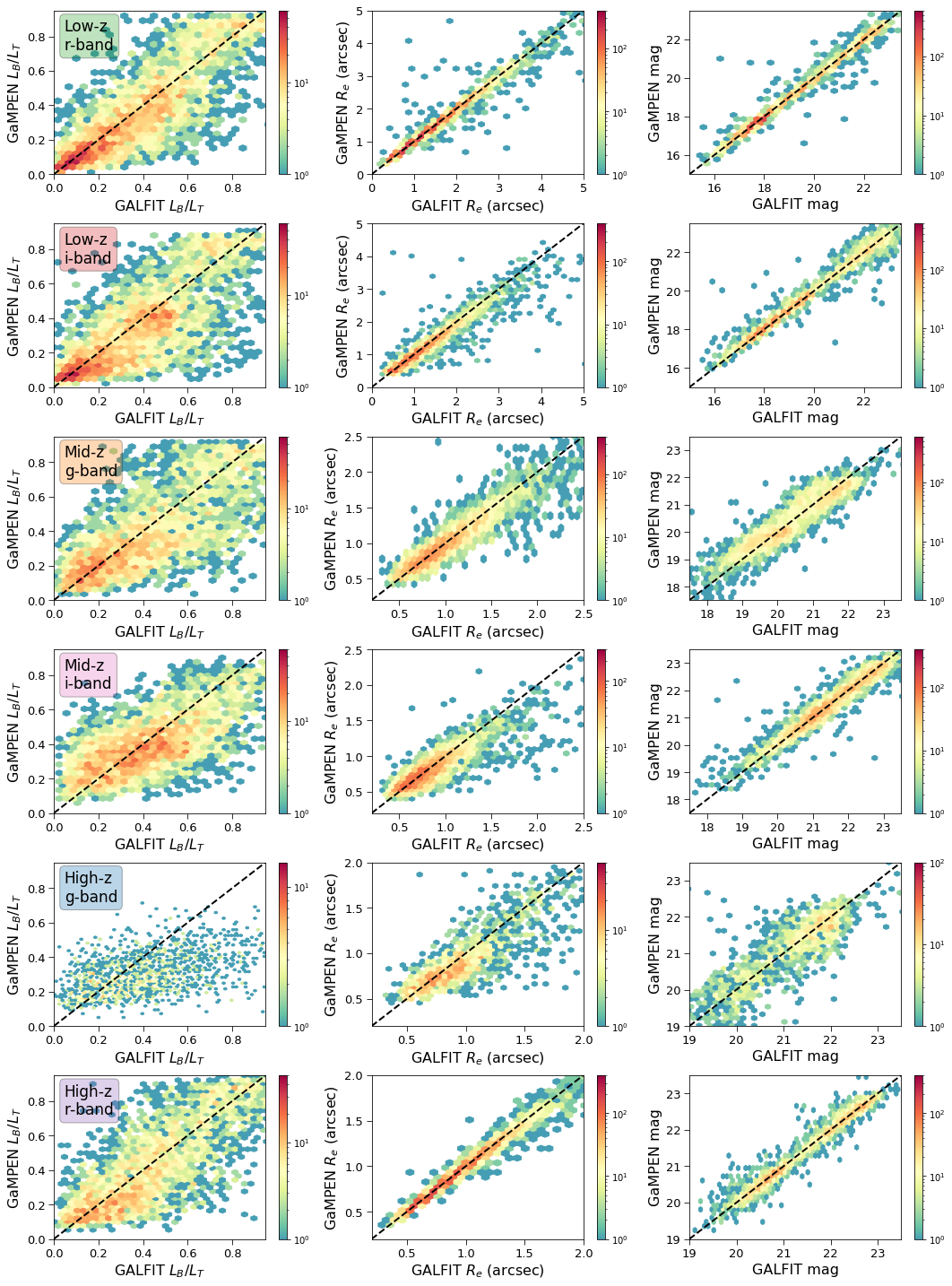}
    \caption{The most probable parameter values predicted by \gampen{} for all galaxies in the test set plotted against the values determined using GALFIT. Galaxies are plotted in hexagonal bins of roughly equal size, and the number of galaxies in each bin is represented according to the logarithmic colorbar to the right of each panel. Each row corresponds to a specific filter in each redshift slice, as mentioned in the colored boxes of the left column. The dashed black $y=x$ line represents the line of equality. Across almost all panels, values predicted by \gampen{} closely mirror the values obtained using light-profile fitting. The only exception is $L_B/L_T$ predictions for the high-$z$ \gb{}-band, which we discuss further in \S\ref{sec:gampen_v_galfit}. For bands not included in this figure, the reader may refer to Figure 17 of \citet{Morpho_Single_Band} for an analogous illustration.}
    \label{fig:pred_true_all_z}
\end{figure*}

Having shown above that the posteriors predicted by \gampen{} are well calibrated, we now examine how closely the modes (most probable values) of the predicted distributions match the values obtained from light-profile fitting. Figure~\ref{fig:pred_true_all_z} shows the most probable \gampen{} predictions for the test set plotted against the corresponding \textsc{GALFIT} measurements, displayed in hexagonal bins of approximately equal area. The number of galaxies in each bin is indicated by the color scale on the right, which is shown in logarithmic units to reveal even low-count bins (down to 1 galaxy/bin). Across all six band--redshift samples (except the high-$z$ \gb-band), the vast majority of galaxies lie close to the line of equality, indicating that the modal values of the predicted posteriors closely track the \textsc{GALFIT} results. The high-$z$ \gb-band sample exhibits noticeably larger scatter, particularly for $L_B/L_T$ compared to the other two parameters, suggesting a degradation in performance. We investigate the origin of this behavior in more detail later in this section.

\begin{figure*}[htb]
    \centering
    \includegraphics[width = 0.81\textwidth]{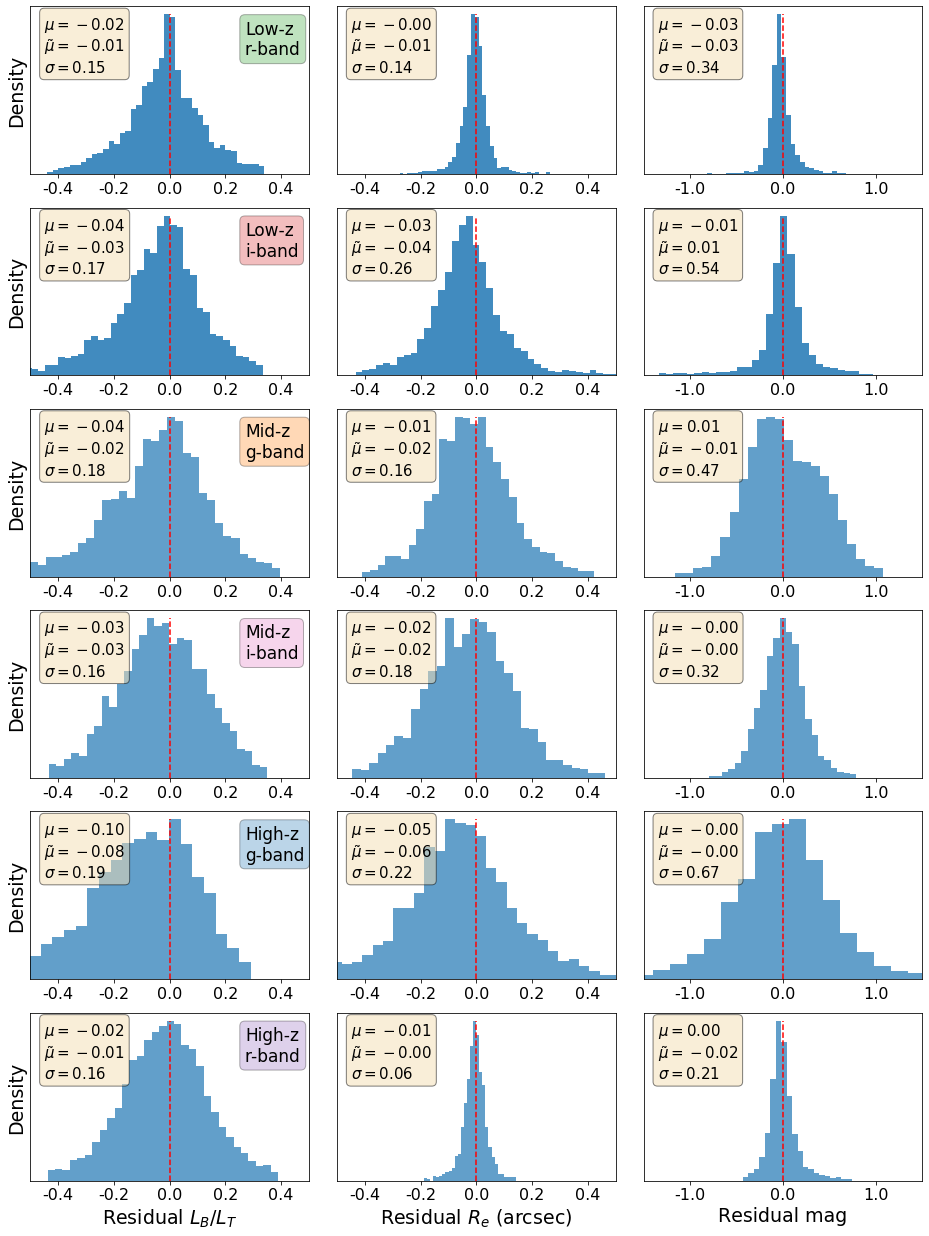}
    \caption{Distributions of residuals for all galaxies in the test set; specifically, the differences between the values predicted by \gampen{} and those obtained via light-profile fitting. Each row corresponds to a specific filter in each redshift slice, as mentioned in the colored boxes of the left column. The yellow boxes in the top-left corner of each panel show the mean ($\mu$), median ($\tilde{\mu}$), and standard deviation ($\sigma$) of each residual distribution. The $\sigma$ of each distribution identifies the typical disagreement for each parameter (e.g., for the first row, in $68.27\%$ of cases, the predicted magnitude is within $\pm0.34$ of the value determined by light-profile fitting). The dashed red vertical line marks  $x=0$. For bands not included in this figure, the reader may refer to Figure 18 of \citet{Morpho_Single_Band} for an analogous illustration.}
    \label{fig:resi_all_z}
\end{figure*}

Figure~\ref{fig:resi_all_z} shows the residual distributions for the \gampen{} output parameters across the six bins. We define the residual for each parameter as the difference between the most probable value predicted by \gampen{} and the value obtained from light-profile fitting. The box in the upper-left corner lists the mean ($\mu$), median ($\tilde{\mu}$), and standard deviation ($\sigma$) of each distribution. All eighteen distributions are approximately Gaussian, with $\mu \approx \tilde{\mu} \approx 0$. The value of $\sigma$ provides a useful measure of the typical disagreement between the two methods (e.g., in the low-$z$ \rb-band bin, $68.27\%$ of galaxies have predicted $L_B/L_T$ values within $\pm0.15$ of the light-profile fitting value). For $R_e$, the residuals, when converted to physical units, correspond to typical disagreements of $0.32$ arcsec, $0.14$ arcsec, and $0.14$ arcsec for the low-, mid-, and high-$z$ bins, respectively. The $L_B/L_T$ residuals remain broadly constant across all bins (except the high-$z$ \gb-band), while the residuals for $R_e$ and $F$ decrease slightly from low to high redshift. This trend may be driven by the improvement in the median HSC seeing from the \gb{}- to the \ib{}-band (\gb{}: $0.\arcsec79$; \rb{}: $0.\arcsec75$; \ib{}: $0.\arcsec61$). In addition, low-$z$ galaxies tend to exhibit more resolved structure (e.g., spiral arms) that is not captured by our disk+bulge \textsc{GALFIT} decomposition, which may contribute to the slightly larger disagreements at low redshift.

\begin{figure*}[htb]
    \centering
    \includegraphics[width = 0.9\textwidth]{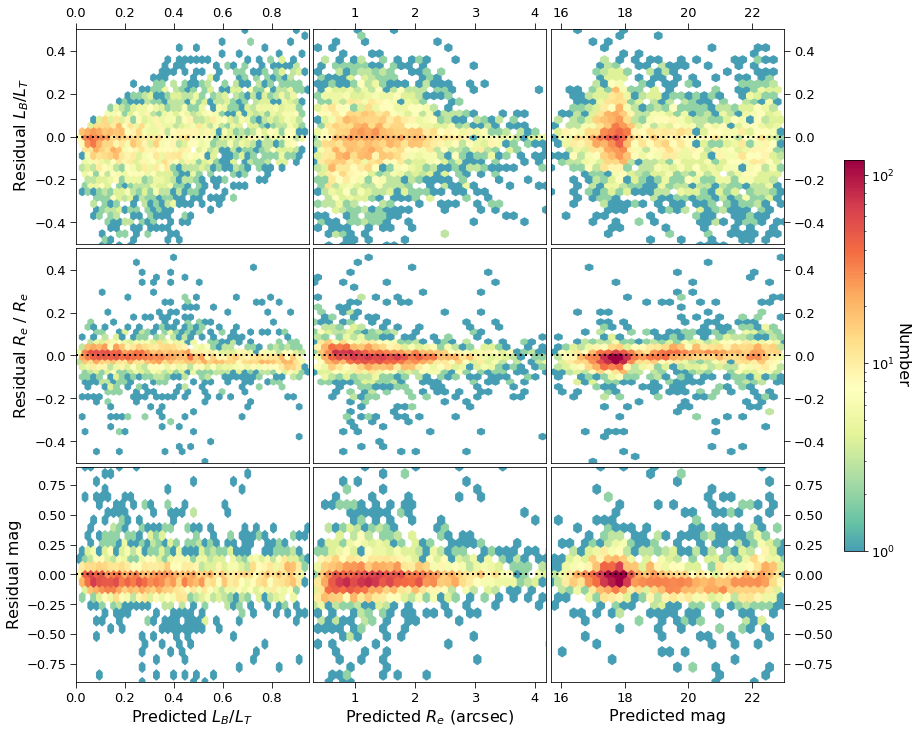}
    \caption{Residuals of the output parameters (difference between \gampen{} and GALFIT predictions) plotted against the values predicted by \gampen{} for all galaxies in the low-$z$ \rb-band test set. To make the y-axis dimensionless for all three parameters, we plot the fractional $R_e$ residuals instead of absolute values. This figure allows us to assign quality labels to \gampen{}'s predictions based on the output values (e.g., flagging regions of the parameter space with high levels of disagreement, as shown by the line-shaded region in the top-left panel). See \S\,\ref{sec:gampen_v_galfit} for details.}
    \label{fig:2d_res_low_zr}
\end{figure*}

While Figures \ref{fig:pred_true_all_z} and \ref{fig:resi_all_z} demonstrate the overall consistency between \gampen{} and GALFIT, they do not capture how this agreement varies across different regions of the parameter space. Understanding this variation is essential, as it highlights where \gampen{} performs particularly well or poorly in comparison to light-profile fitting. Such information allows future users to assess the reliability of predictions in specific regions or to selectively use results tailored to particular scientific applications. In Figure \ref{fig:2d_res_low_zr}, we present the residuals for the three output parameters in the low-redshift bin, plotted against the corresponding \gampen{} predictions. To ensure a consistent, dimensionless y-axis across all parameters, we display fractional residuals for $R_e$ instead of absolute differences. Similar to Figure \ref{fig:pred_true_all_z}, the parameter space is divided into hexagonal bins, with a logarithmic color scale indicating the number of galaxies per bin. Since the residual trends as a function of different parameters are broadly consistent across most redshift bins and optical bands, we adopt the low-$z$ \rb-band as a representative case for discussion in the main text. We also discuss the high-$z$ \gb-band, which shows the weakest performance among all bands considered. 

Across most panels in Figure \ref{fig:2d_res_low_zr}, the majority of galaxies are distributed evenly around the black dashed line, $y = 0$, which represents the ideal scenario of perfectly recovered parameters (under the assumption that GALFIT outputs are accurate). However, several distinct patterns emerge in the top row, corresponding to the $L_B/L_T$ residuals. In the top-left panel, the largest residuals appear near the extreme values of $L_B/L_T$. This behavior was also identified in tests of \gampen{} using simulated galaxies in \citet{gampen_software_paper}, where it was termed the ``edge effect.'' 
When $L_B/L_T$ approaches its boundaries (i.e., when either the disk or bulge component overwhelmingly dominates), accurately recovering $L_B/L_T$ becomes difficult for \gampen{}---a challenge that extends to most image analysis techniques. In such cases, \gampen{} often attributes nearly all the light to the dominant component, producing the characteristic streaks observed at the edges of the plot. Similar difficulties in determining structural parameters for galaxies with $L_B/L_T < 0.2$ and $L_B/L_T > 0.8$ have also been reported in independent studies employing different methodologies \citep[e.g.,][]{euclid_morph, galapagos}. To mitigate this, we recommend, following \citet{Morpho_Single_Band}, that users convert the quantitative $L_B/L_T$ predictions in these edge regions (line-shaded in the top-left panel of Figure~\ref{fig:2d_res_low_zr}) into qualitative labels: ``highly disk-dominated'' for $L_B/L_T \leq 0.2$ and ``highly bulge-dominated'' for $L_B/L_T \geq 0.8$. \citet{gampen_software_paper} found such labels to be $\gtrsim$95\% accurate. The top-middle panel of Figure \ref{fig:2d_res_low_zr} further indicates that galaxies with smaller $R_e$ exhibit larger $L_B/L_T$ residuals, with the disagreement becoming increasingly significant as galaxy sizes approach the seeing limit (\rb-band: $0.75$ arcsec).

\begin{figure*}[htb]
    \centering
    \includegraphics[width = 0.9\textwidth]{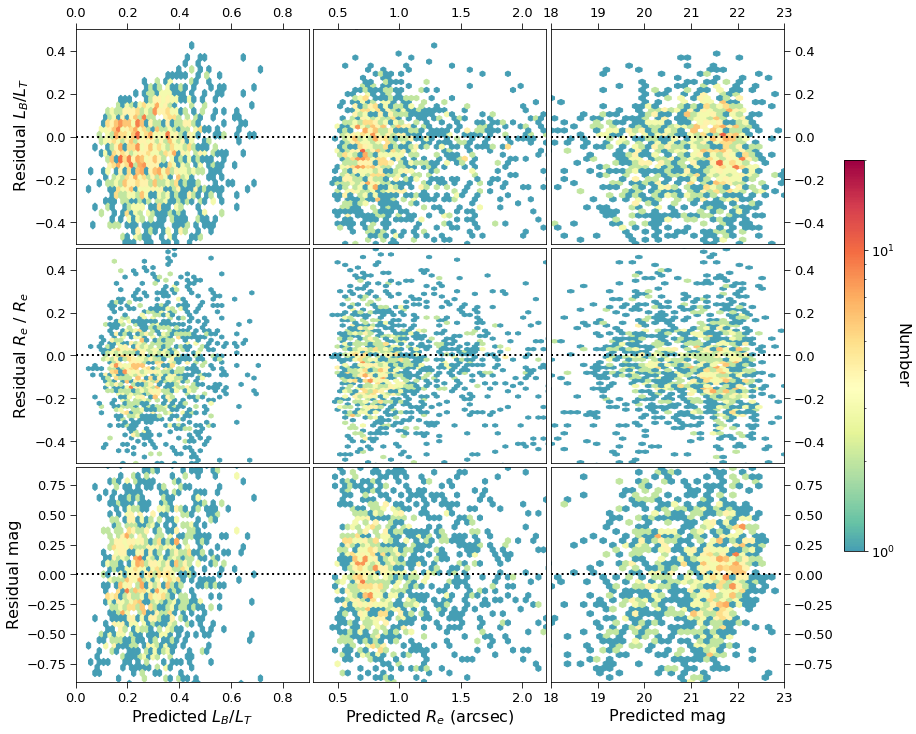}
    \caption{Residuals of the output parameters (difference between \gampen{} and GALFIT predictions) plotted against the values predicted by \gampen{} for all galaxies in the anomalous high-$z$ \gb-band test set. Owing to this degraded performance, the high-$z$ \gb-band measurements are excluded from the public catalog. See \S\,\ref{sec:gampen_v_galfit} for details. }
    \label{fig:2d_res_high_g}
\end{figure*}

Having established the typical behavior using the representative case, we now turn to a contrasting example where the model performance deviates from this general trend. Across most panels in Figure \ref{fig:2d_res_high_g}, the residuals are more broadly dispersed around the black dashed line, $y = 0$, compared to Figure \ref{fig:2d_res_low_zr}, indicating a degradation in parameter recovery in the high-$z$ \gb-band. While the distributions remain approximately centered on zero, the increased scatter reflects larger uncertainties and reduced predictive accuracy of \gampen{} in this regime. 

This behavior can be understood in the context of wavelength-redshift effects (i.e., morphological $k$-corrections). In the redshift range $0.6 \lesssim z \lesssim 0.8$, the observed \gb-band shifts toward bluer rest-frame wavelengths, where galaxy light is increasingly dominated by young, clumpy star-forming regions rather than the smoother, older stellar populations that trace bulge+disk structure. This transition is well known to significantly affect morphological measurements, leading to noisier and less stable structural parameter recovery \citep[e.g.,][]{giavalisco1996, papovich2003, lotz2004, huertas_company2015}.

Consistent with this interpretation, the panels exhibit weaker structural trends and a more diffuse morphology across all rows. In the top row ($L_B/L_T$ residuals), the previously well-defined “edge effect” becomes less distinct, with enhanced scatter across the full parameter range rather than being confined to the boundaries. This suggests that the difficulty is no longer limited to extreme bulge- or disk-dominated systems, but is instead driven by the reduced correspondence between observed light and underlying stellar structure.

The middle and bottom rows further support this picture, showing increased dispersion with little dependence on the predicted parameters. The absence of clear correlations indicates that the model is not strongly biased but is instead limited by the information content of the input images. In addition to bandpass shifting, surface brightness dimming and decreasing signal-to-noise with redshift further exacerbate these effects, making structural decomposition increasingly uncertain \citep[e.g.,][]{ferguson2004, ravindranath2006}.

Overall, because the \gb-band probes progressively shorter rest-frame wavelengths and increasingly traces UV emission at higher redshifts, both traditional profile-fitting techniques and machine-learning-based approaches face growing challenges in recovering robust structural parameters. To avoid the inadvertent use of potentially unreliable measurements, we therefore exclude the high-redshift \gb-band results from the publicly released catalogs.


\subsection{Inspecting the Predicted Uncertainties} \label{sec:uncertainties}

\begin{figure*}[htb]
    \centering
    \includegraphics[width = 0.9\textwidth]{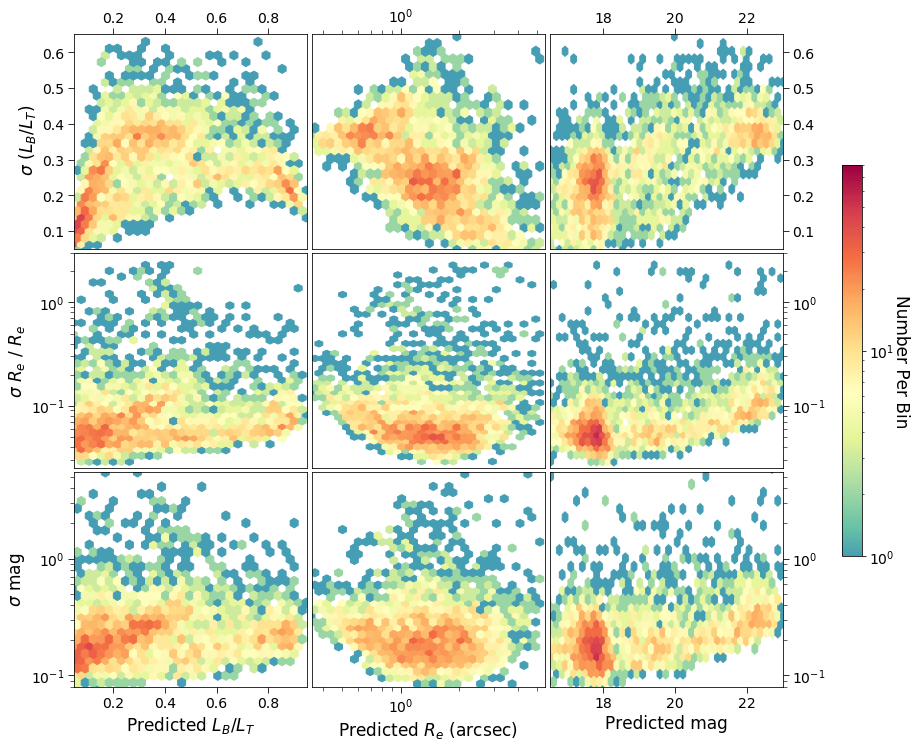}
    \caption{Uncertainties predicted by \gampen{} for each parameter plotted against the predicted values for the low-$z$ \rb-band test set. The $\sigma$ for each parameter is defined as the width of the $68.27\%$ confidence interval. Note that we plot fractional uncertainties for the radius in order to make the y-axis dimensionless for all three rows. (See \S \ref{sec:uncertainties} for details.)}
    \label{fig:2d_uncer_low_r}
\end{figure*}


A key strength of a Bayesian machine learning approach such as \gampen{} lies in its capability to generate full posterior distributions for predicted parameters, rather than limiting outputs to single point estimates. Consequently, one naturally expects the model to produce broader posterior distributions (i.e., larger uncertainties) in regions of parameter space where the residuals are larger.

Figure \ref{fig:2d_uncer_low_r} presents the predicted uncertainties for the three output parameters as a function of their predicted values for the low-redshift \rb-band test sample. For each parameter, we quantify the uncertainty as the width of the $68.27\%$ confidence interval---defined as the range enclosing $68.27\%$ of the highest-probability values in the predicted posterior (see Figure\,\ref{fig:example_pred_dists}). To facilitate comparison across parameters, the y-axis in the middle row is scaled to show fractional (dimensionless) uncertainties. Since the overall structure and behavior of these uncertainty distributions are nearly identical across all redshift bins and optical bands with the exception of high-$z$ \gb-band, we include only the low-$z$ \rb-band results in the main text for brevity, while the corresponding plots for other respective bands and redshift ranges can be made available upon request.

From Figure \ref{fig:2d_res_low_zr}, we previously noted that the $L_B/L_T$ residuals from \gampen{} tend to increase as $R_e$ decreases. Consistent with this, we now observe that \gampen{} assigns larger uncertainties to $L_B/L_T$ in the regime of smaller $R_e$. This adaptive behavior, in which higher-residual regions are accompanied by higher predicted uncertainties, is what enables \gampen{} to reproduce the well-calibrated coverage probabilities.

Examining the right column, it is evident that the predicted uncertainties for all three parameters systematically increase toward fainter magnitudes. This trend aligns with expectations and with earlier findings from \citet{gampen_software_paper}: morphological parameters are inherently more difficult to constrain in faint galaxies compared to bright ones, and thus should be associated with larger uncertainties.

The top-left panel of Figure \ref{fig:2d_uncer_low_r} indicates that \gampen{} maintains a fairly consistent level of confidence in its bulge-to-total ratio predictions across the full parameter range, with slightly reduced uncertainties near the extremes ($L_B/L_T \leq 0.2$ or $L_B/L_T \geq 0.8$). This behavior was also observed in \citet{gampen_software_paper} using simulated data. There, it was shown that these lower uncertainties at the boundaries are predominantly associated with single-component systems, whereas in two-component systems the edge effect is less pronounced (see Figure 15 of \citealp{gampen_software_paper}). The same pattern is evident here: galaxies that can be effectively described by a single disk or bulge component tend to exhibit systematically smaller uncertainties at the edges of $L_B/L_T$. These reduced uncertainties also play a role in amplifying the residuals at the parameter boundaries, as broader uncertainty distributions would otherwise reduce the number of high-residual cases in these regions. Nonetheless, as shown in the top-left panel of Figure \ref{fig:2d_uncer_low_r}, the majority of galaxies with very small uncertainties are located in the edge regions ($L_B/L_T \leq 0.2$ or $L_B/L_T \geq 0.8$), where we recommend converting the quantitative $L_B/L_T$ outputs into qualitative classifications, as discussed in \S \ref{sec:gampen_v_galfit}. 

Because \gampen{} models the joint distribution of the transformed variables in Equation~\ref{eq:transformation_f''} ($\operatorname{logit} L_B/L_T$, $\log R_e$, and $\log F$), its posteriors in physical units are naturally non-Gaussian, which guarantees that every prediction remains in a physically-allowed regime. The effect is strongest for $L_B/L_T$, whose posteriors become increasingly skewed toward the interior of $[0,1]$ as the prediction approaches either boundary (e.g., rows 4 and 5 of Figure~\ref{fig:example_pred_dists}). Since the inverse logit maps a given width in logit space onto a narrower range of $L_B/L_T$ near the boundaries, this transformation also contributes to the smaller uncertainties at the extremes noted above. Such asymmetric posteriors are better summarized by credible intervals than by a single symmetric $\sigma$. For this reason, we quantify uncertainties throughout using the width of the highest-probability $68.27\%$ interval rather than a Gaussian $\sigma$, and the public catalog provides the full marginal posterior of each parameter (Appendix~\ref{sec:ap:data_access}). We recommend that users work with these distributions, or with asymmetric credible intervals derived from them, rather than with a symmetric error.


Overall, the findings presented in this section highlight the central advantage of adopting a Bayesian framework like \gampen{}: even in regimes where predictive accuracy is not perfect, the model is capable of assigning appropriately scaled uncertainties, thereby ensuring that its outputs remain both reliable and well-calibrated.

\section{Comparing Our Predictions to Other Catalogs} \label{sec:compare}

\begin{figure*}[htb]
    \centering
    \includegraphics[width = 0.9\textwidth]{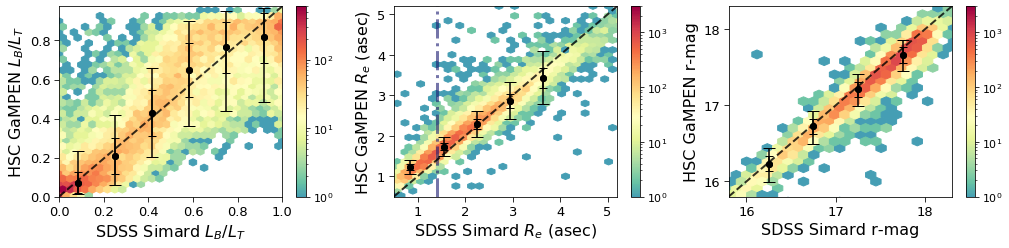}
    \caption{\gampen{} predictions plotted against values estimated by \citet{simard_11} for a cross-matched sample of $\sim19,000$ galaxies with $z < 0.2$ and $m < 18.5$ in \rb-band. The density of points in each histogram is represented according to the logarithmic colorbar on the right. The black dots show the median y values in bins of equal width along the x-axis. The error bars show the $68.27\%$ and $95.45\%$ confidence intervals predicted by \gampen{}, averaged over the galaxies in each bin; they represent the typical per-galaxy uncertainty, not the uncertainty on the median. The dash-dotted vertical line in the middle panel shows the median SDSS \rb{}-band seeing. The agreement is strong across most of the parameter space, with Spearman rank correlation coefficients of 0.87, 0.96, and 0.97 for $L_B/L_T$, $R_e$, and flux, and with the scatter about the median relation largely contained within the predicted uncertainties; the main departures occur for galaxies smaller than the SDSS seeing.}
    \label{fig:gampen_v_simard}
\end{figure*}
The analysis in \S \ref{sec:gampen_v_galfit} established consistency between \gampen{} and parameters derived from our own light-profile modeling. To independently test the accuracy of these predictions, we evaluate them against two external morphological catalogs created using different fitting methodologies and imaging data.

We first consider the catalog of \citet{simard_11}, with the comparison shown in Figure \ref{fig:gampen_v_simard}. An angular matching radius of $0.\arcsec15$ was adopted to identify common objects, resulting in a shared sample of approximately 19,000 galaxies. The overlap is dominated by relatively bright and nearby systems ($m < 18.5$, $z < 0.2$). To examine trends and the spread in the predictions, we divide the reference-parameter axis (i.e., x-axis) in each panel into equal-width bins. Within each bin, we display the median predicted value together with the mean $68.27\%$ and $95.45\%$ confidence intervals of the galaxies in that bin; these reflect the typical per-galaxy uncertainty rather than the uncertainty on the median.

Although SDSS and HSC differ significantly in pixel scale and median atmospheric seeing, the agreement between the two sets of structural measurements is strong across most of the explored parameter space. To quantify the relationship between the two measurements without assuming a linear dependence, we compute their Spearman rank correlation coefficient \citep{spearman}. Spearman rank correlation coefficients of 0.87 for $L_B/L_T$, 0.96 for $R_e$, and 0.97 for $F$ indicate a strong correspondence between the two catalogs. Additionally, the spread of points around the median relation in Figure \ref{fig:gampen_v_simard} is largely contained within the predicted uncertainty envelopes. This suggests that \gampen{}’s error calibration reliably captures the intrinsic scatter.

A notable accumulation of objects near predicted $L_B/L_T \sim 0.9$ reflects the boundary behavior discussed in \S \ref{sec:gampen_v_galfit}. Deviations in effective radius become apparent around $R_e \sim 1\arcsec$, where differences persist even when uncertainties are taken into account. These radii lie approximately $0.\arcsec4$ below the median SDSS seeing (indicated by the dash-dotted vertical line in Figure \ref{fig:gampen_v_simard}), making increased disagreement expected in this regime. 

\begin{figure*}[htb]
    \centering
    \includegraphics[width = 0.9\textwidth]{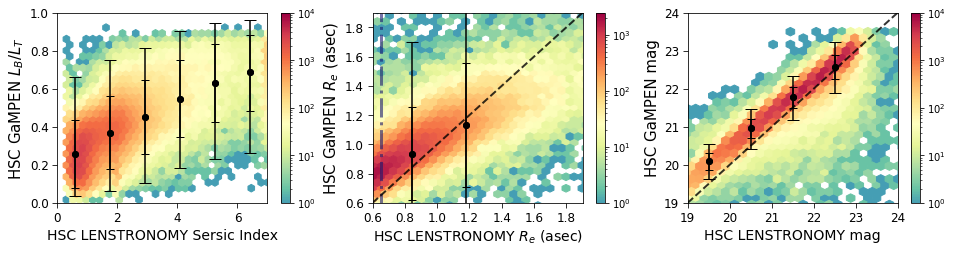}
    \caption{\gampen{} predictions plotted against values estimated by \citet{hsc_sersic} for a cross-matched sample of $\sim170,000$ galaxies with $0 < z \leq 0.5$ and $m \leq 23$ in \ib-band. Similar to Figure \ref{fig:gampen_v_simard}, we show the median y-values and associated error bars in bins of equal width along the x-axis; the error bars are the predicted per-galaxy confidence intervals averaged over each bin, not uncertainties on the median. The dash-dotted vertical line in the middle panel shows the median HSC \ib{}-band seeing. \gampen{} effective radii and magnitudes agree closely with those of \citet{hsc_sersic}, with Spearman coefficients of 0.88 and 0.96, and the left panel shows the expected increase of $L_B/L_T$ with \sersic{} index, with $n \leq 1.5$ galaxies predominantly disk-dominated.}
    \label{fig:gampen_v_lenstronomy}
\end{figure*}

 A second external validation is performed using the catalog of \citet{hsc_sersic}, illustrated in Figure \ref{fig:gampen_v_lenstronomy}. In that study, $1.7 \times 10^6$ galaxies in the HSC \ib{} band were modeled with single-component \sersic{} profiles using \lenstronomy{} \citep{Birrer2018Lenstronomy:Package}. Restricting to objects in common with our low- and mid-$z$ \ib-band samples yields approximately 170,000 galaxies spanning $0 < z \leq 0.5$ and \ib{} $\leq 23$. Median relations and representative uncertainty intervals are presented in the same manner as in the previous comparison.

Although the magnitude predictions agree when considering uncertainties, a small systematic offset is visible at the brighter end. This can be attributed to differences in processing and signal-to-noise between the Deep/Ultra-Deep and Wide layers of HSC. While we use HSC-Wide for this work, \citet{hsc_sersic} use the HSC Deep and Ultra-Deep layers. Effective radius estimates show consistency within the expected uncertainty bounds, though the median relation departs slightly from the one-to-one line at radii smaller than the median HSC seeing (again marked by a dash-dotted vertical line in Figure \ref{fig:gampen_v_lenstronomy}). The difference can also be attributed to the fact that our definition of $R_e$ corresponds to the half-light radius of the composite bulge+disk model, whereas \citet{hsc_sersic} report the semi-major axis enclosing half the total flux of a single \sersic{} fit. The Spearman coefficients are 0.88 for $R_e$ and 0.96 for magnitude.

Because the \citet{hsc_sersic} catalog relies on single-component models, direct comparison of $L_B/L_T$ is not possible. Nevertheless, the left panel of Figure \ref{fig:gampen_v_lenstronomy} illustrates the empirical mapping between fitted \sersic{} index and bulge-to-total light fraction. As expected, increasing $n$ correlates with increasing bulge dominance. Galaxies with $n \leq 1.5$ are overwhelmingly disk-dominated, while systems with $n \geq 3$ typically exhibit substantial bulge contributions, spanning roughly $40\%$--$95\%$ of the total light. This behavior is consistent with trends previously reported between $n$ and $L_B/L_T$ \citep[e.g.,][]{simmons_08, gamornet_paper}.

\section{Discussion \& Conclusions} \label{sec:discussions}

In this work, we have extended the Bayesian deep-learning framework \gampen{} to derive structural parameters and associated posterior distributions for galaxies observed in multiple photometric bands. Using transfer learning from the models developed by \citet{Morpho_Single_Band}, we generated homogeneous estimates of the bulge-to-total light ratio, effective radius, and flux for nearly $10$ million galaxies in the HSC Wide survey at $z\leq0.75$ with $m<23$, corresponding to more than $18$ million individual galaxy images across the $g$, $r$, and $i$ bands. Combined with the catalog of \citet{Morpho_Single_Band}, this represents the first publicly available large-scale HSC structural parameter catalog with homogeneous multi-wavelength structural measurements and robust uncertainties. To our knowledge, it is also the largest multi-band catalog of bulge--disk structural parameters with per-galaxy posteriors, covering $\sim$1400\,deg$^2$ of ground-based imaging.

A central result of this work is that transfer learning allows accurate structural parameter estimation in new imaging bands using only a small fraction of galaxies with traditional light-profile fits. Across all redshift bins and bands, \gampen{} reproduces GALFIT-derived measurements with residual distributions centered near zero and coverage probabilities that remain close to their nominal confidence levels. This demonstrates that the information learned from one photometric band can be efficiently transferred to neighboring bands without requiring the construction of entirely new training datasets. Such behavior is particularly encouraging in the context of forthcoming surveys such as Rubin-LSST, \textit{Euclid}, and the Nancy Grace Roman Space Telescope, where the volume of imaging data will far exceed the practical limits of traditional profile-fitting approaches.

Galaxy structure is intrinsically wavelength dependent, as different stellar populations dominate the emitted light at different wavelengths, while dust attenuation and recent star formation can further modify the observed light distribution \citep[e.g.,][]{giavalisco1996,papovich2003,kelvin_12_gama,vulcani2014_gama,kennedy2016_gama,Conselice_2014}. Consequently, structural parameters measured in different bands contain complementary information regarding stellar population gradients, bulge and disk evolution, and the spatial distribution of star formation. By providing consistent measurements across multiple filters for $\sim$3.9 million galaxies measured in every band released for their redshift (\gb, \rb, and \ib{} at $z \leq 0.5$; \rb{} and \ib{} at $0.5 < z \leq 0.75$), our catalog enables statistical investigations of morphological color gradients, the relationship between stellar populations and galaxy structure, environmental influences on morphological evolution, and the emergence of structural scaling relations across cosmic time \citep[e.g.,][]{Bernardi2013TheProfile,vanderWel20143D-HST+CANDELS:3}. More broadly, the combination of structural parameters and their associated uncertainties for millions of galaxies provides an important resource for connecting observations with theoretical models of galaxy formation and evolution \citep[e.g.,][]{schawinski_14}. A detailed characterization of how structural parameters vary with wavelength, including trends in bulge prominence, galaxy size, and morphological color gradients across different galaxy populations, will be presented in a forthcoming study based on the catalog released in this work. The methodology presented here is also readily extensible to future datasets and to additional morphological quantities beyond those considered in this study.

Our analysis also highlights the inherent limitations of morphological analysis at progressively shorter rest-frame wavelengths. While the majority of bands considered in this study exhibit comparable performance, the high-redshift $g$-band sample shows substantially larger scatter and weaker agreement between \gampen{} and GALFIT. At $z\gtrsim0.5$, the observed $g$-band probes the rest-frame near-ultraviolet, where galaxy light is increasingly dominated by young star-forming regions rather than the older stellar populations that more closely trace the underlying bulge--disk structure \citep[e.g.,][]{giavalisco1996,papovich2003,lotz2004,huertas_company2015}. In addition, surface-brightness dimming and declining signal-to-noise further complicate structural decomposition at these wavelengths \citep{ferguson2004,ravindranath2006}. Importantly, \gampen{} responds to this reduced information content by predicting broader posterior distributions, thereby maintaining approximately calibrated uncertainties even in this regime. Because both traditional profile-fitting methods and machine-learning approaches become increasingly unreliable under these conditions, we exclude the high-redshift $g$-band results from the public catalog.

The Bayesian nature of \gampen{} provides an additional advantage beyond computational efficiency. Rather than producing only point estimates, the framework returns full posterior distributions that naturally encode both aleatoric and epistemic uncertainties. We showed that regions of parameter space associated with larger residuals---such as galaxies with small apparent sizes or faint magnitudes---are accompanied by correspondingly larger predicted uncertainties. This adaptive behavior is crucial for ensuring that downstream scientific analyses properly account for measurement uncertainties and, where the joint posteriors are used, parameter covariances. In particular, it enables the rigorous propagation of uncertainties rather than relying on approximate error estimates, which can be critical for identifying and characterizing subtle correlations in galaxy evolution studies \citep[e.g.,][]{Ghosh_2024_env,igel2026environmentalsculptinggalaxystructure}.

With this work, we publicly release the trained \gampen{} models, source code, documentation, and the resulting multi-wavelength morphology catalog, as outlined in Appendix \ref{sec:ap:data_access}. Together with the previously published rest-frame morphology catalog of \citet{Morpho_Single_Band}, these data products provide a homogeneous view of galaxy structure across both wavelength and cosmic time. As next-generation surveys begin delivering samples containing hundreds of millions of galaxies, methods capable of rapidly producing robust structural measurements and calibrated uncertainties will become increasingly important. The framework developed here demonstrates that Bayesian deep-learning approaches are now capable of meeting this challenge while enabling systematic investigations of the wavelength dependence of galaxy structure across cosmological volumes.

\begin{acknowledgments}

M.S., A.G., and A.J.C. acknowledge support from the DiRAC Institute in the Department of Astronomy at the University of Washington. The DiRAC Institute is supported through generous gifts from the Charles and Lisa Simonyi Fund for Arts and Sciences and the Washington Research Foundation. 

A.G. is supported by an LSST-DA Catalyst Fellowship; this publication was thus made possible through the support of Grant 62192 from the John Templeton Foundation to LSST-DA. The opinions expressed in this publication are those of the author(s) and do not necessarily reflect the views of LSST-DA or the John Templeton Foundation. 

A.G. acknowledges support from the University of Washington eScience Institute for the UW Data Science Postdoctoral Fellowship.

J.S.S. was supported by NSERC Discovery Grant RGPIN-2023-04849.

This material is based upon work supported by the National Aeronautics and Space Administration under Grant No. 80NSSC25K0007.

The galaxy profile fitting and ancillary analyses presented in this work were conducted primarily using the EPYC and Hyak Klone computing resources at the University of Washington. The large-scale training, transfer learning, and inference runs required for the application of \gampen{} to the Hyper Suprime-Cam survey were performed on the Hyak Tillicum supercomputer through Wave 1 of the Tillicum Early Access Program. Access to NVIDIA H200 GPU resources through this program was instrumental in enabling the computationally intensive deep-learning workflows and large-scale inference on millions of galaxies that made this study possible. We gratefully acknowledge the University of Washington Research Computing team for providing, maintaining, and supporting the Hyak computing infrastructure, including the Klone and Tillicum clusters.

This work is based on data collected at the Subaru Telescope and retrieved from the Hyper Suprime-Cam (HSC) data archive system. We respectfully acknowledge that the data utilized in this research were acquired from Maunakea, a location of profound cultural, historical, and natural significance in Hawaii. We acknowledge the efforts of the HSC collaboration in producing and maintaining these high-quality survey data products. We also thank the developers and maintainers of the open-source scientific software packages that contributed to this work.

The Hyper Suprime-Cam (HSC) collaboration includes the astronomical communities of Japan and Taiwan, and Princeton University. The HSC instrumentation and software were developed by the National Astronomical Observatory of Japan (NAOJ), the Kavli Institute for the Physics and Mathematics of the Universe (Kavli IPMU), the University of Tokyo, the High Energy Accelerator Research Organization (KEK), the Academia Sinica Institute for Astronomy and Astrophysics in Taiwan (ASIAA), and Princeton University. Funding was contributed by the FIRST program from the Japanese Cabinet Office, the Ministry of Education, Culture, Sports, Science and Technology (MEXT), the Japan Society for the Promotion of Science (JSPS), Japan Science and Technology Agency (JST), the Toray Science Foundation, NAOJ, Kavli IPMU, KEK, ASIAA, and Princeton University.
\end{acknowledgments}

\appendix
\setcounter{dbltopnumber}{2} 
\renewcommand{\dbltopfraction}{0.9}

\section{Data Access}\label{sec:ap:data_access}
\gampen{}'s source code is publicly available. In addition, we make available the trained models, a catalog of the predicted morphological parameters as well as the corresponding marginal PDFs for the $\sim$10 million galaxies in our sample. The catalog provides the marginal posterior of each parameter. It includes every galaxy for which we provide measurements, whether in one, two, or all of the bands released for its redshift bin (Table~\ref{tab:data}). The joint posteriors are not included, but are available from the authors upon request.

Links to the various components of the public data release mentioned above are available at:
\begin{itemize}[topsep=10pt,partopsep=8pt, parsep=8pt, itemsep=8pt]
    \item \href{https://github.com/aritraghsh09/GaMPEN}{\url{https://github.com/aritraghsh09/GaMPEN}}
    \item \href{https://gampen.readthedocs.io/en/latest/}{\url{https://gampen.readthedocs.io/en/latest/}}
\end{itemize}

Users are encouraged to consult the Public Data Release Handbook, available at
\href{https://gampen.readthedocs.io/en/latest/Public_data.html}
{\url{https://gampen.readthedocs.io/en/latest/Public_data.html}},
for a detailed description of the data products and usage considerations prior to working with the morphological catalogs presented in this paper.

In addition to the source code and trained models, the public data release
includes the set of parameters derived from the light-profile fitting pipeline described in \S\ref{sec:galfitting}. 

\textit{Note: While \gampen{}'s source code is already public, the trained models, the morphological catalog with its posterior distributions, and the light-profile fitting results will be made available once this paper has completed the peer-review process.}

\section{Additional Details About Data}\label{ap:data}

The data-selection procedure used in this work closely follows that described in \citet{Morpho_Single_Band}. As outlined in \S\ref{sec:data}, we use the \texttt{cleanflags\_any} parameter available in HSC PDR2 to exclude objects flagged by the HSC pipeline as having significant imaging issues. A detailed breakdown of the various triggers contributing to this flag and their prevalence within the excluded sample is presented in Figure 24 of \citet{Morpho_Single_Band}. As discussed there, approximately $80\%$ of the exclusions are associated with cosmic-ray hits and interpolated pixels.

For completeness and reproducibility, the SQL query template used to construct the low-$z$, mid-$z$, and high-$z$ parent samples is provided in Listing~\ref{code:sql}. After retrieving the data from the HSC database, additional quality cuts were applied based on the imaging flags described above and the reliability of the photometric redshift estimates (see \S\ref{sec:data}). Users may also choose to employ the various \texttt{blendedness} flags available in HSC PDR2 to further exclude potentially blended or merging systems.

\begin{listing}[H]
\begin{Verbatim}[
frame=lines,
framesep=2mm,
baselinestretch=1.2,
fontsize=\footnotesize,
bgcolor=LightGray,
numbers=left,
breaklines,
]

SELECT
    object_id
      , ra
      , dec
      
FROM
    pdr2_wide.smallcat
    LEFT JOIN pdr2_wide.specz USING (object_id)
    LEFT JOIN pdr2_wide.photoz_mizuki USING (object_id)
    WHERE
    ((photoz_best > (lower_limit) AND 
    photoz_best <= (higher_limit))
    OR 
    (specz_redshift > (lower_limit) AND 
    specz_redshift <=  (higher_limit)))
    AND ((HSC_band)_kronflux_mag < 23 OR (HSC_band)_cmodel_mag < 23)
    AND (HSC_band)_extendedness_value = 1
;
\end{Verbatim}
\caption{Sample SQL query}
\label{code:sql}
\end{listing}

\section{Additional Details About \gampen{}}\label{ap:gampen}

\begin{figure*}[htb]
    \centering
    \includegraphics[width
    =\textwidth]{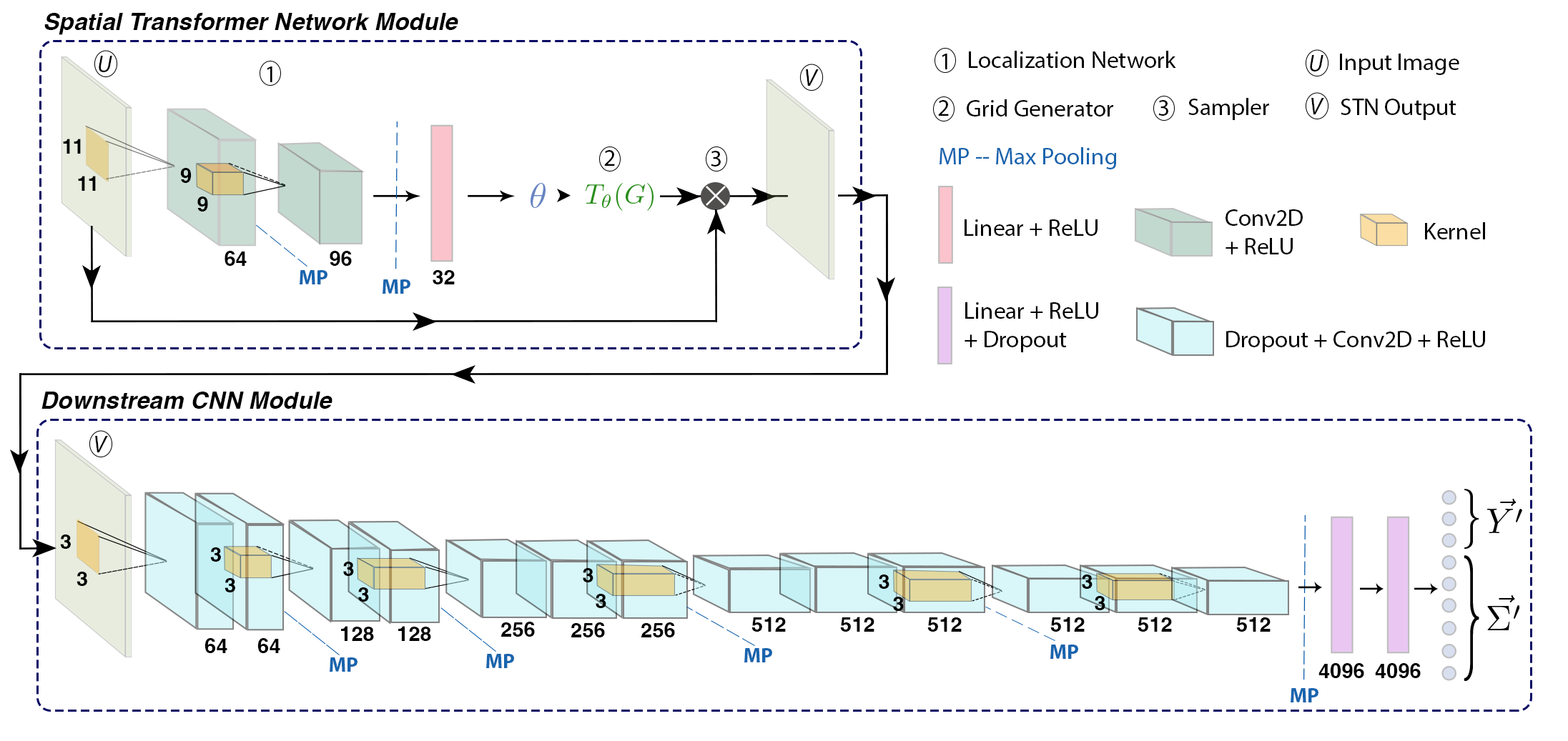}
    \caption{Architecture of \gampen{}, from Figure 3 in \citet{gampen_software_paper}. \gampen{} is designed with two sequential components: an upstream STN and a downstream CNN. The STN autonomously learns cropping transformations of galaxy images, ensuring that the CNN receives optimally framed inputs. The CNN then infers the posterior distributions of morphological parameters. Layer specifications are annotated in the schematic, with the number of filters indicated below each layer and kernel sizes marked alongside the convolutional blocks. For clarity, only a representative kernel is shown per set of convolutional layers, though all layers within a set share the same configuration. Standard operations such as 2D convolutions and rectified linear units (ReLU) form the building blocks of the network. Additional details of the architecture can be found in \citet{gampen_software_paper}.}
    \label{fig:gampen_schematic}
\end{figure*}

To give readers a visual understanding of how \gampen{}'s different architectural components are organized, Figure \ref{fig:gampen_schematic} shows an illustrative diagram (adapted from \citealt{gampen_software_paper}) outlining the structure of both the STN and CNN in \gampen{}. For complete details on individual layers in \gampen{}, please refer to \citet{gampen_software_paper}.

\section{Additional Details About Trained \gampen{} Models}\label{ap:sec:trained_models}

\begin{deluxetable*}{c|cccccc}[htbp]
\tablecaption{Tuned Values of Various Hyperparameters  \label{tab:hyper_para}}
\tablecolumns{6}
\tablehead{
\colhead{Model Name} & \colhead{Learning Rate} & \colhead{Momentum} & \colhead{L2 Regularization ($\lambda$)} & \colhead{Batch Size} & \colhead{Dropout Rate}
}
\startdata
    \hline
    \hline
    Low-z \rb-band & $5\times10^{-8}$ & 0.99 & $10^{-4}$ & 16 & $6\times10^{-5}$\\
    Low-z \ib-band & $5\times10^{-8}$ & 0.99 & $10^{-4}$ & 16 & $1\times10^{-4}$\\
    \hline
    Mid-z \gb-band & $5\times10^{-8}$ & 0.99 & $10^{-4}$ & 16 & $2\times10^{-4}$\\
    Mid-z \ib-band & $5\times10^{-8}$ & 0.99 & $10^{-4}$ & 16 & $8\times10^{-5}$\\
    \hline
    High-z \gb-band & $5\times10^{-6}$ & 0.99 & $10^{-4}$ & 16 & $2\times10^{-4}$\\
    High-z \rb-band & $5\times10^{-6}$ & 0.99 & $10^{-4}$ & 16 & $4\times10^{-4}$\\
\enddata
\end{deluxetable*}

As noted in \S \ref{sec:method}, the training procedure of \gampen{} involves the tuning of various hyperparameters (e.g., learning rate and batch size). These hyperparameters are chosen based on the combination of the values that result in the best performance on the validation set. The final hyperparameters adopted for the fine-tuned models are listed in Table \ref{tab:hyper_para}. Please refer to \S \ref{sec:method} for more details on how we train these models.

\section{Identifying Issues with our Light Profile Fitting Pipeline}\label{ap:sec:galfit_failures}

As discussed in \S\ref{sec:galfitting}, we use a semi-automated GALFIT-based pipeline to derive structural parameters for the galaxies used in our transfer-learning and validation samples. The fitting methodology closely follows that described in \citet{Morpho_Single_Band}, where the pipeline was developed and applied to HSC \gb-band imaging.

To assess the quality of the resulting fits, we visually inspected a randomly selected subset of galaxies from each redshift--band sample. Consistent with the findings of \citet{Morpho_Single_Band}, we identified three primary failure modes: (i) fits in which one of the components was assigned an unphysically small axis ratio, resulting in a highly elongated bulge or disk; (ii) fits in which the centroids of the bulge and disk components were separated by an implausibly large distance; and (iii) fits with large residuals, indicating that the adopted bulge+disk model failed to adequately describe the observed galaxy light distribution. In some cases, multiple failure modes were present simultaneously.

\begin{figure*}[p]
    \centering
    \includegraphics[width=\textwidth,height=0.88\textheight,keepaspectratio]{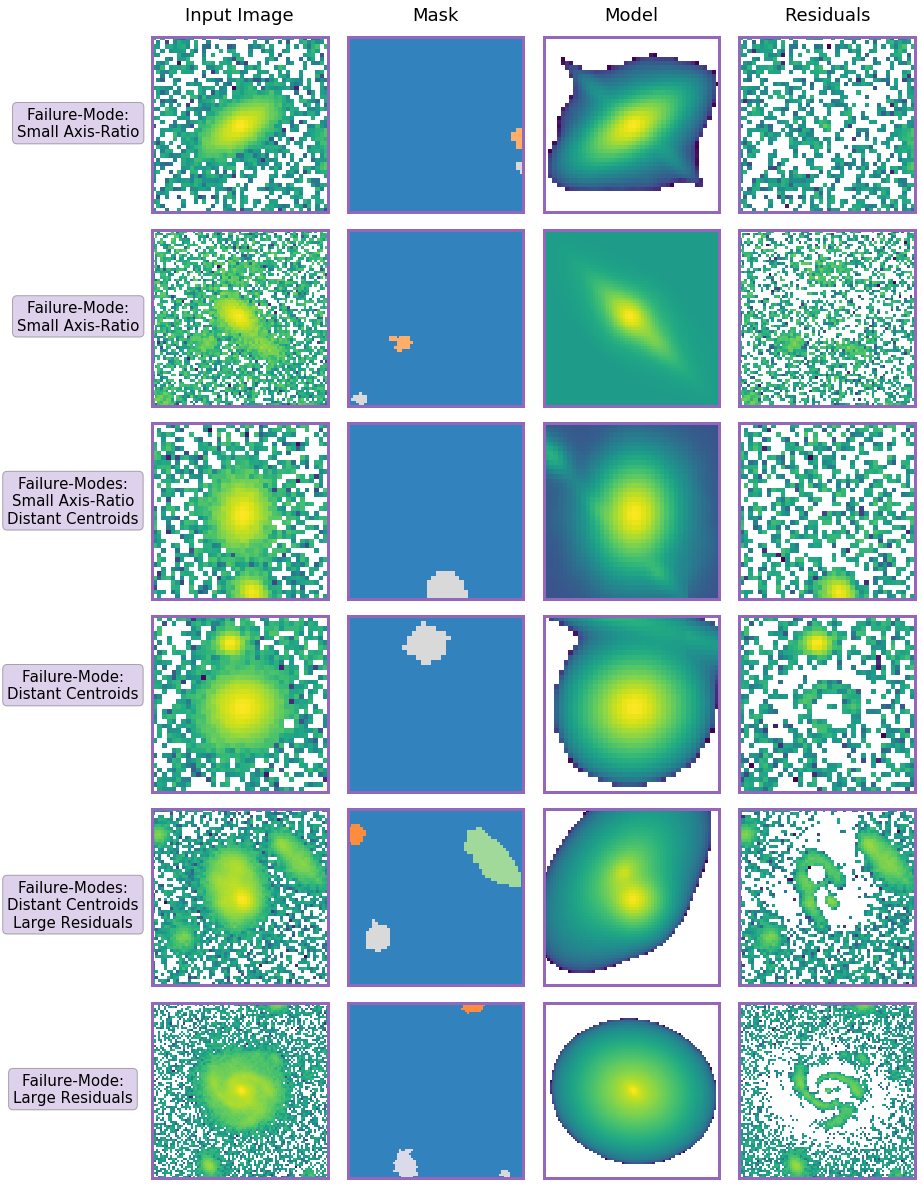}
    \caption{Examples of the GALFIT failure modes identified during visual inspection, drawn from our transfer-learning samples. From left to right, the columns show the input image, the segmentation mask used to exclude secondary sources, the best-fit bulge+disk model, and the residuals. The label to the left of each row gives the failure mode(s) present: an unphysically small axis ratio for one component, an implausibly large separation between the bulge and disk centroids, or large residuals.}
    \label{fig:galfit_failures}
\end{figure*}

The prevalence and appearance of these failure modes were found to be qualitatively similar across the different bands analyzed in this work. Figure~\ref{fig:galfit_failures} shows example galaxies from our transfer-learning samples illustrating each failure mode; a more detailed discussion of their origin is presented in Appendix E of \citet{Morpho_Single_Band}. The quality cuts adopted in \S\ref{sec:galfitting} were specifically designed to identify and remove these problematic fits from the transfer-learning and validation datasets.

\software{PyTorch \citep{pytorch},
          Ignite \citep{ignite},
          MLflow \citep{mlflow},
          NumPy \citep{numpy},
          SciPy \citep{Virtanen2020SciPyPython},
          Astropy \citep{astropy_1,astropy_2},
          Pandas \citep{pandas},
          Scikit-learn \citep{scikitlearn},
          Matplotlib \citep{matplotlib}
          }

\section*{AI Disclosure Statement}
Large language model (LLM) tools, including Claude (Anthropic), were used to assist with editing and rephrasing the text of this manuscript and with writing code. All LLM-assisted text and code were reviewed and revised by the authors, who take full responsibility for the content of this article.

\bibliographystyle{aasjournalv7}
\bibliography{references}

@article{giavalisco1996,
   title={Hubble Space Telescope Imaging of Star-forming Galaxies at Redshifts Z \&gt; 3},
   volume={470},
   ISSN={1538-4357},
   url={http://dx.doi.org/10.1086/177859},
   DOI={10.1086/177859},
   journal={The Astrophysical Journal},
   publisher={American Astronomical Society},
   author={Giavalisco, Mauro and Steidel, Charles C. and Macchetto, F. Duccio},
   year={1996},
   month=Oct, pages={189} }

@article{nedkova_2024,
    author = {Nedkova, Kalina V and Häußler, Boris and Marchesini, Danilo and Brammer, Gabriel B and Feinstein, Adina D and Johnston, Evelyn J and Kartaltepe, Jeyhan S and Koekemoer, Anton M and Martis, Nicholas S and Muzzin, Adam and Rafelski, Marc and Shipley, Heath V and Skelton, Rosalind E and Stefanon, Mauro and van der Wel, Arjen and Whitaker, Katherine E},
    title = {Bulge+disc decomposition of HFF and CANDELS galaxies: UVJ diagrams and stellar mass–size relations of galaxy components at 0.2 ≤ z ≤ 1.5},
    journal = {Monthly Notices of the Royal Astronomical Society},
    volume = {532},
    number = {4},
    pages = {3747-3777},
    year = {2024},
    month = {08},
    issn = {0035-8711},
    doi = {10.1093/mnras/stae1702},
    url = {https://doi.org/10.1093/mnras/stae1702},
    eprint = {https://academic.oup.com/mnras/article-pdf/532/4/3747/58665590/stae1702.pdf},
}

@article{papovich2003,
   title={The Internal Ultraviolet–Optical Color Dispersion: Quantifying the Morphological<i>K</i>‐Correction},
   volume={598},
   ISSN={1538-4357},
   url={http://dx.doi.org/10.1086/378976},
   DOI={10.1086/378976},
   number={2},
   journal={The Astrophysical Journal},
   publisher={American Astronomical Society},
   author={Papovich, Casey and Giavalisco, Mauro and Dickinson, Mark and Conselice, Christopher J. and Ferguson, Henry C.},
   year={2003},
   month=Dec, pages={827–847} }

@article{lotz2004,
   title={A New Nonparametric Approach to Galaxy Morphological Classification},
   volume={128},
   ISSN={1538-3881},
   url={http://dx.doi.org/10.1086/421849},
   DOI={10.1086/421849},
   number={1},
   journal={The Astronomical Journal},
   publisher={American Astronomical Society},
   author={Lotz, Jennifer M. and Primack, Joel and Madau, Piero},
   year={2004},
   month={7}, pages={163–182} }

@article{huertas_company2015,
   title={A CATALOG OF VISUAL-LIKE MORPHOLOGIES IN THE 5 CANDELS FIELDS USING DEEP LEARNING},
   volume={221},
   ISSN={1538-4365},
   url={http://dx.doi.org/10.1088/0067-0049/221/1/8},
   DOI={10.1088/0067-0049/221/1/8},
   number={1},
   journal={The Astrophysical Journal Supplement Series},
   publisher={American Astronomical Society},
   author={Huertas-Company, M. and Gravet, R. and Cabrera-Vives, G. and Pérez-González, P. G. and Kartaltepe, J. S. and Barro, G. and Bernardi, M. and Mei, S. and Shankar, F. and Dimauro, P. and Bell, E. F. and Kocevski, D. and Koo, D. C. and Faber, S. M. and Mcintosh, D. H.},
   year={2015},
   month=Oct, pages={8} }

@article{ferguson2004,
   title={The Size Evolution of High-Redshift Galaxies},
   volume={600},
   ISSN={1538-4357},
   url={http://dx.doi.org/10.1086/378578},
   DOI={10.1086/378578},
   number={2},
   journal={The Astrophysical Journal},
   publisher={American Astronomical Society},
   author={Ferguson, Henry C. and Dickinson, Mark and Giavalisco, Mauro and Kretchmer, Claudia and Ravindranath, Swara and Idzi, Rafal and Taylor, Edward and Conselice, Christopher J. and Fall, S. Michael and Gardner, Jonathan P. and Livio, Mario and Madau, Piero and Moustakas, Leonidas A. and Papovich, Casey M. and Somerville, Rachel S. and Spinrad, Hyron and Stern, Daniel},
   year={2004},
   month=jan, pages={L107–L110} }

@article{ravindranath2006,
   title={The Morphological Diversities among Star‐forming Galaxies at High Redshifts in the Great Observatories Origins Deep Survey},
   volume={652},
   ISSN={1538-4357},
   url={http://dx.doi.org/10.1086/507016},
   DOI={10.1086/507016},
   number={2},
   journal={The Astrophysical Journal},
   publisher={American Astronomical Society},
   author={Ravindranath, Swara and Giavalisco, Mauro and Ferguson, Henry C. and Conselice, Christopher and Katz, Neal and Weinberg, Martin and Lotz, Jennifer and Dickinson, Mark and Fall, S. Michael and Mobasher, Bahram and Papovich, Casey},
   year={2006},
   month=Dec, pages={963–980} }

@article{Morpho_Single_Band,
   title={Morphological Parameters and Associated Uncertainties for 8 Million Galaxies in the Hyper Suprime-Cam Wide Survey},
   volume={953},
   ISSN={1538-4357},
   url={http://dx.doi.org/10.3847/1538-4357/acd546},
   DOI={10.3847/1538-4357/acd546},
   number={2},
   journal={The Astrophysical Journal},
   publisher={American Astronomical Society},
   author={Ghosh, Aritra and Urry, C. Megan and Mishra, Aayush and Perreault-Levasseur, Laurence and Natarajan, Priyamvada and Sanders, David B. and Nagai, Daisuke and Tian, Chuan and Cappelluti, Nico and Kartaltepe, Jeyhan S. and Powell, Meredith C. and Rau, Amrit and Treister, Ezequiel},
   year={2023},
   month=aug, pages={134} }

@article{simard_11,
    title = {{A catalog of bulge+disk decompositions and updated photometry for 1.12 million galaxies in the Sloan digital sky survey}},
    year = {2011},
    journal = {Astrophysical Journal, Supplement Series},
    author = {Simard, Luc and Trevor Mendel, J. and Patton, David R. and Ellison, Sara L. and McConnachie, Alan W.},
    number = {1},
    month = {9},
    pages = {11},
    volume = {196},
    publisher = {IOP Publishing},
    url = {ttp://stacks.iop.org/0067-0049/196/i=1/a=11},
    doi = {10.1088/0067-0049/196/1/11},
    issn = {00670049}
}

@article{tarsitano_18,
    title = {{A catalogue of structural and morphological measurements for DES Y1}},
    year = {2018},
    journal = {Monthly Notices of the Royal Astronomical Society},
    author = {Tarsitano, F and Hartley, W G and Amara, A and Bluck, A and Bruderer, C and Carollo, M and Conselice, C and Melchior, P and Moraes, B and Refregier, A and Sevilla-Noarbe, I and Woo, J and Abbott, T M C and Allam, S and Annis, J and Avila, S and Banerji, M and Bertin, E and Brooks, D and Burke, D L and Carnero Rosell, A and Carrasco Kind, M and Carretero, J and Cunha, C E and D’Andrea, C B and da Costa, L N and Davis, C and De Vicente, J and Desai, S and Doel, P and Estrada, J and Frieman, J and Garc{\'{i}}a-Bellido, J and Gruen, D and Gruendl, R A and Gutierrez, G and Hollowood, D and Honscheid, K and James, D J and Jeltema, T and Krause, E and Kuehn, K and Kuropatkin, N and Lahav, O and Maia, M A G and Menanteau, F and Miquel, R and Plazas, A A and Romer, A K and Roodman, A and Sanchez, E and Santiago, B and Schindler, R and Smith, M and Smith, R C and Soares-Santos, M and Sobreira, F and Suchyta, E and Swanson, M E C and Tarle, G and Thomas, D and Vikram, V and Walker, A R},
    number = {2},
    month = {12},
    pages = {2018--2040},
    volume = {481},
    doi = {10.1093/mnras/sty1970},
    issn = {0035-8711}
}

@article{zanisi_21,
    title = {{A deep learning approach to test the small-scale galaxy morphology and its relationship with star formation activity in hydrodynamical simulations}},
    year = {2021},
    journal = {Monthly Notices of the Royal Astronomical Society},
    author = {Zanisi, Lorenzo and Huertas-Company, Marc and Lanusse, François and Bottrell, Connor and Pillepich, Annalisa and Nelson, Dylan and Rodriguez-Gomez, Vicente and Shankar, Francesco and Hernquist, Lars and Dekel, Avishai and Margalef-Bentabol, Berta and Vogelsberger, Mark and Primack, Joel},
    number = {3},
    month = {1},
    pages = {4359--4382},
    volume = {501},
    doi = {10.1093/mnras/staa3864},
    issn = {0035-8711}
}

@article{numpy,
    title = {{Array programming with NumPy}},
    year = {2020},
    journal = {Nature 2020 585:7825},
    author = {Harris, Charles R. and Millman, K. Jarrod and van der Walt, Stéfan J. and Gommers, Ralf and Virtanen, Pauli and Cournapeau, David and Wieser, Eric and Taylor, Julian and Berg, Sebastian and Smith, Nathaniel J. and Kern, Robert and Picus, Matti and Hoyer, Stephan and van Kerkwijk, Marten H. and Brett, Matthew and Haldane, Allan and del R{\'{i}}o, Jaime Fernández and Wiebe, Mark and Peterson, Pearu and G{\'{e}}rard-Marchant, Pierre and Sheppard, Kevin and Reddy, Tyler and Weckesser, Warren and Abbasi, Hameer and Gohlke, Christoph and Oliphant, Travis E.},
    number = {7825},
    month = {9},
    pages = {357--362},
    volume = {585},
    publisher = {Nature Publishing Group},
    url = {https://www.nature.com/articles/s41586-020-2649-2},
    doi = {10.1038/s41586-020-2649-2},
    issn = {1476-4687},
    pmid = {32939066},
    arxivId = {2006.10256}
}

@article{astropy_1,
    title = {{Astropy: A community Python package for astronomy}},
    year = {2013},
    journal = {Astronomy {\&} Astrophysics},
    author = {Robitaille, Thomas P. and Tollerud, Erik J. and Greenfield, Perry and Droettboom, Michael and Bray, Erik and Aldcroft, Tom and Davis, Matt and Ginsburg, Adam and Price-Whelan, Adrian M. and Kerzendorf, Wolfgang E. and Conley, Alexander and Crighton, Neil and Barbary, Kyle and Muna, Demitri and Ferguson, Henry and Grollier, Frédéric and Parikh, Madhura M. and Nair, Prasanth H. and G{\"{u}}nther, Hans M. and Deil, Christoph and Woillez, Julien and Conseil, Simon and Kramer, Roban and Turner, James E.H. and Singer, Leo and Fox, Ryan and Weaver, Benjamin A. and Zabalza, Victor and Edwards, Zachary I. and Azalee Bostroem, K. and Burke, D. J. and Casey, Andrew R. and Crawford, Steven M. and Dencheva, Nadia and Ely, Justin and Jenness, Tim and Labrie, Kathleen and Lim, Pey Lian and Pierfederici, Francesco and Pontzen, Andrew and Ptak, Andy and Refsdal, Brian and Servillat, Mathieu and Streicher, Ole},
    month = {10},
    pages = {A33},
    volume = {558},
    publisher = {EDP Sciences},
    url = {https://www.aanda.org/articles/aa/full_html/2013/10/aa22068-13/aa22068-13.html https://www.aanda.org/articles/aa/abs/2013/10/aa22068-13/aa22068-13.html},
    doi = {10.1051/0004-6361/201322068},
    issn = {0004-6361},
    arxivId = {1307.6212}
}

@article{importance_sampling,
    title = {{Bayesian Estimates of Equation System Parameters: An Application of Integration by Monte Carlo}},
    year = {1978},
    journal = {Econometrica},
    author = {Kloek, T. and van Dijk, H. K.},
    number = {1},
    month = {1},
    pages = {1},
    volume = {46},
    doi = {10.2307/1913641},
    issn = {00129682}
}

@inproceedings{blum_1998,
    title = {{Combining labeled and unlabeled data with co-training}},
    year = {1998},
    booktitle = {Proceedings of the eleventh annual conference on Computational learning theory  - COLT' 98},
    author = {Blum, Avrim and Mitchell, Tom},
    pages = {92--100},
    publisher = {ACM Press},
    address = {New York, New York, USA},
    isbn = {1581130570},
    doi = {10.1145/279943.279962}
}

@inproceedings{pandas,
    title = {{Data Structures for Statistical Computing in Python}},
    year = {2010},
    booktitle = {Proceedings of the 9th Python in Science Conference},
    author = {McKinney, Wes},
    pages = {56--61},
    url = {https://conference.scipy.org/proceedings/scipy2010/mckinney.html},
    doi = {10.25080/Majora-92bf1922-00a}
}

@book{candela_09,
    title = {{Dataset Shift in Machine Learning}},
    year = {2009},
    author = {Qui{\~{n}}onero-Candela, Joaqu{\'{i}}n and Sugiyama, Masashi and Schwaighofer, Anton and Lawrence, Neil D},
    publisher = {The MIT Press},
    isbn = {0262170051}
}

@article{galfit,
    title = {{Detailed Structural Decomposition of Galaxy Images}},
    year = {2002},
    journal = {The Astronomical Journal},
    author = {Peng, Chien Y. and Ho, Luis C. and Impey, Chris D. and Rix, Hans-Walter},
    number = {1},
    month = {7},
    pages = {266--293},
    volume = {124},
    publisher = {IOP Publishing},
    url = {http://stacks.iop.org/1538-3881/124/i=1/a=266},
    doi = {10.1086/340952},
    issn = {00046256}
}

@inproceedings{mlflow,
    title = {{Developments in MLflow}},
    year = {2020},
    booktitle = {Proceedings of the Fourth International Workshop on Data Management for End-to-End Machine Learning},
    author = {Chen, Andrew and Chow, Andy and Davidson, Aaron and DCunha, Arjun and Ghodsi, Ali and Hong, Sue Ann and Konwinski, Andy and Mewald, Clemens and Murching, Siddharth and Nykodym, Tomas and Ogilvie, Paul and Parkhe, Mani and Singh, Avesh and Xie, Fen and Zaharia, Matei and Zang, Richard and Zheng, Juntai and Zumar, Corey},
    month = {6},
    pages = {1--4},
    publisher = {ACM},
    address = {New York, NY, USA},
    isbn = {9781450380232},
    doi = {10.1145/3399579.3399867}
}

@inproceedings{recht_19,
    title = {{Do ImageNet Classifiers Generalize to ImageNet?}},
    year = {2019},
    author = {Recht, Benjamin and Roelofs, Rebecca and Schmidt, Ludwig and Shankar, Vaishaal},
    booktitle = {Proceedings of the 36th International Conference on Machine Learning},
    series = {Proceedings of Machine Learning Research},
    volume = {97},
    pages = {5389--5400},
    url = {http://arxiv.org/abs/1902.10811},
    arxivId = {1902.10811}
}

@inproceedings{gal_2016,
    title = {{Dropout as a Bayesian approximation: Representing model uncertainty in deep learning}},
    year = {2016},
    booktitle = {33rd International Conference on Machine Learning, ICML 2016},
    author = {Gal, Yarin and Ghahramani, Zoubin},
    month = {6},
    pages = {1651--1660},
    volume = {3},
    publisher = {PMLR},
    url = {https://proceedings.mlr.press/v48/gal16.html},
    isbn = {9781510829008},
    issn = {1938-7228},
    arxivId = {1506.02142}
}

@article{euclid_morph,
    title = {{Euclid preparation XXVI: The Euclid Morphology Challenge. Towards structural parameters for billions of galaxies}},
    year = {2023},
    author = {{Euclid Collaboration} and Bretonni{\`{e}}re, H. and Kuchner, U. and Huertas-Company, M. and Merlin, E. and Castellano, M. and Tuccillo, D. and Buitrago, F. and Conselice, C. J. and Boucaud, A. and H{\"{a}}u{\ss}ler, B. and K{\"{u}}mmel, M. and Hartley, W. G. and Ayllon, A. Alvarez and Bertin, E. and Ferrari, F. and Ferreira, L. and Gavazzi, R. and Hern{\'{a}}ndez-Lang, D. and Lucatelli, G. and Robotham, A. S. G. and Schefer, M. and Wang, L. and Cabanac, R. and Duc, P. -A. and Fotopoulou, S. and Kruk, S. and La Marca, A. and Margalef-Bentabol, B. and Marleau, F. R. and Tortora, C. and Aghanim, N. and Amara, A. and Auricchio, N. and Azzollini, R. and Baldi, M. and Bender, R. and Bodendorf, C. and Branchini, E. and Brescia, M. and Brinchmann, J. and Camera, S. and Capobianco, V. and Carbone, C. and Carretero, J. and Castander, F. J. and Cavuoti, S. and Cimatti, A. and Cledassou, R. and Congedo, G. and Conversi, L. and Copin, Y. and Corcione, L. and Courbin, F. and Cropper, M. and Da Silva, A. and Degaudenzi, H. and Dinis, J. and Dubath, F. and Duncan, C. A. J. and Dupac, X. and Dusini, S. and Farrens, S. and Ferriol, S. and Frailis, M. and Franceschi, E. and Fumana, M. and Galeotta, S. and Garilli, B. and Gillis, B. and Giocoli, C. and Grazian, A. and Grupp, F. and Haugan, S. V. H. and Hoekstra, H. and Holmes, W. and Hormuth, F. and Hornstrup, A. and Hudelot, P. and Jahnke, K. and Kermiche, S. and Kiessling, A. and Kohley, R. and Kunz, M. and Kurki-Suonio, H. and Ligori, S. and Lilje, P. B. and Lloro, I. and Mansutti, O. and Marggraf, O. and Markovic, K. and Marulli, F. and Massey, R. and McCracken, H. J. and Medinaceli, E. and Melchior, M. and Meneghetti, M. and Meylan, G. and Moresco, M. and Moscardini, L. and Munari, E. and Niemi, S. M. and Padilla, C. and Paltani, S. and Pasian, F. and Pedersen, K. and Percival, W. and Pettorino, V. and Polenta, G. and Poncet, M. and Pozzetti, L. and Raison, F. and Rebolo, R. and Renzi, A. and Rhodes, J. and Riccio, G. and Romelli, E. and Rosset, C. and Rossetti, E. and Saglia, R. and Sapone, D. and Sartoris, B. and Schneider, P. and Secroun, A. and Seidel, G. and Sirignano, C. and Sirri, G. and Skottfelt, J. and Starck, J. -L. and Tallada-Cresp{\'{i}}, P. and Taylor, A. N. and Tereno, I. and Toledo-Moreo, R. and Tutusaus, I. and Valentijn, E. A. and Valenziano, L. and Vassallo, T. and Wang, Y. and Weller, J. and Zamorani, G. and Zoubian, J. and Andreon, S. and Bardelli, S. and Colodro-Conde, C. and Di Ferdinando, D. and Graci{\'{a}}-Carpio, J. and Lindholm, V. and Mauri, N. and Mei, S. and Scottez, V. and Zucca, E. and Baccigalupi, C. and Ballardini, M. and Bernardeau, F. and Biviano, A. and Borgani, S. and Borlaff, A. S. and Burigana, C. and Cappi, A. and Carvalho, C. S. and Casas, S. and Castignani, G. and Cooray, A. R. and Coupon, J. and Courtois, H. M. and Davini, S. and De Lucia, G. and Desprez, G. and Escartin, J. A. and Escoffier, S. and Fabricius, M. and Farina, M. and Fontana, A. and Ganga, K. and Garcia-Bellido, J. and George, K. and Gozaliasl, G. and Hildebrandt, H. and Hook, I. and Ilbert, O. and Ili{\'{c}}, S. and Joachimi, B. and Kansal, V. and Keihanen, E. and Kirkpatrick, C. C. and Loureiro, A. and Macias-Perez, J. and Magliocchetti, M. and Maoli, R. and Marcin, S. and Martinelli, M. and Martinet, N. and Maturi, M. and Monaco, P. and Morgante, G. and Nadathur, S. and Nucita, A. A. and Patrizii, L. and Popa, V. and Porciani, C. and Potter, D. and Pourtsidou, A. and P{\"{o}}ntinen, M. and Reimberg, P. and S{\'{a}}nchez, A. G. and Sakr, Z. and Schirmer, M. and Sefusatti, E. and Sereno, M. and Stadel, J. and Teyssier, R. and Valiviita, J. and van Mierlo, S. E. and Veropalumbo, A. and Viel, M. and Weaver, J. R. and Scott, D.},
    journal = {Astronomy and Astrophysics},
    volume = {671},
    pages = {A102},
    doi = {10.1051/0004-6361/202245042},
    url = {http://arxiv.org/abs/2209.12907},
    arxivId = {2209.12907}
}

@article{galapagos,
    title = {{Galapagos-2/Galfitm/Gama – Multi-wavelength measurement of galaxy structure: Separating the properties of spheroid and disk components in modern surveys}},
    year = {2022},
    journal = {Astronomy {\&} Astrophysics},
    author = {H{\"{a}}u{\ss}ler, Boris and Vika, Marina and Bamford, Steven P. and Johnston, Evelyn J. and Brough, Sarah and Casura, Sarah and Holwerda, Benne W. and Kelvin, Lee S. and Popescu, Cristina},
    month = {8},
    pages = {A92},
    volume = {664},
    doi = {10.1051/0004-6361/202142935},
    issn = {0004-6361}
}

@article{kelvin_12_gama,
    author = {Kelvin, Lee S. and Driver, Simon P. and Robotham, Aaron S. G. and Hill, David T. and Alpaslan, Mehmet and Baldry, Ivan K. and Bamford, Steven P. and Bland-Hawthorn, Joss and Brough, Sarah and Graham, Alister W. and H{\"{a}}u{\ss}ler, Boris and Hopkins, Andrew M. and Liske, Jochen and Loveday, Jon and Norberg, Peder and Phillipps, Steven and Popescu, Cristina C. and Prescott, Matthew and Taylor, Edward N. and Tuffs, Richard J.},
    title = {Galaxy And Mass Assembly (GAMA): Structural Investigation of Galaxies via Model Analysis},
    journal = {Monthly Notices of the Royal Astronomical Society},
    volume = {421},
    number = {2},
    pages = {1007-1039},
    year = {2012},
    month = {04},
    issn = {0035-8711},
    doi = {10.1111/j.1365-2966.2012.20355.x},
    url = {https://doi.org/10.1111/j.1365-2966.2012.20355.x},
    eprint = {https://academic.oup.com/mnras/article-pdf/421/2/1007/3915366/mnras0421-1007.pdf},
}

@article{vika2013_megamorph,
    author = {Vika, Marina and Bamford, Steven P. and H{\"{a}}u{\ss}ler, Boris and Rojas, Alex L. and Borch, Andrea and Nichol, Robert C.},
    title = {MegaMorph – multiwavelength measurement of galaxy structure. Sérsic profile fits to galaxies near and far},
    journal = {Monthly Notices of the Royal Astronomical Society},
    volume = {435},
    number = {1},
    pages = {623-649},
    year = {2013},
    month = {10},
    issn = {0035-8711},
    doi = {10.1093/mnras/stt1320},
    url = {https://doi.org/10.1093/mnras/stt1320},
    eprint = {https://academic.oup.com/mnras/article-pdf/435/1/623/3892243/stt1320.pdf},
}

@article{haubler_2013_megamorph,
    author = {H{\"{a}}u{\ss}ler, Boris and Bamford, Steven P. and Vika, Marina and Rojas, Alex L. and Barden, Marco and Kelvin, Lee S. and Alpaslan, Mehmet and Robotham, Aaron S. G. and Driver, Simon P. and Baldry, I. K. and Brough, Sarah and Hopkins, Andrew M. and Liske, Jochen and Nichol, Robert C. and Popescu, Cristina C. and Tuffs, Richard J.},
    title = {MegaMorph – multiwavelength measurement of galaxy structure: complete Sérsic profile information from modern surveys},
    journal = {Monthly Notices of the Royal Astronomical Society},
    volume = {430},
    number = {1},
    pages = {330-369},
    year = {2013},
    month = {03},
    issn = {0035-8711},
    doi = {10.1093/mnras/sts633},
    url = {https://doi.org/10.1093/mnras/sts633},
    eprint = {https://academic.oup.com/mnras/article-pdf/430/1/330/3081005/sts633.pdf},
}

@misc{igel2026environmentalsculptinggalaxystructure,
      title={Environmental Sculpting of Galaxy Structure at Fixed Stellar Mass: A Multi-Scale Analysis Across Cosmic Time using 3 Million HSC Galaxies}, 
      author={Caitlin Igel and Aritra Ghosh and Andrew J. Connolly and Brant Robertson and C. Megan Urry and Louise O. V. Edwards and Rhythm Shimakawa},
      year={2026},
      eprint={2605.18963},
      archivePrefix={arXiv},
      primaryClass={astro-ph.GA},
      url={https://arxiv.org/abs/2605.18963}, 
}

@article{Ghosh_2024_env,
doi = {0.3847/1538-4357/ad596f},
url = {https://doi.org/10.3847/1538-4357/ad596f},
year = {2024},
month = {aug},
publisher = {The American Astronomical Society},
volume = {971},
number = {2},
pages = {142},
author = {Ghosh, Aritra and Urry, C. Megan and Powell, Meredith C. and Shimakawa, Rhythm and van den Bosch, Frank C. and Nagai, Daisuke and Mitra, Kaustav and Connolly, Andrew J.},
title = {Denser Environments Cultivate Larger Galaxies: A Comprehensive Study beyond the Local Universe with 3 Million Hyper Suprime-Cam Galaxies},
journal = {The Astrophysical Journal}
}

@article{kennedy2016_gama,
    author = {Kennedy, Rebecca and Bamford, Steven P. and H{\"{a}}u{\ss}ler, Boris and Baldry, Ivan and Bremer, Malcolm and Brough, Sarah and Brown, Michael J. I. and Driver, Simon and Duncan, Kenneth and Graham, Alister W. and Holwerda, Benne W. and Hopkins, Andrew M. and Kelvin, Lee S. and Lange, Rebecca and Phillipps, Steven and Vika, Marina and Vulcani, Benedetta},
    title = {Galaxy And Mass Assembly (GAMA): understanding the wavelength dependence of galaxy structure with bulge-disc decompositions},
    journal = {Monthly Notices of the Royal Astronomical Society},
    volume = {460},
    number = {4},
    pages = {3458-3471},
    year = {2016},
    month = {08},
    issn = {0035-8711},
    doi = {10.1093/mnras/stw1176},
    url = {https://doi.org/10.1093/mnras/stw1176},
    eprint = {https://academic.oup.com/mnras/article-pdf/460/4/3458/8116876/stw1176.pdf},
}

@article{vulcani2014_gama,
    author = {Vulcani, Benedetta and Bamford, Steven P. and H{\"{a}}u{\ss}ler, Boris and Vika, Marina and Rojas, Alex and Agius, Nicola K. and Baldry, Ivan and Bauer, Amanda E. and Brown, Michael J. I. and Driver, Simon and Graham, Alister W. and Kelvin, Lee S. and Liske, Jochen and Loveday, Jon and Popescu, Cristina C. and Robotham, Aaron S. G. and Tuffs, Richard J.},
    title = {Galaxy And Mass Assembly (GAMA): the wavelength-dependent sizes and profiles of galaxies revealed by MegaMorph},
    journal = {Monthly Notices of the Royal Astronomical Society},
    volume = {441},
    number = {2},
    pages = {1340-1362},
    year = {2014},
    month = {06},
    issn = {0035-8711},
    doi = {10.1093/mnras/stu632},
    url = {https://doi.org/10.1093/mnras/stu632},
    eprint = {https://academic.oup.com/mnras/article-pdf/441/2/1340/3693312/stu632.pdf},
}

@article{gamornet_paper,
    title = {{Galaxy Morphology Network: A Convolutional Neural Network Used to Study Morphology and Quenching in ∼100,000 SDSS and ∼20,000 CANDELS Galaxies}},
    year = {2020},
    journal = {The Astrophysical Journal},
    author = {Ghosh, Aritra and Urry, C. Megan and Wang, Zhengdong and Schawinski, Kevin and Turp, Dennis and Powell, Meredith C.},
    number = {2},
    month = {6},
    pages = {112},
    volume = {895},
    url = {https://iopscience.iop.org/article/10.3847/1538-4357/ab8a47},
    doi = {10.3847/1538-4357/ab8a47},
    issn = {1538-4357}
}

@article{wuyts_11,
    title = {{GALAXY STRUCTURE AND MODE OF STAR FORMATION IN THE SFR-MASS PLANE FROM z ∼ 2.5 TO z ∼ 0.1}},
    year = {2011},
    journal = {The Astrophysical Journal},
    author = {Wuyts, Stijn and F{\"{o}}rster Schreiber, Natascha M. and van der Wel, Arjen and Magnelli, Benjamin and Guo, Yicheng and Genzel, Reinhard and Lutz, Dieter and Aussel, Hervé and Barro, Guillermo and Berta, Stefano and Cava, Antonio and Graci{\'{a}}-Carpio, Javier and Hathi, Nimish P. and Huang, Kuang-Han and Kocevski, Dale D. and Koekemoer, Anton M. and Lee, Kyoung-Soo and Le Floc'h, Emeric and McGrath, Elizabeth J. and Nordon, Raanan and Popesso, Paola and Pozzi, Francesca and Riguccini, Laurie and Rodighiero, Giulia and Saintonge, Amelie and Tacconi, Linda},
    number = {2},
    month = {12},
    pages = {96},
    volume = {742},
    url = {https://iopscience.iop.org/article/10.1088/0004-637X/742/2/96},
    doi = {10.1088/0004-637X/742/2/96},
    issn = {0004-637X}
}

@article{walmsley_decals,
    title = {{Galaxy Zoo DECaLS: Detailed visual morphology measurements from volunteers and deep learning for 314 000 galaxies}},
    year = {2021},
    journal = {Monthly Notices of the Royal Astronomical Society},
    author = {Walmsley, Mike and Lintott, Chris and G{\'{e}}ron, Tobias and Kruk, Sandor and Krawczyk, Coleman and Willett, Kyle W and Bamford, Steven and Kelvin, Lee S and Fortson, Lucy and Gal, Yarin and Keel, William and Masters, Karen L and Mehta, Vihang and Simmons, Brooke D and Smethurst, Rebecca and Smith, Lewis and Baeten, Elisabeth M and Macmillan, Christine},
    number = {3},
    month = {12},
    pages = {3966--3988},
    volume = {509},
    doi = {10.1093/mnras/stab2093},
    issn = {0035-8711}
}

@article{gampen_software_paper,
    title = {{GaMPEN: A Machine-learning Framework for Estimating Bayesian Posteriors of Galaxy Morphological Parameters}},
    year = {2022},
    journal = {The Astrophysical Journal},
    author = {Ghosh, Aritra and Urry, C. Megan and Rau, Amrit and Perreault-Levasseur, Laurence and Cranmer, Miles and Schawinski, Kevin and Stark, Dominic and Tian, Chuan and Ofman, Ryan and Ananna, Tonima Tasnim and Auge, Connor and Cappelluti, Nico and Sanders, David B. and Treister, Ezequiel},
    number = {2},
    month = {8},
    pages = {138},
    volume = {935},
    url = {https://iopscience.iop.org/article/10.3847/1538-4357/ac7f9e},
    doi = {10.3847/1538-4357/ac7f9e},
    issn = {0004-637X}
}

@misc{ignite,
    title = {{High-level library to help with training neural networks in PyTorch}},
    year = {2020},
    booktitle = {GitHub repository},
    author = {Fomin, V and Anmol, J and Desroziers, S and Kriss, J and Tejani, A},
    publisher = {GitHub},
    howpublished = {https://github.com/pytorch/ignite}
}

@article{hsc_sersic,
    title = {{Hyper Suprime-Cam Subaru Strategic Program: A Mass-dependent Slope of the Galaxy Size−Mass Relation at z {\&}lt; 1}},
    year = {2021},
    journal = {The Astrophysical Journal},
    author = {Kawinwanichakij, Lalitwadee and Silverman, John D. and Ding, Xuheng and George, Angelo and Damjanov, Ivana and Sawicki, Marcin and Tanaka, Masayuki and Taranu, Dan S. and Birrer, Simon and Huang, Song and Li, Junyao and Onodera, Masato and Shibuya, Takatoshi and Yasuda, Naoki},
    number = {1},
    month = {11},
    pages = {38},
    volume = {921},
    url = {https://iopscience.iop.org/article/10.3847/1538-4357/ac1f21},
    doi = {10.3847/1538-4357/ac1f21},
    issn = {0004-637X}
}

@article{lsst,
    title = {{LSST: From Science Drivers to Reference Design and Anticipated Data Products}},
    year = {2019},
    journal = {The Astrophysical Journal},
    author = {Ivezi{\'{c}}, Zeljko and Kahn, Steven M. and Tyson, J. Anthony and Abel, Bob and Acosta, Emily and Allsman, Robyn and Alonso, David and AlSayyad, Yusra and Anderson, Scott F. and Andrew, John and P. Angel, James Roger and Angeli, George Z. and Ansari, Reza and Antilogus, Pierre and Araujo, Constanza and Armstrong, Robert and Arndt, Kirk T. and Astier, Pierre and Aubourg, Éric and Auza, Nicole and Axelrod, Tim S. and Bard, Deborah J. and Barr, Jeff D. and Barrau, Aurelian and Bartlett, James G. and Bauer, Amanda E. and Bauman, Brian J. and Baumont, Sylvain and Bechtol, Ellen and Bechtol, Keith and Becker, Andrew C. and Becla, Jacek and Beldica, Cristina and Bellavia, Steve and Bianco, Federica B. and Biswas, Rahul and Blanc, Guillaume and Blazek, Jonathan and Blandford, Roger D. and Bloom, Josh S. and Bogart, Joanne and Bond, Tim W. and Booth, Michael T. and Borgland, Anders W. and Borne, Kirk and Bosch, James F. and Boutigny, Dominique and Brackett, Craig A. and Bradshaw, Andrew and Brandt, William Nielsen and Brown, Michael E. and Bullock, James S. and Burchat, Patricia and Burke, David L. and Cagnoli, Gianpietro and Calabrese, Daniel and Callahan, Shawn and Callen, Alice L. and Carlin, Jeffrey L. and Carlson, Erin L. and Chandrasekharan, Srinivasan and Charles-Emerson, Glenaver and Chesley, Steve and Cheu, Elliott C. and Chiang, Hsin-Fang and Chiang, James and Chirino, Carol and Chow, Derek and Ciardi, David R. and Claver, Charles F. and Cohen-Tanugi, Johann and Cockrum, Joseph J. and Coles, Rebecca and Connolly, Andrew J. and Cook, Kem H. and Cooray, Asantha and Covey, Kevin R. and Cribbs, Chris and Cui, Wei and Cutri, Roc and Daly, Philip N. and Daniel, Scott F. and Daruich, Felipe and Daubard, Guillaume and Daues, Greg and Dawson, William and Delgado, Francisco and Dellapenna, Alfred and Peyster, Robert de and Val-Borro, Miguel de and Digel, Seth W. and Doherty, Peter and Dubois, Richard and Dubois-Felsmann, Gregory P. and Durech, Josef and Economou, Frossie and Eifler, Tim and Eracleous, Michael and Emmons, Benjamin L. and Neto, Angelo Fausti and Ferguson, Henry and Figueroa, Enrique and Fisher-Levine, Merlin and Focke, Warren and Foss, Michael D. and Frank, James and Freemon, Michael D. and Gangler, Emmanuel and Gawiser, Eric and Geary, John C. and Gee, Perry and Geha, Marla and Gessner, Charles J. B. and Gibson, Robert R. and Gilmore, D. Kirk and Glanzman, Thomas and Glick, William and Goldina, Tatiana and Goldstein, Daniel A. and Goodenow, Iain and Graham, Melissa L. and Gressler, William J. and Gris, Philippe and Guy, Leanne P. and Guyonnet, Augustin and Haller, Gunther and Harris, Ron and Hascall, Patrick A. and Haupt, Justine and Hernandez, Fabio and Herrmann, Sven and Hileman, Edward and Hoblitt, Joshua and Hodgson, John A. and Hogan, Craig and Howard, James D. and Huang, Dajun and Huffer, Michael E. and Ingraham, Patrick and Innes, Walter R. and Jacoby, Suzanne H. and Jain, Bhuvnesh and Jammes, Fabrice and Jee, James and Jenness, Tim and Jernigan, Garrett and Jevremovi{\'{c}}, Darko and Johns, Kenneth and Johnson, Anthony S. and Johnson, Margaret W. G. and Jones, R. Lynne and Juramy-Gilles, Claire and Juri{\'{c}}, Mario and Kalirai, Jason S. and Kallivayalil, Nitya J. and Kalmbach, Bryce and Kantor, Jeffrey P. and Karst, Pierre and Kasliwal, Mansi M. and Kelly, Heather and Kessler, Richard and Kinnison, Veronica and Kirkby, David and Knox, Lloyd and Kotov, Ivan V. and Krabbendam, Victor L. and Krughoff, K. Simon and Kub{\'{a}}nek, Petr and Kuczewski, John and Kulkarni, Shri and Ku, John and Kurita, Nadine R. and Lage, Craig S. and Lambert, Ron and Lange, Travis and Langton, J. Brian and Guillou, Laurent Le and Levine, Deborah and Liang, Ming and Lim, Kian-Tat and Lintott, Chris J. and Long, Kevin E. and Lopez, Margaux and Lotz, Paul J. and Lupton, Robert H. and Lust, Nate B. and MacArthur, Lauren A. and Mahabal, Ashish and Mandelbaum, Rachel and Markiewicz, Thomas W. and Marsh, Darren S. and Marshall, Philip J. and Marshall, Stuart and May, Morgan and McKercher, Robert and McQueen, Michelle and Meyers, Joshua and Migliore, Myriam and Miller, Michelle and Mills, David J. and Miraval, Connor and Moeyens, Joachim and Moolekamp, Fred E. and Monet, David G. and Moniez, Marc and Monkewitz, Serge and Montgomery, Christopher and Morrison, Christopher B. and Mueller, Fritz and Muller, Gary P. and Arancibia, Freddy Muñoz and Neill, Douglas R. and Newbry, Scott P. and Nief, Jean-Yves and Nomerotski, Andrei and Nordby, Martin and O’Connor, Paul and Oliver, John and Olivier, Scot S. and Olsen, Knut and O’Mullane, William and Ortiz, Sandra and Osier, Shawn and Owen, Russell E. and Pain, Reynald and Palecek, Paul E. and Parejko, John K. and Parsons, James B. and Pease, Nathan M. and Peterson, J. Matt and Peterson, John R. and Petravick, Donald L. and Petrick, M. E. Libby and Petry, Cathy E. and Pierfederici, Francesco and Pietrowicz, Stephen and Pike, Rob and Pinto, Philip A. and Plante, Raymond and Plate, Stephen and Plutchak, Joel P. and Price, Paul A. and Prouza, Michael and Radeka, Veljko and Rajagopal, Jayadev and Rasmussen, Andrew P. and Regnault, Nicolas and Reil, Kevin A. and Reiss, David J. and Reuter, Michael A. and Ridgway, Stephen T. and Riot, Vincent J. and Ritz, Steve and Robinson, Sean and Roby, William and Roodman, Aaron and Rosing, Wayne and Roucelle, Cecille and Rumore, Matthew R. and Russo, Stefano and Saha, Abhijit and Sassolas, Benoit and Schalk, Terry L. and Schellart, Pim and Schindler, Rafe H. and Schmidt, Samuel and Schneider, Donald P. and Schneider, Michael D. and Schoening, William and Schumacher, German and Schwamb, Megan E. and Sebag, Jacques and Selvy, Brian and Sembroski, Glenn H. and Seppala, Lynn G. and Serio, Andrew and Serrano, Eduardo and Shaw, Richard A. and Shipsey, Ian and Sick, Jonathan and Silvestri, Nicole and Slater, Colin T. and Smith, J. Allyn and Smith, R. Chris and Sobhani, Shahram and Soldahl, Christine and Storrie-Lombardi, Lisa and Stover, Edward and Strauss, Michael A. and Street, Rachel A. and Stubbs, Christopher W. and Sullivan, Ian S. and Sweeney, Donald and Swinbank, John D. and Szalay, Alexander and Takacs, Peter and Tether, Stephen A. and Thaler, Jon J. and Thayer, John Gregg and Thomas, Sandrine and Thornton, Adam J. and Thukral, Vaikunth and Tice, Jeffrey and Trilling, David E. and Turri, Max and Berg, Richard Van and Berk, Daniel Vanden and Vetter, Kurt and Virieux, Francoise and Vucina, Tomislav and Wahl, William and Walkowicz, Lucianne and Walsh, Brian and Walter, Christopher W. and Wang, Daniel L. and Wang, Shin-Yawn and Warner, Michael and Wiecha, Oliver and Willman, Beth and Winters, Scott E. and Wittman, David and Wolff, Sidney C. and Wood-Vasey, W. Michael and Wu, Xiuqin and Xin, Bo and Yoachim, Peter and Zhan, Hu},
    number = {2},
    month = {3},
    pages = {111},
    volume = {873},
    doi = {10.3847/1538-4357/ab042c},
    issn = {1538-4357}
}

@article{matplotlib,
    title = {{Matplotlib: A 2D Graphics Environment}},
    year = {2007},
    journal = {Computing in Science {\&} Engineering},
    author = {Hunter, John D.},
    number = {3},
    pages = {90--95},
    volume = {9},
    doi = {10.1109/MCSE.2007.55},
    issn = {1521-9615}
}

@article{ml_pz,
    title = {{Measuring photometric redshifts using galaxy images and Deep Neural Networks}},
    year = {2016},
    journal = {Astronomy and Computing},
    author = {Hoyle, B.},
    month = {7},
    pages = {34--40},
    volume = {16},
    publisher = {Elsevier},
    url = {https://www.sciencedirect.com/science/article/pii/S221313371630021X},
    doi = {10.1016/j.ascom.2016.03.006},
    issn = {22131337}
}

@inproceedings{taori_20,
    title = {{Measuring Robustness to Natural Distribution Shifts in Image Classification}},
    year = {2020},
    booktitle = {Advances in Neural Information Processing Systems},
    author = {Taori, Rohan and Dave, Achal and Shankar, Vaishaal and Carlini, Nicholas and Recht, Benjamin and Schmidt, Ludwig},
    editor = {Larochelle, H and Ranzato, M and Hadsell, R and Balcan, M F and Lin, H},
    pages = {18583--18599},
    volume = {33},
    publisher = {Curran Associates, Inc.},
    url = {https://proceedings.neurips.cc/paper/2020/file/d8330f857a17c53d217014ee776bfd50-Paper.pdf},
    arxivId = {2007.00644}
}

@article{powell_17,
    title = {{MORPHOLOGY AND THE COLOR–MASS DIAGRAM AS CLUES TO GALAXY EVOLUTION AT z ∼ 1}},
    year = {2017},
    journal = {The Astrophysical Journal},
    author = {Powell, Meredith C. and Urry, C. Megan and Cardamone, Carolin N. and Simmons, Brooke D. and Schawinski, Kevin and Young, Sydney and Kawakatsu, Mari},
    number = {1},
    month = {1},
    pages = {22},
    volume = {835},
    url = {https://iopscience.iop.org/article/10.3847/1538-4357/835/1/22},
    doi = {10.3847/1538-4357/835/1/22},
    issn = {1538-4357}
}

@article{mizuki,
    title = {{PHOTOMETRIC REDSHIFT WITH BAYESIAN PRIORS ON PHYSICAL PROPERTIES OF GALAXIES}},
    year = {2015},
    journal = {The Astrophysical Journal},
    author = {Tanaka, Masayuki},
    number = {1},
    month = {2},
    pages = {20},
    volume = {801},
    doi = {10.1088/0004-637X/801/1/20},
    issn = {1538-4357}
}

@article{photoz_hsc_pdr2,
    title = {{Photometric Redshifts for the Hyper Suprime-Cam Subaru Strategic Program Data Release 2}},
    year = {2020},
    journal = {Publ. Astron. Soc. Japan},
    author = {Nishizawa, Atsushi J. and Hsieh, Bau-Ching and Tanaka, Masayuki and Takata, Tadafumi},
    number = {0},
    month = {2},
    pages = {1--12},
    url = {https://arxiv.org/abs/2003.01511v2},
    doi = {10.48550/arxiv.2003.01511},
    arxivId = {2003.01511}
}

@incollection{pytorch,
    title = {{PyTorch: An Imperative Style, High-Performance Deep Learning Library}},
    year = {2019},
    booktitle = {Advances in Neural Information Processing Systems 32},
    author = {Paszke, Adam and Gross, Sam and Massa, Francisco and Lerer, Adam and Bradbury, James and Chanan, Gregory and Killeen, Trevor and Lin, Zeming and Gimelshein, Natalia and Antiga, Luca and Desmaison, Alban and Kopf, Andreas and Yang, Edward and DeVito, Zachary and Raison, Martin and Tejani, Alykhan and Chilamkurthy, Sasank and Steiner, Benoit and Fang, Lu and Bai, Junjie and Chintala, Soumith},
    editor = {Wallach, H and Larochelle, H and Beygelzimer, A and d Alch{\'{e}}-Buc, F and Fox, E and Garnett, R},
    pages = {8024--8035},
    publisher = {Curran Associates, Inc.}
}

@article{scikitlearn,
    title = {{Scikit-learn: Machine Learning in Python}},
    year = {2011},
    journal = {Journal of Machine Learning Research},
    author = {Pedregosa, F. and Varoquaux, G. and Gramfort, A. and Michel, V. and Thirion, B. and Grisel, O.},
    pages = {2825--2830},
    volume = {12}
}

@article{hsc_pdr2,
    title = {{Second data release of the Hyper Suprime-Cam Subaru Strategic Program}},
    year = {2019},
    journal = {Publications of the Astronomical Society of Japan},
    author = {Aihara, Hiroaki and Alsayyad, Yusra and Ando, Makoto and Armstrong, Robert and Bosch, James and Egami, Eiichi and Furusawa, Hisanori and Furusawa, Junko and Goulding, Andy and Harikane, Yuichi and Hikage, Chiaki and Ho, Paul T.P. and Hsieh, Bau Ching and Huang, Song and Ikeda, Hiroyuki and Imanishi, Masatoshi and Ito, Kei and Iwata, Ikuru and Jaelani, Anton T. and Kakuma, Ryota and Kawana, Kojiro and Kikuta, Satoshi and Kobayashi, Umi and Koike, Michitaro and Komiyama, Yutaka and Li, Xiangchong and Liang, Yongming and Lin, Yen Ting and Luo, Wentao and Lupton, Robert and Lust, Nate B. and Macarthur, Lauren A. and Matsuoka, Yoshiki and Mineo, Sogo and Miyatake, Hironao and Miyazaki, Satoshi and More, Surhud and Murata, Ryoma and Namiki, Shigeru V. and Nishizawa, Atsushi J. and Oguri, Masamune and Okabe, Nobuhiro and Okamoto, Sakurako and Okura, Yuki and Ono, Yoshiaki and Onodera, Masato and Onoue, Masafusa and Osato, Ken and Ouchi, Masami and Shibuya, Takatoshi and Strauss, Michael A. and Sugiyama, Naoshi and Suto, Yasushi and Takada, Masahiro and Takagi, Yuhei and Takata, Tadafumi and Takita, Satoshi and Tanaka, Masayuki and Terai, Tsuyoshi and Toba, Yoshiki and Uchiyama, Hisakazu and Utsumi, Yousuke and Wang, Shiang Yu and Wang, Wenting and Yamada, Yoshihiko},
    number = {6},
    month = {12},
    pages = {114--115},
    volume = {71},
    publisher = {Oxford Academic},
    url = {https://academic.oup.com/pasj/article/71/6/114/5602617},
    doi = {10.1093/PASJ/PSZ103},
    issn = {0004-6264},
    arxivId = {1905.12221}
}

@article{s_extract,
    title = {{SExtractor: Software for source extraction}},
    year = {1996},
    journal = {Astronomy and Astrophysics Supplement Series},
    author = {Bertin, E. and Arnouts, S.},
    number = {2},
    month = {6},
    pages = {393--404},
    volume = {117},
    url = {http://aas.aanda.org/10.1051/aas:1996164},
    doi = {10.1051/aas:1996164},
    issn = {03650138}
}

@incollection{spearman,
    title = {{Spearman Rank Correlation Coefficient}},
    year = {2008},
    booktitle = {The Concise Encyclopedia of Statistics},
    author = {Dodge, Yadolah},
    pages = {502--505},
    publisher = {Springer New York},
    url = {https://doi.org/10.1007/978-0-387-32833-1_379},
    address = {New York, NY},
    isbn = {978-0-387-32833-1},
    doi = {10.1007/978-0-387-32833-1{\_}379}
}

@article{ml_sz,
    title = {{Star-galaxy classification using deep convolutional neural networks}},
    year = {2017},
    journal = {Monthly Notices of the Royal Astronomical Society},
    author = {Kim, Edward J. and Brunner, Robert J.},
    number = {4},
    month = {2},
    pages = {4463--4475},
    volume = {464},
    publisher = {Oxford University Press},
    url = {https://academic.oup.com/mnras/article-lookup/doi/10.1093/mnras/stw2672},
    doi = {10.1093/mnras/stw2672},
    issn = {13652966}
}

@article{vdw_12,
    title = {{Structural parameters of galaxies in candels}},
    year = {2012},
    journal = {Astrophysical Journal, Supplement Series},
    author = {van der Wel, A. and Bell, E. F. and H{\"{a}}u{\ss}ler, B. and McGrath, E. J. and Chang, Yu Yen and Guo, Yicheng and McIntosh, D. H. and Rix, H. W. and Barden, M. and Cheung, E. and Faber, S. M. and Ferguson, H. C. and Galametz, A. and Grogin, N. A. and Hartley, W. and Kartaltepe, J. S. and Kocevski, D. D. and Koekemoer, A. M. and Lotz, J. and Mozena, M. and Peth, M. A. and Peng, Chien Y.},
    number = {2},
    month = {12},
    pages = {24},
    volume = {203},
    publisher = {IOP Publishing},
    url = {http://stacks.iop.org/0067-0049/203/i=2/a=24},
    doi = {10.1088/0067-0049/203/2/24},
    issn = {00670049}
}

@article{shimakawa_2021,
    title = {{Subaru Hyper Suprime-Cam revisits the large-scale environmental dependence on galaxy morphology over 360 deg2 at <i>z</i> = 0.3–0.6}},
    year = {2021},
    journal = {Publications of the Astronomical Society of Japan},
    author = {Shimakawa, Rhythm and Tanaka, Takumi S and Toshikage, Seiji and Tanaka, Masayuki},
    number = {6},
    month = {12},
    pages = {1575--1588},
    volume = {73},
    url = {https://academic.oup.com/pasj/article/73/6/1575/6406719},
    doi = {10.1093/pasj/psab097},
    issn = {0004-6264}
}

@article{simmons_08,
    title = {{The Accuracy of Morphological Decomposition of Active Galactic Nucleus Host Galaxies}},
    year = {2008},
    journal = {The Astrophysical Journal},
    author = {Simmons, B. D. and Urry, C. M.},
    number = {2},
    month = {8},
    pages = {644--658},
    volume = {683},
    publisher = {IOP Publishing},
    url = {http://stacks.iop.org/0004-637X/683/i=2/a=644},
    doi = {10.1086/589827},
    issn = {0004-637X}
}

@article{astropy_2,
    title = {{The Astropy Project: Building an inclusive, open-science project and status of the v2.0 core package}},
    year = {2018},
    journal = {The Astronomical Journal},
    author = {Price-Whelan, A. M. and Sip{\H{o}}cz, B. M. and G{\"{u}}nther, H. M. and Lim, P. L. and Crawford, S. M. and Conseil, S. and Shupe, D. L. and Craig, M. W. and Dencheva, N. and Ginsburg, A. and VanderPlas, J. T. and Bradley, L. D. and P{\'{e}}rez-Su{\'{a}}rez, D. and de Val-Borro, M. and Aldcroft, T. L. and Cruz, K. L. and Robitaille, T. P. and Tollerud, E. J. and Ardelean, C. and Babej, T. and Bach, Y. P. and Bachetti, M. and Bakanov, A. V. and Bamford, S. P. and Barentsen, G. and Barmby, P. and Baumbach, A. and Berry, K. L. and Biscani, F. and Boquien, M. and Bostroem, K. A. and Bouma, L. G. and Brammer, G. B. and Bray, E. M. and Breytenbach, H. and Buddelmeijer, H. and Burke, D. J. and Calderone, G. and Rodr{\'{i}}guez, J. L. Cano and Cara, M. and Cardoso, J. V. M. and Cheedella, S. and Copin, Y. and Corrales, L. and Crichton, D. and D’Avella, D. and Deil, C. and Depagne, É. and Dietrich, J. P. and Donath, A. and Droettboom, M. and Earl, N. and Erben, T. and Fabbro, S. and Ferreira, L. A. and Finethy, T. and Fox, R. T. and Garrison, L. H. and Gibbons, S. L. J. and Goldstein, D. A. and Gommers, R. and Greco, J. P. and Greenfield, P. and Groener, A. M. and Grollier, F. and Hagen, A. and Hirst, P. and Homeier, D. and Horton, A. J. and Hosseinzadeh, G. and Hu, L. and Hunkeler, J. S. and Ivezi{\'{c}}, Ž. and Jain, A. and Jenness, T. and Kanarek, G. and Kendrew, S. and Kern, N. S. and Kerzendorf, W. E. and Khvalko, A. and King, J. and Kirkby, D. and Kulkarni, A. M. and Kumar, A. and Lee, A. and Lenz, D. and Littlefair, S. P. and Ma, Z. and Macleod, D. M. and Mastropietro, M. and McCully, C. and Montagnac, S. and Morris, B. M. and Mueller, M. and Mumford, S. J. and Muna, D. and Murphy, N. A. and Nelson, S. and Nguyen, G. H. and Ninan, J. P. and N{\"{o}}the, M. and Ogaz, S. and Oh, S. and Parejko, J. K. and Parley, N. and Pascual, S. and Patil, R. and Patil, A. A. and Plunkett, A. L. and Prochaska, J. X. and Rastogi, T. and Janga, V. Reddy and Sabater, J. and Sakurikar, P. and Seifert, M. and Sherbert, L. E. and Sherwood-Taylor, H. and Shih, A. Y. and Sick, J. and Silbiger, M. T. and Singanamalla, S. and Singer, L. P. and Sladen, P. H. and Sooley, K. A. and Sornarajah, S. and Streicher, O. and Teuben, P. and Thomas, S. W. and Tremblay, G. R. and Turner, J. E. H. and Terr{\'{o}}n, V. and Kerkwijk, M. H. van and de la Vega, A. and Watkins, L. L. and Weaver, B. A. and Whitmore, J. B. and Woillez, J. and Zabalza, V.},
    number = {3},
    month = {1},
    pages = {123},
    volume = {156},
    publisher = {American Astronomical Society},
    url = {https://arxiv.org/abs/1801.02634v2},
    doi = {10.3847/1538-3881/aabc4f},
    issn = {00046256},
    arxivId = {1801.02634}
}

@inproceedings{euclid,
    title = {{The Euclid mission design}},
    year = {2016},
    booktitle = {Proceedings of SPIE Astronomical Telescopes + Instrumentation},
    author = {Racca, Giuseppe D. and Laureijs, René and Stagnaro, Luca and Salvignol, Jean-Christophe and Lorenzo Alvarez, José and Saavedra Criado, Gonzalo and Gaspar Venancio, Luis and Short, Alex and Strada, Paolo and B{\"{o}}nke, Tobias and Colombo, Cyril and Calvi, Adriano and Maiorano, Elena and Piersanti, Osvaldo and Prezelus, Sylvain and Rosato, Pierluigi and Pinel, Jacques and Rozemeijer, Hans and Lesna, Valentina and Musi, Paolo and Sias, Marco and Anselmi, Alberto and Cazaubiel, Vincent and Vaillon, Ludovic and Mellier, Yannick and Amiaux, Jérôme and Berth{\'{e}}, Michel and Sauvage, Marc and Azzollini, Ruyman and Cropper, Mark and Pottinger, Sabrina and Jahnke, Knud and Ealet, Anne and Maciaszek, Thierry and Pasian, Fabio and Zacchei, Andrea and Scaramella, Roberto and Hoar, John and Kohley, Ralf and Vavrek, Roland and Rudolph, Andreas and Schmidt, Micha},
    editor = {MacEwen, Howard A. and Fazio, Giovanni G. and Lystrup, Makenzie and Batalha, Natalie and Siegler, Nicholas and Tong, Edward C.},
    month = {7},
    pages = {99040O},
    url = {http://proceedings.spiedigitallibrary.org/proceeding.aspx?doi=10.1117/12.2230762},
    doi = {10.1117/12.2230762}
}

@article{hsc_pipeline,
    title = {{The Hyper Suprime-Cam software pipeline}},
    year = {2018},
    journal = {Publications of the Astronomical Society of Japan},
    author = {Bosch, James and Armstrong, Robert and Bickerton, Steven and Furusawa, Hisanori and Ikeda, Hiroyuki and Koike, Michitaro and Lupton, Robert and Mineo, Sogo and Price, Paul and Takata, Tadafumi and Tanaka, Masayuki and Yasuda, Naoki and AlSayyad, Yusra and Becker, Andrew C and Coulton, William and Coupon, Jean and Garmilla, Jose and Huang, Song and Krughoff, K Simon and Lang, Dustin and Leauthaud, Alexie and Lim, Kian-Tat and Lust, Nate B and MacArthur, Lauren A and Mandelbaum, Rachel and Miyatake, Hironao and Miyazaki, Satoshi and Murata, Ryoma and More, Surhud and Okura, Yuki and Owen, Russell and Swinbank, John D and Strauss, Michael A and Yamada, Yoshihiko and Yamanoi, Hitomi},
    number = {SP1},
    month = {1},
    volume = {70},
    doi = {10.1093/pasj/psx080},
    issn = {0004-6264}
}

@article{tanaka_22,
    title = {{Where’s Swimmy?: Mining unique color features buried in galaxies by deep anomaly detection using Subaru Hyper Suprime-Cam data}},
    year = {2022},
    journal = {Publications of the Astronomical Society of Japan},
    author = {Tanaka, Takumi S and Shimakawa, Rhythm and Shimasaku, Kazuhiro and Toba, Yoshiki and Kashikawa, Nobunari and Tanaka, Masayuki and Inoue, Akio K},
    number = {1},
    month = {2},
    pages = {1--23},
    volume = {74},
    doi = {10.1093/pasj/psab105},
    issn = {0004-6264}
}

@article{ngrst,
    title = {{Wide-Field InfraRed Survey Telescope-Astrophysics Focused Telescope Assets WFIRST-AFTA Final Report}},
    year = {2013},
    author = {Spergel, D. and Gehrels, N. and Breckinridge, J. and Donahue, M. and Dressler, A. and Gaudi, B. S. and Greene, T. and Guyon, O. and Hirata, C. and Kalirai, J. and Kasdin, N. J. and Moos, W. and Perlmutter, S. and Postman, M. and Rauscher, B. and Rhodes, J. and Wang, Y. and Weinberg, D. and Centrella, J. and Traub, W. and Baltay, C. and Colbert, J. and Bennett, D. and Kiessling, A. and Macintosh, B. and Merten, J. and Mortonson, M. and Penny, M. and Rozo, E. and Savransky, D. and Stapelfeldt, K. and Zu, Y. and Baker, C. and Cheng, E. and Content, D. and Dooley, J. and Foote, M. and Goullioud, R. and Grady, K. and Jackson, C. and Kruk, J. and Levine, M. and Melton, M. and Peddie, C. and Ruffa, J. and Shaklan, S.},
    journal = {arXiv e-prints},
    eprint = {1305.5422},
    archivePrefix = {arXiv},
    url = {http://arxiv.org/abs/1305.5422},
    arxivId = {1305.5422}
}

@article{pozzetti_10,
    title = {{zCOSMOS – 10k-bright spectroscopic sample}},
    year = {2010},
    journal = {Astronomy {\&} Astrophysics},
    author = {Pozzetti, L. and Bolzonella, M. and Zucca, E. and Zamorani, G. and Lilly, S. and Renzini, A. and Moresco, M. and Mignoli, M. and Cassata, P. and Tasca, L. and Lamareille, F. and Maier, C. and Meneux, B. and Halliday, C. and Oesch, P. and Vergani, D. and Caputi, K. and Kova{\v{c}}, K. and Cimatti, A. and Cucciati, O. and Iovino, A. and Peng, Y. and Carollo, M. and Contini, T. and Kneib, J.-P. and Le F{\'{e}}vre, O. and Mainieri, V. and Scodeggio, M. and Bardelli, S. and Bongiorno, A. and Coppa, G. and de la Torre, S. and de Ravel, L. and Franzetti, P. and Garilli, B. and Kampczyk, P. and Knobel, C. and Le Borgne, J.-F. and Le Brun, V. and Pell{\`{o}}, R. and Perez Montero, E. and Ricciardelli, E. and Silverman, J. D. and Tanaka, M. and Tresse, L. and Abbas, U. and Bottini, D. and Cappi, A. and Guzzo, L. and Koekemoer, A. M. and Leauthaud, A. and Maccagni, D. and Marinoni, C. and McCracken, H. J. and Memeo, P. and Porciani, C. and Scaramella, R. and Scarlata, C. and Scoville, N.},
    month = {11},
    pages = {A13},
    volume = {523},
    url = {http://www.aanda.org/10.1051/0004-6361/200913020},
    doi = {10.1051/0004-6361/200913020},
    issn = {0004-6361}
}

@article{vanderWel20143D-HST+CANDELS:3,
    title = {{3D-HST+CANDELS: THE EVOLUTION OF THE GALAXY SIZE-MASS DISTRIBUTION SINCE z = 3}},
    year = {2014},
    journal = {The Astrophysical Journal},
    author = {van der Wel, A. and Franx, M. and van Dokkum, P. G. and Skelton, R. E. and Momcheva, I. G. and Whitaker, K. E. and Brammer, G. B. and Bell, E. F. and Rix, H.-W. and Wuyts, S. and Ferguson, H. C. and Holden, B. P. and Barro, G. and Koekemoer, A. M. and Chang, Yu-Yen and McGrath, E. J. and H{\"{a}}u{\ss}ler, B. and Dekel, A. and Behroozi, P. and Fumagalli, M. and Leja, J. and Lundgren, B. F. and Maseda, M. V. and Nelson, E. J. and Wake, D. A. and Patel, S. G. and Labb{\'{e}}, I. and Faber, S. M. and Grogin, N. A. and Kocevski, D. D.},
    number = {1},
    month = {5},
    pages = {28},
    volume = {788},
    url = {https://iopscience.iop.org/article/10.1088/0004-637X/788/1/28},
    doi = {10.1088/0004-637X/788/1/28},
    issn = {0004-637X}
}

@article{csst,
   title={Introduction to the Chinese Space Station Survey Telescope (CSST)},
   volume={69},
   ISSN={1869-1927},
   url={http://dx.doi.org/10.1007/s11433-025-2809-0},
   DOI={10.1007/s11433-025-2809-0},
   number={3},
   journal={Science China Physics, Mechanics \& Astronomy},
   publisher={Springer Science and Business Media LLC},
   author={{CSST Collaboration} and Gong, Yan and Miao, Haitao and Zhan, Hu and Li, Zhao-Yu and Shangguan, Jinyi and Li, Haining and Liu, Chao and Chen, Xuefei and Yuan, Haibo and Zhou, Jilin and Liu, Hui-Gen and Yu, Cong and Ji, Jianghui and Qi, Zhaoxiang and Liu, Jiacheng and Dai, Zigao and Wang, Xiaofeng and Zheng, Zhenya and Hao, Lei and Dou, Jiangpei and Ao, Yiping and Lin, Zhenhui and Zhang, Kun and Wang, Wei and Sun, Guotong and Li, Ran and Li, Guoliang and Xu, Youhua and Li, Xinfeng and Li, Shengyang and Wu, Peng and Zhang, Jiuxing and Wang, Bo and Bai, Jinming and Cai, Yi-Fu and Cai, Zheng and Cao, Jie and Chan, Kwan Chuen and Chang, Jin and Chen, Xiaodian and Chen, Xuelei and Chen, Yuqin and Chen, Yun and Cui, Wei and Dong, Subo and Du, Pu and Duan, Wenying and Fan, Junhui and Fan, LuLu and Fan, Zhou and Fan, Zuhui and Fang, Taotao and Fu, Jianning and Fu, Liping and Fu, Zhensen and Gao, Jian and Gu, Shenghong and Gu, Yidong and Guo, Qi and Han, Zhanwen and Hu, Bin and Huang, Zhiqi and Ho, Luis C. and Jiang, Linhua and Jiang, Ning and Jing, Yipeng and Kang, Xi and Kong, Xu and Li, Cheng and Li, Chengyuan and Li, Di and Li, Jing and Li, Nan and Li, Yang A. and Liao, Shilong and Lin, Weipeng and Liu, Fengshan and Liu, Jifeng and Liu, Xiangkun and Liu, Zhuokai and Mao, Ruiqing and Mao, Shude and Meng, Xianmin and Pang, Xiaoying and Peng, Xiyan and Peng, Yingjie and Shan, Huanyuan and Shen, Juntai and Shen, Shiyin and Shen, Zhiqiang and Shi, Sheng-Cai and Shi, Yong and Tan, Siyuan and Tian, Hao and Wang, Jianmin and Wang, Jun-Xian and Wang, Xin and Wang, Yuting and Wu, Hong and Wu, Jingwen and Wu, Xuebing and Xu, Chun and Xue, Xiang-Xiang and Xue, Yongquan and Yang, Ji and Yang, Xiaohu and Yao, Qijun and Yuan, Fangting and Yuan, Zhen and Zhang, Jun and Zhang, Pengjie and Zhang, Tianmeng and Zhang, Wei and Zhang, Xin and Zhao, Gang and Zhao, Gongbo and Zhong, Hongen and Zhong, Jing and Zhou, Liyong and Zhu, Wei and Zu, Ying},
   year={2026},
   month=Jan }

@misc{akeson2019widefieldinfraredsurvey,
      title={The Wide Field Infrared Survey Telescope: 100 Hubbles for the 2020s}, 
      author={Rachel Akeson and Lee Armus and Etienne Bachelet and Vanessa Bailey and Lisa Bartusek and Andrea Bellini and Dominic Benford and David Bennett and Aparna Bhattacharya and Ralph Bohlin and Martha Boyer and Valerio Bozza and Geoffrey Bryden and Sebastiano Calchi Novati and Kenneth Carpenter and Stefano Casertano and Ami Choi and David Content and Pratika Dayal and Alan Dressler and Olivier Doré and S. Michael Fall and Xiaohui Fan and Xiao Fang and Alexei Filippenko and Steven Finkelstein and Ryan Foley and Steven Furlanetto and Jason Kalirai and B. Scott Gaudi and Karoline Gilbert and Julien Girard and Kevin Grady and Jenny Greene and Puragra Guhathakurta and Chen Heinrich and Shoubaneh Hemmati and David Hendel and Calen Henderson and Thomas Henning and Christopher Hirata and Shirley Ho and Eric Huff and Anne Hutter and Rolf Jansen and Saurabh Jha and Samson Johnson and David Jones and Jeremy Kasdin and Patrick Kelly and Robert Kirshner and Anton Koekemoer and Jeffrey Kruk and Nikole Lewis and Bruce Macintosh and Piero Madau and Sangeeta Malhotra and Kaisey Mandel and Elena Massara and Daniel Masters and Julie McEnery and Kristen McQuinn and Peter Melchior and Mark Melton and Bertrand Mennesson and Molly Peeples and Matthew Penny and Saul Perlmutter and Alice Pisani and Andrés Plazas and Radek Poleski and Marc Postman and Clément Ranc and Bernard Rauscher and Armin Rest and Aki Roberge and Brant Robertson and Steven Rodney and James Rhoads and Jason Rhodes and Russell Ryan Jr. and Kailash Sahu and David Sand and Dan Scolnic and Anil Seth and Yossi Shvartzvald and Karelle Siellez and Arfon Smith and David Spergel and Keivan Stassun and Rachel Street and Louis-Gregory Strolger and Alexander Szalay and John Trauger and M. A. Troxel and Margaret Turnbull and Roeland van der Marel and Anja von der Linden and Yun Wang and David Weinberg and Benjamin Williams and Rogier Windhorst and Edward Wollack and Hao-Yi Wu and Jennifer Yee and Neil Zimmerman},
      year={2019},
      eprint={1902.05569},
      archivePrefix={arXiv},
      primaryClass={astro-ph.IM},
      url={https://arxiv.org/abs/1902.05569}, 
}

@misc{laureijs2011eucliddefinitionstudyreport,
      title={Euclid Definition Study Report}, 
      author={R. Laureijs and J. Amiaux and S. Arduini and J. -L. Auguères and J. Brinchmann and R. Cole and M. Cropper and C. Dabin and L. Duvet and A. Ealet and B. Garilli and P. Gondoin and L. Guzzo and J. Hoar and H. Hoekstra and R. Holmes and T. Kitching and T. Maciaszek and Y. Mellier and F. Pasian and W. Percival and J. Rhodes and G. Saavedra Criado and M. Sauvage and R. Scaramella and L. Valenziano and S. Warren and R. Bender and F. Castander and A. Cimatti and O. Le Fèvre and H. Kurki-Suonio and M. Levi and P. Lilje and G. Meylan and R. Nichol and K. Pedersen and V. Popa and R. Rebolo Lopez and H. -W. Rix and H. Rottgering and W. Zeilinger and F. Grupp and P. Hudelot and R. Massey and M. Meneghetti and L. Miller and S. Paltani and S. Paulin-Henriksson and S. Pires and C. Saxton and T. Schrabback and G. Seidel and J. Walsh and N. Aghanim and L. Amendola and J. Bartlett and C. Baccigalupi and J. -P. Beaulieu and K. Benabed and J. -G. Cuby and D. Elbaz and P. Fosalba and G. Gavazzi and A. Helmi and I. Hook and M. Irwin and J. -P. Kneib and M. Kunz and F. Mannucci and L. Moscardini and C. Tao and R. Teyssier and J. Weller and G. Zamorani and M. R. Zapatero Osorio and O. Boulade and J. J. Foumond and A. Di Giorgio and P. Guttridge and A. James and M. Kemp and J. Martignac and A. Spencer and D. Walton and T. Blumchen and C. Bonoli and F. Bortoletto and C. Cerna and L. Corcione and C. Fabron and K. Jahnke and S. Ligori and F. Madrid and L. Martin and G. Morgante and T. Pamplona and E. Prieto and M. Riva and R. Toledo and M. Trifoglio and F. Zerbi and F. Abdalla and M. Douspis and C. Grenet and S. Borgani and R. Bouwens and F. Courbin and J. -M. Delouis and P. Dubath and A. Fontana and M. Frailis and A. Grazian and J. Koppenhofer and O. Mansutti and M. Melchior and M. Mignoli and J. Mohr and C. Neissner and K. Noddle and M. Poncet and M. Scodeggio and S. Serrano and N. Shane and J. -L. Starck and C. Surace and A. Taylor and G. Verdoes-Kleijn and C. Vuerli and O. R. Williams and A. Zacchei and B. Altieri and I. Escudero Sanz and R. Kohley and T. Oosterbroek and P. Astier and D. Bacon and S. Bardelli and C. Baugh and F. Bellagamba and C. Benoist and D. Bianchi and A. Biviano and E. Branchini and C. Carbone and V. Cardone and D. Clements and S. Colombi and C. Conselice and G. Cresci and N. Deacon and J. Dunlop and C. Fedeli and F. Fontanot and P. Franzetti and C. Giocoli and J. Garcia-Bellido and J. Gow and A. Heavens and P. Hewett and C. Heymans and A. Holland and Z. Huang and O. Ilbert and B. Joachimi and E. Jennins and E. Kerins and A. Kiessling and D. Kirk and R. Kotak and O. Krause and O. Lahav and F. van Leeuwen and J. Lesgourgues and M. Lombardi and M. Magliocchetti and K. Maguire and E. Majerotto and R. Maoli and F. Marulli and S. Maurogordato and H. McCracken and R. McLure and A. Melchiorri and A. Merson and M. Moresco and M. Nonino and P. Norberg and J. Peacock and R. Pello and M. Penny and V. Pettorino and C. Di Porto and L. Pozzetti and C. Quercellini and M. Radovich and A. Rassat and N. Roche and S. Ronayette and E. Rossetti and B. Sartoris and P. Schneider and E. Semboloni and S. Serjeant and F. Simpson and C. Skordis and G. Smadja and S. Smartt and P. Spano and S. Spiro and M. Sullivan and A. Tilquin and R. Trotta and L. Verde and Y. Wang and G. Williger and G. Zhao and J. Zoubian and E. Zucca},
      year={2011},
      eprint={1110.3193},
      archivePrefix={arXiv},
      primaryClass={astro-ph.CO},
      url={https://arxiv.org/abs/1110.3193}, 
}

@article{schawinski_14,
    author = {Schawinski, Kevin and Urry, C. Megan and Simmons, Brooke D. and Fortson, Lucy and Kaviraj, Sugata and Keel, William C. and Lintott, Chris J. and Masters, Karen L. and Nichol, Robert C. and Sarzi, Marc and Skibba, Ramin and Treister, Ezequiel and Willett, Kyle W. and Wong, O. Ivy and Yi, Sukyoung K.},
    title = {The green valley is a red herring: Galaxy Zoo reveals two evolutionary pathways towards quenching of star formation in early- and late-type galaxies},
    journal = {Monthly Notices of the Royal Astronomical Society},
    volume = {440},
    number = {1},
    pages = {889-907},
    year = {2014},
    month = {05},
    issn = {0035-8711},
    doi = {10.1093/mnras/stu327},
    url = {https://doi.org/10.1093/mnras/stu327},
    eprint = {https://academic.oup.com/mnras/article-pdf/440/1/889/9388724/stu327.pdf},
}

@article{Conselice_2014,
   title={The Evolution of Galaxy Structure Over Cosmic Time},
   volume={52},
   ISSN={1545-4282},
   url={http://dx.doi.org/10.1146/annurev-astro-081913-040037},
   DOI={10.1146/annurev-astro-081913-040037},
   number={1},
   journal={Annual Review of Astronomy and Astrophysics},
   publisher={Annual Reviews},
   author={Conselice, Christopher J.},
   year={2014},
   month=Aug, pages={291–337} }

@article{Sharma2020ApplicationClassification,
    title = {{Application of convolutional neural networks for stellar spectral classification}},
    year = {2020},
    journal = {Monthly Notices of the Royal Astronomical Society},
    author = {Sharma, Kaushal and Kembhavi, Ajit and Kembhavi, Aniruddha and Sivarani, T. and Abraham, Sheelu and Vaghmare, Kaustubh},
    number = {2},
    month = {1},
    pages = {2280--2300},
    volume = {491},
    publisher = {Oxford Academic},
    url = {https://academic.oup.com/mnras/article/491/2/2280/5613963},
    doi = {10.1093/MNRAS/STZ3100},
    issn = {0035-8711},
    arxivId = {1909.05459}
}

@article{Dimauro2022CoincidenceGrowth,
    title = {{Coincidence between morphology and star-formation activity through cosmic time: The impact of the bulge growth}},
    year = {2022},
    journal = {Monthly Notices of the Royal Astronomical Society},
    author = {Dimauro, Paola and Daddi, Emanuele and Shankar, Francesco and Cattaneo, Andrea and Huertas-Company, Marc and Bernardi, Mariangela and Caro, Fernando and Dupke, Renato and H{\"{a}}u{\ss}ler, Boris and Johnston, Johnston and Evelyn, Evelyn and Cortesi, Arianna and Mei, Simona and Peletier, Reynier},
    url = {https://academic.oup.com/mnras/advance-article/doi/10.1093/mnras/stac884/6563709},
    doi = {10.1093/MNRAS/STAC884}
}

@article{Tuccillo2018DeepFitting,
    title = {{Deep learning for galaxy surface brightness profile fitting}},
    year = {2018},
    journal = {Monthly Notices of the Royal Astronomical Society},
    author = {Tuccillo, D and Huertas-Company, M and Decenci{\`{e}}re, E and Velasco-Forero, S and Dom{\'{i}}nguez S{\'{a}}nchez, H and Dimauro, P},
    number = {1},
    month = {3},
    pages = {894--909},
    volume = {475},
    url = {http://academic.oup.com/mnras/article/475/1/894/4725057},
    doi = {10.1093/mnras/stx3186},
    issn = {0035-8711}
}

@article{Srivastava2014Dropout:Overfitting,
    title = {{Dropout: A simple way to prevent neural networks from overfitting}},
    year = {2014},
    journal = {Journal of Machine Learning Research},
    author = {Srivastava, Nitish and Hinton, Geoffrey and Krizhevsky, Alex and Sutskever, Ilya and Salakhutdinov, Ruslan},
    pages = {1929--1958},
    volume = {15},
    doi = {10.5555/2627435},
    issn = {15337928}
}

@article{Bender1992DynamicallyProperties,
    title = {{Dynamically hot galaxies. I - Structural properties}},
    year = {1992},
    journal = {The Astrophysical Journal},
    author = {Bender, Ralf and Burstein, David and Faber, S. M.},
    month = {11},
    pages = {462},
    volume = {399},
    url = {http://adsabs.harvard.edu/doi/10.1086/171940},
    doi = {10.1086/171940},
    issn = {0004-637X}
}

@article{Cheng2021GalaxyNetworks,
    title = {{Galaxy morphological classification catalogue of the Dark Energy Survey Year 3 data with convolutional neural networks}},
    year = {2021},
    journal = {Monthly Notices of the Royal Astronomical Society},
    author = {Cheng, Ting-Yun and Conselice, Christopher J and Arag{\'{o}}n-Salamanca, Alfonso and Aguena, M and Allam, S and Andrade-Oliveira, F and Annis, J and Bluck, A F L and Brooks, D and Burke, D L and Carrasco Kind, M and Carretero, J and Choi, A and Costanzi, M and da Costa, L N and Pereira, M E S and De Vicente, J and Diehl, H T and Drlica-Wagner, A and Eckert, K and Everett, S and Evrard, A E and Ferrero, I and Fosalba, P and Frieman, J and Garc{\'{i}}a-Bellido, J and Gerdes, D W and Giannantonio, T and Gruen, D and Gruendl, R A and Gschwend, J and Gutierrez, G and Hinton, S R and Hollowood, D L and Honscheid, K and James, D J and Krause, E and Kuehn, K and Kuropatkin, N and Lahav, O and Maia, M A G and March, M and Menanteau, F and Miquel, R and Morgan, R and Paz-Chinch{\'{o}}n, F and Pieres, A and Plazas Malag{\'{o}}n, A A and Roodman, A and Sanchez, E and Scarpine, V and Serrano, S and Sevilla-Noarbe, I and Smith, M and Soares-Santos, M and Suchyta, E and Swanson, M E C and Tarle, G and Thomas, D and To, C},
    number = {3},
    month = {9},
    pages = {4425--4444},
    volume = {507},
    url = {https://academic.oup.com/mnras/article/507/3/4425/6327560},
    doi = {10.1093/mnras/stab2142},
    issn = {0035-8711}
}

@article{Walmsley2020GalaxyLearning,
    title = {{Galaxy Zoo: probabilistic morphology through Bayesian CNNs and active learning}},
    year = {2020},
    journal = {Monthly Notices of the Royal Astronomical Society},
    author = {Walmsley, Mike and Smith, Lewis and Lintott, Chris and Gal, Yarin and Bamford, Steven and Dickinson, Hugh and Fortson, Lucy and Kruk, Sandor and Masters, Karen and Scarlata, Claudia and Simmons, Brooke and Smethurst, Rebecca and Wright, Darryl},
    number = {2},
    month = {1},
    pages = {1554--1574},
    volume = {491},
    url = {https://academic.oup.com/mnras/article/491/2/1554/5583078},
    doi = {10.1093/mnras/stz2816},
    issn = {0035-8711}
}

@article{Shallue2018IdentifyingKepler-90,
    title = {{Identifying Exoplanets with Deep Learning: A Five-planet Resonant Chain around Kepler-80 and an Eighth Planet around Kepler-90}},
    year = {2018},
    journal = {The Astronomical Journal},
    author = {Shallue, Christopher J. and Vanderburg, Andrew},
    number = {2},
    month = {1},
    pages = {94},
    volume = {155},
    publisher = {American Astronomical Society},
    url = {https://ui.adsabs.harvard.edu/abs/2018AJ....155...94S/abstract},
    doi = {10.3847/1538-3881/aa9e09},
    issn = {00046256},
    arxivId = {1712.05044}
}

@article{Tarsitano2022ImageLearning,
    title = {{Image feature extraction and galaxy classification: a novel and efficient approach with automated machine learning}},
    year = {2022},
    journal = {Monthly Notices of the Royal Astronomical Society},
    author = {Tarsitano, F and Bruderer, C and Schawinski, K and Hartley, W G},
    number = {3},
    month = {2},
    pages = {3330--3338},
    volume = {511},
    doi = {10.1093/mnras/stac233},
    issn = {0035-8711}
}

@article{Birrer2018Lenstronomy:Package,
    title = {{lenstronomy: Multi-purpose gravitational lens modelling software package}},
    year = {2018},
    journal = {Physics of the Dark Universe},
    author = {Birrer, Simon and Amara, Adam},
    month = {12},
    pages = {189--201},
    volume = {22},
    doi = {10.1016/j.dark.2018.11.002},
    issn = {22126864}
}

@article{Huertas-Company2016MassCANDELS,
    title = {{Mass assembly and morphological transformations since z ∼ 3 from CANDELS}},
    year = {2016},
    journal = {Monthly Notices of the Royal Astronomical Society},
    author = {Huertas-Company, M. and Bernardi, M. and P{\'{e}}rez-Gonz{\'{a}}lez, P. G. and Ashby, M. L. N. and Barro, G. and Conselice, C. and Daddi, E. and Dekel, A. and Dimauro, P. and Faber, S. M. and Grogin, N. A. and Kartaltepe, J. S. and Kocevski, D. D. and Koekemoer, A. M. and Koo, D. C. and Mei, S. and Shankar, F.},
    number = {4},
    month = {11},
    pages = {4495--4516},
    volume = {462},
    url = {https://academic.oup.com/mnras/article-lookup/doi/10.1093/mnras/stw1866},
    doi = {10.1093/mnras/stw1866},
    issn = {0035-8711}
}

@article{Hausen2020MorpheusData,
    title = {{Morpheus : A Deep Learning Framework for the Pixel-level Analysis of Astronomical Image Data}},
    year = {2020},
    journal = {The Astrophysical Journal Supplement Series},
    author = {Hausen, Ryan and Robertson, Brant E.},
    number = {1},
    month = {5},
    pages = {20},
    volume = {248},
    url = {https://iopscience.iop.org/article/10.3847/1538-4365/ab8868},
    doi = {10.3847/1538-4365/ab8868},
    issn = {1538-4365}
}

@article{Weinmann2006PropertiesMass,
    title = {{Properties of galaxy groups in the Sloan Digital Sky Survey -- I. The dependence of colour, star formation and morphology on halo mass}},
    year = {2006},
    journal = {Monthly Notices of the Royal Astronomical Society},
    author = {Weinmann, S. M. and van den Bosch, F. C. and Yang, X. and Mo, H. J.},
    number = {1},
    month = {2},
    pages = {2--28},
    volume = {366},
    doi = {10.1111/j.1365-2966.2005.09865.x},
    issn = {0035-8711}
}

@article{Vega-Ferrero2021PushingSurvey,
    title = {{Pushing automated morphological classifications to their limits with the Dark Energy Survey}},
    year = {2021},
    journal = {Monthly Notices of the Royal Astronomical Society},
    author = {Vega-Ferrero, J and Dom{\'{i}}nguez S{\'{a}}nchez, H and Bernardi, M and Huertas-Company, M and Morgan, R and Margalef, B and Aguena, M and Allam, S and Annis, J and Avila, S and Bacon, D and Bertin, E and Brooks, D and Carnero Rosell, A and Carrasco Kind, M and Carretero, J and Choi, A and Conselice, C and Costanzi, M and da Costa, L N and Pereira, M E S and De Vicente, J and Desai, S and Ferrero, I and Fosalba, P and Frieman, J and Garc{\'{i}}a-Bellido, J and Gruen, D and Gruendl, R A and Gschwend, J and Gutierrez, G and Hartley, W G and Hinton, S R and Hollowood, D L and Honscheid, K and Hoyle, B and Jarvis, M and Kim, A G and Kuehn, K and Kuropatkin, N and Lima, M and Maia, M A G and Menanteau, F and Miquel, R and Ogando, R L C and Palmese, A and Paz-Chinch{\'{o}}n, F and Plazas, A A and Romer, A K and Sanchez, E and Scarpine, V and Schubnell, M and Serrano, S and Sevilla-Noarbe, I and Smith, M and Suchyta, E and Swanson, M E C and Tarle, G and Tarsitano, F and To, C and Tucker, D L and Varga, T N and Wilkinson, R D},
    number = {2},
    month = {7},
    pages = {1927--1943},
    volume = {506},
    url = {https://academic.oup.com/mnras/article/506/2/1927/6156625},
    doi = {10.1093/mnras/stab594},
    issn = {0035-8711}
}

@article{Natarajan2021QuasarNet:Holes,
    title = {{QuasarNet: A new research platform for the data-driven investigation of black holes}},
    year = {2021},
    journal = {arXiv preprint arXiv:2103.13932},
    author = {Natarajan, Priyamvada and Tang, Kwok Sun and Khochfar, Sadegh and Nord, Brian and Sigurdsson, Steinn and Tricot, Joe and Cappelluti, Nico and George, Daniel and Hidary, Jack},
    month = {3},
    url = {https://arxiv.org/abs/2103.13932v1},
    isbn = {2103.13932v1},
    doi = {10.48550/arxiv.2103.13932},
    arxivId = {2103.13932}
}

@article{Dieleman2015Rotation-invariantPrediction,
    title = {{Rotation-invariant convolutional neural networks for galaxy morphology prediction}},
    year = {2015},
    journal = {Monthly Notices of the Royal Astronomical Society},
    author = {Dieleman, Sander and Willett, Kyle W. and Dambre, Joni},
    number = {2},
    month = {6},
    pages = {1441--1459},
    volume = {450},
    url = {http://academic.oup.com/mnras/article/450/2/1441/979677/Rotationinvariant-convolutional-neural-networks},
    doi = {10.1093/mnras/stv632},
    issn = {1365-2966}
}

@article{Virtanen2020SciPyPython,
    title = {{SciPy 1.0: fundamental algorithms for scientific computing in Python}},
    year = {2020},
    journal = {Nature Methods},
    author = {Virtanen, Pauli and Gommers, Ralf and Oliphant, Travis E. and Haberland, Matt and Reddy, Tyler and Cournapeau, David and Burovski, Evgeni and Peterson, Pearu and Weckesser, Warren and Bright, Jonathan and van der Walt, Stéfan J. and Brett, Matthew and Wilson, Joshua and Millman, K. Jarrod and Mayorov, Nikolay and Nelson, Andrew R. J. and Jones, Eric and Kern, Robert and Larson, Eric and Carey, C J and Polat, İlhan and Feng, Yu and Moore, Eric W. and VanderPlas, Jake and Laxalde, Denis and Perktold, Josef and Cimrman, Robert and Henriksen, Ian and Quintero, E. A. and Harris, Charles R. and Archibald, Anne M. and Ribeiro, Antônio H. and Pedregosa, Fabian and van Mulbregt, Paul and Vijaykumar, Aditya and Bardelli, Alessandro Pietro and Rothberg, Alex and Hilboll, Andreas and Kloeckner, Andreas and Scopatz, Anthony and Lee, Antony and Rokem, Ariel and Woods, C. Nathan and Fulton, Chad and Masson, Charles and H{\"{a}}ggstr{\"{o}}m, Christian and Fitzgerald, Clark and Nicholson, David A. and Hagen, David R. and Pasechnik, Dmitrii V. and Olivetti, Emanuele and Martin, Eric and Wieser, Eric and Silva, Fabrice and Lenders, Felix and Wilhelm, Florian and Young, G. and Price, Gavin A. and Ingold, Gert-Ludwig and Allen, Gregory E. and Lee, Gregory R. and Audren, Hervé and Probst, Irvin and Dietrich, Jörg P. and Silterra, Jacob and Webber, James T and Slavi{\v{c}}, Janko and Nothman, Joel and Buchner, Johannes and Kulick, Johannes and Sch{\"{o}}nberger, Johannes L. and de Miranda Cardoso, José Vinícius and Reimer, Joscha and Harrington, Joseph and Rodr{\'{i}}guez, Juan Luis Cano and Nunez-Iglesias, Juan and Kuczynski, Justin and Tritz, Kevin and Thoma, Martin and Newville, Matthew and K{\"{u}}mmerer, Matthias and Bolingbroke, Maximilian and Tartre, Michael and Pak, Mikhail and Smith, Nathaniel J. and Nowaczyk, Nikolai and Shebanov, Nikolay and Pavlyk, Oleksandr and Brodtkorb, Per A. and Lee, Perry and McGibbon, Robert T. and Feldbauer, Roman and Lewis, Sam and Tygier, Sam and Sievert, Scott and Vigna, Sebastiano and Peterson, Stefan and More, Surhud and Pudlik, Tadeusz and Oshima, Takuya and Pingel, Thomas J. and Robitaille, Thomas P. and Spura, Thomas and Jones, Thouis R. and Cera, Tim and Leslie, Tim and Zito, Tiziano and Krauss, Tom and Upadhyay, Utkarsh and Halchenko, Yaroslav O. and V{\'{a}}zquez-Baeza, Yoshiki},
    number = {3},
    month = {3},
    pages = {261--272},
    volume = {17},
    doi = {10.1038/s41592-019-0686-2},
    issn = {1548-7091}
}

@article{vanderWel2008TheMass,
    title = {{The Dependence of Galaxy Morphology and Structure on Environment and Stellar Mass}},
    year = {2008},
    journal = {The Astrophysical Journal},
    author = {van der Wel, Arjen},
    number = {1},
    month = {3},
    pages = {L13-L16},
    volume = {675},
    doi = {10.1086/529432},
    issn = {0004-637X}
}

@article{Schawinski2007TheGalaxies,
    title = {{The Effect of Environment on the Ultraviolet Color‐Magnitude Relation of Early‐Type Galaxies}},
    year = {2007},
    journal = {The Astrophysical Journal Supplement Series},
    author = {Schawinski, K. and Kaviraj, S. and Khochfar, S. and Yoon, S.‐J. and Yi, S. K. and Deharveng, J.‐M. and Boselli, A. and Barlow, T. and Conrow, T. and Forster, K. and Friedman, P. G. and Martin, D. C. and Morrissey, P. and Neff, S. and Schiminovich, D. and Seibert, M. and Small, T. and Wyder, T. and Bianchi, L. and Donas, J. and Heckman, T. and Lee, Y.‐W. and Madore, B. and Milliard, B. and Rich, R. M. and Szalay, A.},
    number = {2},
    month = {12},
    pages = {512--523},
    volume = {173},
    doi = {10.1086/516631},
    issn = {0067-0049}
}

@article{Kauffmann2004TheGalaxies,
    title = {{The environmental dependence of the relations between stellar mass, structure, star formation and nuclear activity in galaxies}},
    year = {2004},
    journal = {Monthly Notices of the Royal Astronomical Society},
    author = {Kauffmann, Guinevere and White, Simon D. M. and Heckman, Timothy M. and M{\'{e}}nard, Brice and Brinchmann, Jarle and Charlot, Stéphane and Tremonti, Christy and Brinkmann, Jon},
    number = {3},
    month = {8},
    pages = {713--731},
    volume = {353},
    doi = {10.1111/j.1365-2966.2004.08117.x},
    issn = {00358711}
}

@article{Bernardi2013TheProfile,
    title = {{The massive end of the luminosity and stellar mass functions: dependence on the fit to the light profile}},
    year = {2013},
    journal = {Monthly Notices of the Royal Astronomical Society},
    author = {Bernardi, M. and Meert, A. and Sheth, R. K. and Vikram, V. and Huertas-Company, M. and Mei, S. and Shankar, F.},
    number = {1},
    month = {11},
    pages = {697--704},
    volume = {436},
    url = {https://academic.oup.com/mnras/article-lookup/doi/10.1093/mnras/stt1607},
    doi = {10.1093/mnras/stt1607},
    issn = {0035-8711}
}

@article{Tremaine2002TheCorrelation,
    title = {{The Slope of the Black Hole Mass versus Velocity Dispersion Correlation}},
    year = {2002},
    journal = {The Astrophysical Journal},
    author = {Tremaine, Scott and Gebhardt, Karl and Bender, Ralf and Bower, Gary and Dressler, Alan and Faber, S. M. and Filippenko, Alexei V. and Green, Richard and Grillmair, Carl and Ho, Luis C. and Kormendy, John and Lauer, Tod R. and Magorrian, John and Pinkney, Jason and Richstone, Douglas},
    number = {2},
    month = {8},
    pages = {740--753},
    volume = {574},
    url = {https://iopscience.iop.org/article/10.1086/341002},
    doi = {10.1086/341002},
    issn = {0004-637X}
}

@article{yang_2026_cosmosweb,
      title={COSMOS-Web: A Multi-wavelength Morphological Catalog of ~780,000 Galaxies}, 
      author={Lilan Yang and Jeyhan S. Kartaltepe and Nicole E. Drakos and Andreas L. Faisst and Carter Flayhart and Maximilien Franco and Anton M. Koekemoer and Olivier Ilbert and Hollis B. Akins and Caitlin M. Casey and Xuheng Ding and Ali Hadi and Santosh Harish and Ronaldo Laishram and Daizhong Liu and Georgios E. Magdis and Felix Martinez III and Henry Joy McCracken and Louise Paquereau and Jason Rhodes and Brant E. Robertson and Marko Shuntov and Greta Toni},
      year={2026},
      eprint={2606.14869},
      archivePrefix={arXiv},
      journal={arXiv e-prints},
      primaryClass={astro-ph.GA},
      url={https://arxiv.org/abs/2606.14869}, 
}

@article{quilley_2025_euclid_q1,
      title={Euclid Quick Data Release (Q1). Exploring galaxy morphology across cosmic time through Sersic fits}, 
      author={{Euclid Collaboration} and L. Quilley and I. Damjanov and V. de Lapparent and A. Paulino-Afonso and H. Domínguez Sánchez and A. Ferré-Mateu and M. Huertas-Company and M. Kümmel and D. Delley and C. Spiniello and M. Baes and L. Wang and U. Kuchner and F. Tarsitano and R. Ragusa and M. Siudek and C. Tortora and N. Aghanim and B. Altieri and A. Amara and S. Andreon and N. Auricchio and H. Aussel and C. Baccigalupi and M. Baldi and A. Balestra and S. Bardelli and P. Battaglia and R. Bender and A. Biviano and A. Bonchi and D. Bonino and E. Branchini and M. Brescia and J. Brinchmann and S. Camera and G. Cañas-Herrera and V. Capobianco and C. Carbone and J. Carretero and S. Casas and F. J. Castander and M. Castellano and G. Castignani and S. Cavuoti and K. C. Chambers and A. Cimatti and C. Colodro-Conde and G. Congedo and C. J. Conselice and L. Conversi and Y. Copin and A. Costille and F. Courbin and H. M. Courtois and M. Cropper and A. Da Silva and H. Degaudenzi and G. De Lucia and A. M. Di Giorgio and C. Dolding and H. Dole and C. A. J. Duncan and X. Dupac and S. Dusini and S. Escoffier and M. Fabricius and M. Farina and R. Farinelli and F. Faustini and S. Ferriol and P. Fosalba and S. Fotopoulou and M. Frailis and E. Franceschi and S. Galeotta and K. George and B. Gillis and C. Giocoli and J. Gracia-Carpio and B. R. Granett and A. Grazian and F. Grupp and S. V. H. Haugan and H. Hoekstra and W. Holmes and I. M. Hook and F. Hormuth and A. Hornstrup and P. Hudelot and K. Jahnke and M. Jhabvala and B. Joachimi and E. Keihänen and S. Kermiche and A. Kiessling and B. Kubik and M. Kunz and H. Kurki-Suonio and Q. Le Boulc'h and A. M. C. Le Brun and D. Le Mignant and P. Liebing and S. Ligori and P. B. Lilje and V. Lindholm and I. Lloro and G. Mainetti and D. Maino and E. Maiorano and O. Mansutti and S. Marcin and O. Marggraf and M. Martinelli and N. Martinet and F. Marulli and R. Massey and S. Maurogordato and H. J. McCracken and E. Medinaceli and S. Mei and M. Melchior and Y. Mellier and M. Meneghetti and E. Merlin and G. Meylan and A. Mora and M. Moresco and L. Moscardini and R. Nakajima and C. Neissner and S. -M. Niemi and J. W. Nightingale and C. Padilla and S. Paltani and F. Pasian and K. Pedersen and W. J. Percival and V. Pettorino and S. Pires and G. Polenta and M. Poncet and L. A. Popa and L. Pozzetti and F. Raison and R. Rebolo and A. Renzi and J. Rhodes and G. Riccio and E. Romelli and M. Roncarelli and R. Saglia and Z. Sakr and A. G. Sánchez and D. Sapone and B. Sartoris and J. A. Schewtschenko and P. Schneider and M. Scodeggio and A. Secroun and G. Seidel and M. Seiffert and S. Serrano and P. Simon and C. Sirignano and G. Sirri and L. Stanco and J. Steinwagner and P. Tallada-Crespí and A. N. Taylor and H. I. Teplitz and I. Tereno and N. Tessore and S. Toft and R. Toledo-Moreo and F. Torradeflot and I. Tutusaus and L. Valenziano and J. Valiviita and T. Vassallo and G. Verdoes Kleijn and A. Veropalumbo and Y. Wang and J. Weller and A. Zacchei and G. Zamorani and F. M. Zerbi and I. A. Zinchenko and E. Zucca and V. Allevato and M. Ballardini and M. Bolzonella and E. Bozzo and C. Burigana and R. Cabanac and A. Cappi and D. Di Ferdinando and J. A. Escartin Vigo and L. Gabarra and W. G. Hartley and J. Martín-Fleitas and S. Matthew and N. Mauri and R. B. Metcalf and A. Pezzotta and M. Pöntinen and C. Porciani and I. Risso and V. Scottez and M. Sereno and M. Tenti and M. Viel and M. Wiesmann and Y. Akrami and I. T. Andika and S. Anselmi and M. Archidiacono and F. Atrio-Barandela and C. Benoist and K. Benson and D. Bertacca and M. Bethermin and L. Bisigello and A. Blanchard and L. Blot and S. Borgani and M. L. Brown and S. Bruton and F. Buitrago and A. Calabro and B. Camacho Quevedo and F. Caro and C. S. Carvalho and T. Castro and Y. Charles and F. Cogato and T. Contini and A. R. Cooray and O. Cucciati and S. Davini and F. De Paolis and G. Desprez and A. Díaz-Sánchez and J. J. Diaz and S. Di Domizio and J. M. Diego and P. -A. Duc and A. Enia and Y. Fang and A. G. Ferrari and P. G. Ferreira and A. Finoguenov and A. Franco and K. Ganga and J. García-Bellido and T. Gasparetto and V. Gautard and R. Gavazzi and E. Gaztanaga and F. Giacomini and G. Gozaliasl and M. Guidi and C. M. Gutierrez and A. Hall and C. Hernández-Monteagudo and H. Hildebrandt and J. Hjorth and O. Ilbert and J. J. E. Kajava and Y. Kang and V. Kansal and D. Karagiannis and K. Kiiveri and C. C. Kirkpatrick and S. Kruk and L. Legrand and M. Lembo and F. Lepori and G. Leroy and G. F. Lesci and J. Lesgourgues and L. Leuzzi and T. I. Liaudat and A. Loureiro and J. Macias-Perez and G. Maggio and M. Magliocchetti and E. A. Magnier and C. Mancini and F. Mannucci and R. Maoli and C. J. A. P. Martins and L. Maurin and M. Miluzio and P. Monaco and C. Moretti and G. Morgante and K. Naidoo and A. Navarro-Alsina and S. Nesseris and F. Passalacqua and K. Paterson and L. Patrizii and A. Pisani and D. Potter and S. Quai and M. Radovich and P. -F. Rocci and S. Sacquegna and M. Sahlén and D. B. Sanders and E. Sarpa and C. Scarlata and J. Schaye and A. Schneider and D. Sciotti and E. Sellentin and F. Shankar and L. C. Smith and S. A. Stanford and K. Tanidis and G. Testera and R. Teyssier and S. Tosi and A. Troja and M. Tucci and C. Valieri and A. Venhola and D. Vergani and G. Verza and P. Vielzeuf and N. A. Walton and J. G. Sorce and E. Soubrie and D. Scott},
      year={2025},
      eprint={2503.15309},
      archivePrefix={arXiv},
      journal={arXiv e-prints},
      primaryClass={astro-ph.GA},
      url={https://arxiv.org/abs/2503.15309}, 
}

@article{merchant_2026_canucs,
      title={CANUCS/Technicolor Data Release 2: A Catalogue of Galaxy Structural Parameters in up to 29 HST+JWST bands and a Multi-Wavelength Exploration of the Galaxy Size-Mass Relation at $0.6 < z \leq 4$}, 
      author={Maya Merchant and Lamiya A. Mowla and Georgios E. Magdis and Adam Muzzin and Chris J. Willott and Roberto Abraham and Yoshihisa Asada and Maruša Bradač and Gabriel B. Brammer and Guillaume Desprez and Kartheik G. Iyer and Nicholas S. Martis and Gaël Noirot and Gregor Rihtaršič and Marcin Sawicki and Ghassan T. E. Sarrouh and Sunna Withers and Natalie Allen and Jacqueline Antwi-Danso and Westley Brown and Jon Judež and Danilo Marchesini and Rosa M. Mérida and Katherine Myers and Luke Robbins and Visal Sok},
      year={2026},
      eprint={2606.13781},
      archivePrefix={arXiv},
      journal={arXiv e-prints},
      primaryClass={astro-ph.GA},
      url={https://arxiv.org/abs/2606.13781}, 
}

@article{tanoglidis_2022_lsbg_bnn,
      title={Inferring Structural Parameters of Low-Surface-Brightness-Galaxies with Uncertainty Quantification using Bayesian Neural Networks}, 
      author={Dimitrios Tanoglidis and Aleksandra Ćiprijanović and Alex Drlica-Wagner},
      year={2022},
      eprint={2207.03471},
      archivePrefix={arXiv},
      journal={arXiv e-prints},
      primaryClass={astro-ph.IM},
      url={https://arxiv.org/abs/2207.03471}, 
}

@article{walmsley_2023_gz_desi,
      title={Galaxy Zoo DESI: Detailed Morphology Measurements for 8.7M Galaxies in the DESI Legacy Imaging Surveys}, 
      author={Mike Walmsley and Tobias Géron and Sandor Kruk and Anna M. M. Scaife and Chris Lintott and Karen L. Masters and James M. Dawson and Hugh Dickinson and Lucy Fortson and Izzy L. Garland and Kameswara Mantha and David O'Ryan and Jürgen Popp and Brooke Simmons and Elisabeth M. Baeten and Christine Macmillan},
      year={2023},
      eprint={2309.11425},
      archivePrefix={arXiv},
      journal={arXiv e-prints},
      primaryClass={astro-ph.GA},
      url={https://arxiv.org/abs/2309.11425}, 
}

@article{ting_2026_review,
      title={Deep Learning in Astrophysics}, 
      author={Yuan-Sen Ting},
      year={2026},
      eprint={2510.10713},
      archivePrefix={arXiv},
      journal={Annual Review of Astronomy and Astrophysics},
      volume={64},
      primaryClass={astro-ph.IM},
      doi={10.1146/annurev-astro-051024-021708},
      url={https://arxiv.org/abs/2510.10713}, 
}

@inproceedings{jaderberg_2015_stn,
      title={Spatial Transformer Networks},
      author={Jaderberg, Max and Simonyan, Karen and Zisserman, Andrew and Kavukcuoglu, Koray},
      booktitle={Advances in Neural Information Processing Systems 28},
      year={2015},
      eprint={1506.02025},
      archivePrefix={arXiv},
      url={https://arxiv.org/abs/1506.02025}
}

@article{lecun_1998_cnn,
      title={Gradient-based learning applied to document recognition},
      author={{LeCun}, Y. and Bottou, L. and Bengio, Y. and Haffner, P.},
      journal={Proceedings of the IEEE},
      volume={86},
      number={11},
      pages={2278--2324},
      year={1998},
      doi={10.1109/5.726791}
}

@article{Huertas-Company2023TheSurveys,
  title = {{The Dawes Review 10: The impact of deep learning for the analysis of galaxy surveys}},
  year = {2023},
  journal = {Publications of the Astronomical Society of Australia},
  author = {Huertas-Company, M. and Lanusse, F.},
  month = {1},
  pages = {e001},
  volume = {40},
  doi = {10.1017/pasa.2022.55},
  issn = {1323-3580},
}

\end{document}